\documentclass[superscriptaddress,onecolumn,letterpaper,accepted=2020-07-09]{quantumarticle} %{revtex4-1} % letterpaper,accepted=2020-07-09]{quantumarticle}
\pdfoutput=1 % added for quantum
\usepackage[numbers, sort&compress, merge]{natbib}
\usepackage{graphicx}
\usepackage{amsthm}
\usepackage{thmtools}
\usepackage{mathtools}
\usepackage{thm-restate}
\usepackage{amsmath}
\usepackage{amssymb}
\usepackage{comment}
\usepackage{placeins}
\usepackage{subfig}
\usepackage[colorlinks]{hyperref}
\usepackage{tikz}
\usetikzlibrary{calc,backgrounds}

\usepackage{pgffor}
\usepackage{url}
\usepackage{hypcap}
\usepackage{dsfont}
\usepackage[ruled,vlined]{algorithm2e}
\usepackage{soul}
\usepackage{diagbox}

\newcommand{\eq}[1]{Eq.~\hyperref[eq:#1]{(\ref*{eq:#1})}}
\renewcommand{\sec}[1]{\hyperref[sec:#1]{Section~\ref*{sec:#1}}}
\newcommand{\app}[1]{\hyperref[app:#1]{Appendix~\ref*{app:#1}}}
\newcommand{\tab}[1]{\hyperref[tab:#1]{Table~\ref*{tab:#1}}}
\newcommand{\fig}[1]{\hyperref[fig:#1]{Figure~\ref*{fig:#1}}}
\newcommand{\figa}[2]{\hyperref[fig:#1]{Figure~\ref*{fig:#1}#2}}
\newcommand{\figx}[2]{\hyperref[fig:#1]{Figure~\ref*{fig:#1}(#2)}}
\newcommand{\thm}[1]{\hyperref[thm:#1]{Theorem~\ref*{thm:#1}}}
\newcommand{\lem}[1]{\hyperref[lem:#1]{Lemma~\ref*{lem:#1}}}
\newcommand{\cor}[1]{\hyperref[cor:#1]{Corollary~\ref*{cor:#1}}}
\newcommand{\defn}[1]{\hyperref[def:#1]{Definition~\ref*{def:#1}}}
\newcommand{\alg}[1]{\hyperref[alg:#1]{Algorithm~\ref*{alg:#1}}}
\newcommand{\tvar}{u}

\def\ket#1{\mathinner{|{#1}\rangle}}

\newcommand{\be}{\begin{equation}}
\newcommand{\ee}{\end{equation}}
\newcommand{\ba}{\begin{eqnarray}}
\newcommand{\ea}{\end{eqnarray}}

\newcommand{\nn}{\nonumber \\}

\usepackage{color}
\usepackage{xcolor}
\newcommand{\blu}{\color{blue}}

\usepackage{colortbl}
\makeatletter

    \def\CT@@do@color{%
      \global\let\CT@do@color\relax
            \@tempdima\wd\z@
            \advance\@tempdima\@tempdimb
            \advance\@tempdima\@tempdimc
    \advance\@tempdimb\tabcolsep
    \advance\@tempdimc\tabcolsep
    \advance\@tempdima2\tabcolsep
            \kern-\@tempdimb
            \leaders\vrule
                    \hskip\@tempdima\@plus  1fill
            \kern-\@tempdimc
            \hskip-\wd\z@ \@plus -1fill }
    \makeatother

\input{Qcircuit}

\begin{document}

\title{Improved Fault-Tolerant Quantum Simulation \protect\newline of Condensed-Phase Correlated Electrons \protect\newline via Trotterization}

\date{July 10, 2020}
\author{Ian D.~Kivlichan}
\email[Corresponding author: ]{ian.kivlichan@gmail.com}
\affiliation{Google Research, Venice, CA 90291, USA}
\affiliation{Department of Physics, Harvard University, Cambridge, MA 02138, USA}
\author{Craig Gidney}
\affiliation{Google Research, Santa Barbara, CA 93117, USA}
\author{Dominic W.~Berry}
\affiliation{Department of Physics and Astronomy, Macquarie University, Sydney, NSW 2113, Australia}
\author{Nathan Wiebe}
\affiliation{Institute for Nuclear Theory, University of Washington, Seattle, WA 98195, USA}
\author{Jarrod McClean}
\affiliation{Google Research, Venice, CA 90291, USA}
\author{Wei Sun}
\affiliation{Google Research, Mountain View, CA 94043, USA}
\author{Zhang Jiang}
\affiliation{Google Research, Venice, CA 90291, USA}
\author{Nicholas Rubin}
\affiliation{Google Research, Venice, CA 90291, USA}
\author{Austin Fowler}
\affiliation{Google Research, Santa Barbara, CA 93117, USA}
\author{Al\'{a}n Aspuru-Guzik}
\affiliation{Department of Chemistry, University of Toronto, Toronto, Ontario M5G 1Z8, Canada}
\affiliation{Department of Computer Science, University of Toronto, Toronto, Ontario M5G 1Z8, Canada}
\author{Hartmut Neven}
\affiliation{Google Research, Venice, CA 90291, USA}
\author{Ryan Babbush}
\email[Corresponding author: ]{ryanbabbush@gmail.com}
\affiliation{Google Research, Venice, CA 90291, USA}

\begin{abstract}
Recent work has deployed linear combinations of unitaries techniques to reduce the cost of
fault-tolerant quantum simulations of correlated electron models. Here, we show that one can sometimes improve upon those results with optimized implementations of Trotter-Suzuki-based product formulas. We show that low-order Trotter methods perform surprisingly well when used with phase estimation to compute relative precision quantities (e.g.~energies per unit cell), as is often the goal for condensed-phase systems. In this context, simulations of the Hubbard and plane-wave electronic structure models with $N < 10^5$ fermionic modes can be performed with roughly ${\cal O}(1)$ and ${\cal O}(N^2)$ T complexities. We perform numerics revealing tradeoffs between the error and gate complexity of a Trotter step; e.g.,~we show that split-operator techniques have less Trotter error than popular alternatives. 
By compiling to surface code fault-tolerant gates and assuming error rates of one part per thousand, we show that one can error-correct quantum simulations of interesting, classically intractable instances with a few hundred thousand physical qubits.
\end{abstract}
\maketitle

\section{Introduction}
\label{sec:intro}

The physics of interacting electrons in the presence of external fields predicts the properties of many materials as well as the dynamics of most chemical reactions. While the dynamics of many such systems appear intractable for classical computers, quantum computers were originally introduced as universal simulators capable of efficient quantum dynamical simulation \cite{feynman1982simulating,lloyd1996universal}. 
Later work \cite{abrams1999quantum} combined the quantum phase estimation algorithm \cite{kitaev1995quantum} with quantum simulation of fermions \cite{abrams1997simulation} to show that quantum computers can efficiently prepare ground states and estimate ground state energies of these fermionic systems whenever an initial state can be prepared with non-vanishing overlap on the ground state (as is often the case for systems of interest \cite{tubman2018postponing}).

Of particular interest has been quantum simulations of the molecular electronic structure problem \cite{aspuru2005simulated}. The first quantum circuits for these simulations \cite{whitfield2010simulation} had ${\cal O}(N^{10})$ scaling where $N$ is the number of single-particle basis functions \cite{wecker2014gate}. Most early work in this area focused on performing time evolution of Hamiltonians in a Gaussian basis, by means of Trotter-Suzuki-based methods \cite{trotter1959product,suzuki1991general}. By tightening bounds on the Trotter error \cite{wecker2014gate,poulin2015trotter,mcclean2014exploiting,babbush2015chemical} and optimizing the implementation of Trotter steps \cite{hastings2015improving,motzoi2017linear} this cost was brought down to an empirically observed ${\cal O}(N^6)$ scaling, which would likely be even lower using the techniques of \cite{motta2018low} and \cite{campbell2018random}.

Recent work has further reduced these scalings via application of linear combinations of unitaries \cite{childs2012hamiltonian} methods such as Taylor series  \cite{berry2015simulating,babbush2016exponentially} and qubitization \cite{low2016hamiltonian} either in combination with quantum signal processing \cite{low2017optimal} or used directly in phase estimation \cite{poulin2017quantum,berry2018improved,babbush2018encoding,berry2019qubitization}. Another important innovation in the quantum simulation context has been the use of plane wave, rather than Gaussian, bases \cite{babbush2018low,kivlichan2018quantum}. 
Using plane waves simplifies the quantum computation and also extends simulations to the condensed-phase. 
The best quantum algorithms for plane wave basis electronic structure achieve either $\widetilde{\cal O}(N^2)$ gates with ${\cal O}(N \log N)$ space \cite{low2018hamiltonian} or ${\cal O}(N^3)$ gates with ${\cal O}(N)$ space \cite{babbush2018encoding}. 
There are also many papers which focus on these simulations in first quantization \cite{zalka1998efficient,toloui2013quantum,babbush2018exponentially,kassal2008polynomial,kivlichan2016bounding}, the most efficient of which \cite{babbush2018quantum} obtains ${\cal O}(\eta^{8/3} N^{1/3})$ gate complexity and ${\cal O}(\eta \log N)$ space complexity, where $\eta$ is number of particles.

Here, we will also explore quantum simulations of the Hubbard model \cite{hubbard1963electron}, one of the simplest models of interacting electrons. Despite this simplicity, the Hubbard model demonstrates a range of correlated electron behaviors, and is considered a candidate model of high-temperature superconductivity in cuprates \cite{lee2006doping,leblanc2015solutions}. 
Work focusing on quantum computations of the Hubbard model includes \cite{abrams1997simulation,ortiz2001quantum,verstraete2005mapping,verstraete2009quantum,wecker2015solving,jiang2018quantum,babbush2018encoding,haah2018quantum,childs2019nearly}. Of these, \cite{haah2018quantum,childs2019nearly} achieve very close to ${\cal O}(N)$ scaling, but via techniques that appear less practical than the ${\cal O}(N^2)$ scaling achieved in \cite{babbush2018encoding} owing to worse constant factors. However the bounds in \cite{childs2019nearly} also apply to low-order Trotter methods such as ones studied here.

Only a few papers have assessed the viability of these simulations within a fault-tolerant cost model \cite{reiher2017elucidating,babbush2018encoding,berry2019qubitization,BabbushSYK}. We focus on the surface code \cite{fowler2012surface} using lattice surgery \cite{horsman2012surface,fowler2018low} due to its low qubit requirement for logical operations, and high threshold error rate with experimentally accessible planar connectivities. 
The relevant quantities for the surface code are the number of T and Toffoli gates in the circuit, and the maximum number of logical qubits required in the algorithm. Implementing T and Toffoli gates within the surface code requires the consumption of magic states.  Generating these magic states is far more expensive than any Clifford gate. 
The work of \cite{babbush2018encoding} employs linear combination of unitaries techniques to show that classically intractable instances of plane wave electronic structure and the Hubbard model can be error-corrected with fewer than a million physical qubits in a matter of hours, assuming $10^{-3}$ physical error rates. Other recent work has demonstrated that many-body localization in the Heisenberg model may be simulated with fewer than a billion T gates \cite{childs2017toward}, or under ten million T gates on specialized graphs \cite{nam2018low}.

We focus on simulating simple solid-state materials such as lithium hydride, graphite, diamond, etc., as well as the uniform electron gas and the Hubbard model, both in classically difficult regimes. We focus on the latter two problems due to their scientific importance, the wide range of phases they demonstrate, and their history as benchmarks for classical simulations \cite{leblanc2015solutions,corboz2016improved,shepherd2013many,shepherd2014range,loos2016the,mcclain2016spectral}. Ultimately, we show that classically intractable instances of these simulations can be performed in the surface code with a few hundred thousand physical qubits in less than an hour.

We achieve these results using second-order Trotter formulas. These formulas are further optimized through improved quantum circuits for the Trotter steps introduced in \cite{babbush2018low} and \cite{kivlichan2018quantum}, and then we combine those Trotter steps with phase estimation techniques from \cite{berry2009how,higgins2007entanglement}. We show how Hamming weight phasing \cite{gidney2018halving} can be used to asymptotically improve the T complexity of the split-operator step from ${\cal O}(N^2 \log(1/\epsilon))$ to ${\cal O}(N^2 + N\log N \log(1/\epsilon))$, where $\epsilon$ is the rotation synthesis precision. These are thus the most T-efficient electronic structure Trotter steps in the literature. Using a bound on Trotter error that is similar to (but tighter than) a bound derived in \cite{wecker2014gate}, we perform extensive numerics requiring high-performance computing resources to bound the number of Trotter steps required.

Previous work on quantum simulation has analyzed computations which seek a fixed additive (intensive) error in the energy, $\Delta E = \Theta(1)$, rather than a fixed relative (extensive) error in the energy, $\Delta E = \Theta(N)$. But unlike with small molecule quantum chemistry, for condensed-phase systems one is often interested in the energy per unit cell of a system; after all, absolute energies are meaningless in the thermodynamic limit. For instance, when computing response properties (e.g., the Young's modulus of a solid) the allowable error is plainly extensive with system size.

Because the Bardeen-Cooper-Schrieffer theory of superconductivity predicts an intensive excitation gap \cite{bardeen1957theory}, one would need to seek a fixed additive error in the energy to prepare superconducting ground states of the Hubbard model. However, since the quasiparticle density grows linearly with system size, the energy difference between superconducting and normal metal phases is extensive (were this not the case, then an infinitesimal fluctuation would destroy the superconducting state). 
Put differently, for sufficiently large systems, we expect that a single quasiparticle excitation will not affect most local observables (e.g.,~correlation functions), implying that size extensive error is acceptable in this context. 
In this paper, we take the perspective that for condensed-phase quantum simulations there are valid reasons for targeting either intensive or extensive errors, and we analyze both situations.

We show that low-order Trotter methods perform especially well when used with phase estimation to target an extensive error in the energy. Allowing the target error to grow with system size significantly ameliorates the poor scaling of low-order Trotter methods with respect to time and precision. In this context, we show that practical simulations of the Hubbard model and plane wave electronic structure models can be performed with roughly ${\cal O}(1)$ and ${\cal O}(N^2)$ T complexities until the models are so large (e.g., $N > 10^5$ orbitals) that only one Trotter step is required to obtain target precision, or other assumptions of our analysis break down.

In \sec{trottersteps}, we introduce the Hubbard and plane wave electronic structure Hamiltonians, and briefly discuss the two Trotter step algorithms that we study---the fermionic swap network \cite{kivlichan2018quantum} and split-operator Trotter steps \cite{babbush2018low}---as well as how we compute their fault-tolerant costs. We review a technique for combining repeated arbitrary rotations by the same angle \cite{gidney2018halving} which we use in both Trotter steps, and discuss several extensions, in \app{combining}.
We review the fermionic swap network Trotter step of \cite{kivlichan2018quantum} in greater detail in \app{fermionicswaptrotternetwork} and the split-operator Trotter step of \cite{babbush2018low} in \app{FFTtrotterstep}, including several improvements to the algorithms. In \app{HWP_potential}, in particular, we introduce a scheme for asymptotically more efficiently simulating the potential energy operator in the split-operator Trotter step.

We then proceed to compute upper bounds on the fault-tolerant costs of accurately simulating the ground state energy of the uniform electron gas and the Hubbard model using phase estimation. 
We expand on our discussion of the fault-tolerant costs of each Trotter step in terms of the number of arbitrary rotations, Toffoli, and T gates from \sec{trottersteps} in \app{costs_per_step}, including upper bounds on the numbers of arbitrary rotations and T/Toffoli gates. 
We present an improved bound on the second-order Trotter error, and on the ground state energy shift, in \app{new_bounds}; this bound does not require the product of the evolution time and norm of the Hamiltonian to be small.

In \app{phase_estimation}, we construct circuit primitives for phase estimation from Trotter steps for the electronic structure problem.
In \app{trotter_error_numerics}, we numerically evaluate the commutators which contribute to the Trotter error, allowing us to upper bound the shift in the ground state energy using the results of \app{new_bounds}.
We then combine these results in \sec{resourceanalysis} to compute upper bounds on the fault-tolerant costs of simulation. We use our Trotter analysis to place a numerical upper bound on the number of T gates required for fault-tolerant phase estimation for a range of system sizes. 
Together with the maximum number of logical qubits required, this allows us to determine the number of surface code physical qubits needed for distilling T states, the number of physical qubits to store and route data, and the execution time of these algorithms.
Additional details on the precision targets we use, further numerical data, as well as some nuances of the minimization procedure, are detailed in \app{t_count}.
Finally, we discuss and summarize our results in \sec{conclusions}. We summarize the flow of the paper and how the different sections play into the broader whole in \alg{model}.

\begin{algorithm} 
\SetKwFor{Do}{do}{}{end}
 \KwData{System parameters (electronic Hamiltonian, number of electrons $\eta$; \sec{trottersteps})}
Map physical Hamiltonian to qubit Hamiltonian using Jordan-Wigner transformation (\sec{trottersteps})\;
\textbf{Begin} Phase estimation cost optimizations \\
\Indp
Use split-operator or fermionic swap network Trotter step for evolution within phase estimation (\sec{trottersteps})\;
Optimize Trotter step circuits to maximize simultaneous rotations for Hamming weight phasing (\sec{gatecostsperTrotterstep})\;
Bound Trotter errors and maximum shift in eigenphases (\sec{phaseestimation})\;
Determine precision target---either relative to total energy, or absolute (\sec{t_gates})\;
Numerically minimize number of costly fault-tolerant gates (\sec{t_gates})\;
Estimate surface code resource requirements (\sec{surfacecoderesourceanalysis})\;
\Indm
\textbf{End}\\
 \KwResult{Fault-tolerant costs of estimating system ground state energy (\sec{surfacecoderesourceanalysis}, \sec{conclusions})}
\caption{Determine fault-tolerant costs of estimating electronic structure problem ground state energies}
\label{alg:model}
\end{algorithm}

\section{Hamiltonians and Trotter Steps}
\label{sec:trottersteps}

This paper will analyze and reduce the cost of several approaches to simulating fermionic Hamiltonians of the form 
\begin{equation}
\label{eq:hamiltonian}
H = \sum_{p q} T_{pq} a^\dagger_{p} a_{q}
+ \sum_p U_p n_p
+ \sum_{p\neq q} V_{pq} n_{p} n_{q},
\end{equation}
where $a^\dagger_p$ and $a_p$ are fermionic creation and annihilation operators and $n_p = a^\dagger_p a_p$ is the number operator for the corresponding spin-orbital. Mapping to qubits under the Jordan-Wigner transformation \cite{jordan1928uber,somma2002simulating}, \eq{hamiltonian} becomes
\begin{align}
\label{eq:qubit_ham}
H  & = \sum_{p < q} \frac{T_{pq}}{2} \left( X_{p} Z_{p+1} \cdots Z_{q-1} X_{q} + Y_{p} Z_{p+1} \cdots Z_{q-1} Y_{q} \right)
+ \sum_{p < q} \frac{V_{pq}}{4} Z_{p} Z_{q} \\
& - \sum_{p} \left(\frac{T_{pp} + U_{p}}{2} + \sum_{q} \frac{\left(1 - \delta_{pq} \right) V_{pq}}{2} \right) Z_{p} 
+ \sum_{p} \left(\frac{T_{pp} + U_{p}}{2} + \sum_{q} \frac{\left(1 - \delta_{pq} \right) V_{pq}}{4} \right) \openone. \nonumber
\end{align}
This form includes a range of Hamiltonians. Note that there has also been a large body of work on using different fermionic encodings \cite{bravyi2002fermionic,seeley2012bravyi,tranter2015bravyi,bravyi2017tapering,whitfield2016local,havlicek2017operator,jiang2018majorana}; however, these methods do not appear to offer significant advantages in terms of the T complexity of simulations because they do not reduce the number of rotations that must be simulated.

First, we consider this Hamiltonian when the coefficients correspond to the plane wave electronic structure Hamiltonian. In the plane wave dual basis \cite{babbush2018low}, the coefficients of \eq{hamiltonian} are
\begin{align}
\label{eq:dualbasishamiltonian}
T_{pq} = \delta_{\sigma_p, \sigma_q} &\sum_{\nu} \frac{k_\nu^2 \cos(k_\nu \! \cdot \! r_q - k_\nu \! \cdot \! r_p)}{2 \, N} 
\qquad
U_p =\!  -\!\! \sum_{j, \nu \neq 0} \frac{4 \pi \, \zeta_j \cos (k_\nu \! \cdot \! R_j - k_\nu \! \cdot \! r_p)}{\Omega \, k_\nu^2}
\\ 
& ~~~~\qquad\quad V_{pq} = \sum_{\nu \neq 0} \frac{2 \pi \cos(k_\nu \! \cdot \! r_p - k_\nu \! \cdot \! r_q)}{\Omega \, k_\nu^2} \nonumber
\end{align}
where each orbital index $p$ is associated with a spin label $\sigma \in \{ \uparrow, \downarrow \}$ and an orbital centroid $r_p = p \, (2 \,\Omega / N)^{1/d}$ defined in terms of the number of spatial dimensions $d$ and the computational cell volume $\Omega$ \cite{babbush2018low}. The momentum modes are defined as $k_\nu = 2\pi \nu / \Omega^{1/d}$ with $\nu \in [-(N/4)^{1/d}, (N/4)^{1/d})^d$. 
Throughout, $N$ is the number of spin-orbitals.

We will focus on instances of this Hamiltonian corresponding to periodic materials such as lithium hydride metal, diamond, graphite, crystalline silicon, etc., as well as the special case of $U_p = 0$, which is the uniform electron gas (``jellium''). When dealing with molecular potentials, $R_j$ and $\zeta_j$ are the position and charge, respectively, of the $j^\text{th}$ nucleus. 
Because these only change \eq{qubit_ham} through the terms $U_p Z_p$, they enter our algorithms as a layer of single-qubit rotations per Trotter step, adding essentially no cost to the Trotter step circuit. However, this layer changes the Trotter error and hence the total simulation cost.

The Hubbard model corresponds to the case where $T_{pq}$ are nonzero only for nearest neighbors on a lattice, $V_{pq}$ is the on-site interaction $U$ if $p$ and $q$ correspond to the same orbital (with opposite spin), and $U_p = \mu$ is the chemical potential. Ignoring the chemical potential, the Hubbard Hamiltonian is \cite{hubbard1963electron}
\be
H = -\tau \sum_{\langle i, j\rangle, \sigma} (a^\dagger_{i,\sigma} a_{j,\sigma} + \text{h.c.}) + U \sum_i n_{i,\uparrow} n_{i,\downarrow}.
\ee 
The first term describes hopping between adjacent lattice sites ($\langle i, j\rangle$ denotes a sum over nearest neighbor lattice sites) and the second is the on-site interaction. Despite its simplicity, the Hubbard model demonstrates a wide range of correlated electron behaviors, including metal-insulator transitions and a superconducting phase, and further is believed to be a candidate model of high-temperature superconductivity in cuprates \cite{lee2006doping,leblanc2015solutions}.

Both these Hamiltonians can be simulated using the algorithms of \cite{babbush2018low} and \cite{kivlichan2018quantum}. 
The first of these is a split-operator method: it uses a network of Givens rotations or the fast fermionic Fourier transform (FFFT) to alternate between simulating evolution under the kinetic energy terms in the momentum basis and simulating evolution under the potential energy terms in the position basis. 
(The FFFT was first described in \cite{verstraete2009quantum} and has also been studied under the name of the ``spectral tensor network'' \cite{ferris2014fourier,chandran2015spectral}; the Givens rotation network was first studied in \cite{wecker2015solving} and further developed in \cite{kivlichan2018quantum}).
In the momentum basis, the kinetic energy terms can be simulated using only single-qubit rotations; in the position basis, the potential energy terms can be simulated in linear depth using a swap network \cite{babbush2018low}. This algorithm operates on a planar array of qubits. 
By comparison, the second algorithm uses a linear array of qubits, and constructs a swap network which simulates the kinetic and potential operators without the FFFT \cite{kivlichan2018quantum}.

We discuss the split-operator algorithm, including the generalization of the fermionic Fourier transform using Givens rotations introduced in \cite{kivlichan2018quantum}, in greater depth in \app{FFTtrotterstep}, and the fermionic swap network in \app{fermionicswaptrotternetwork}. 
In \app{HWP_potential} we show how to reduce the cost of simulating the potential energy operator within the split-operator method from ${\cal O}(N^2 \log(1/\epsilon))$ to ${\cal O}(N^2 + N\log N \log(1/\epsilon))$, where $\epsilon$ is the rotation synthesis precision, using Hamming weight phasing.
We will discuss in the next subsection the fault-tolerant costs of each Trotter step in terms of the number of arbitrary rotations, Toffoli, and T gates; we describe this in greater detail in \app{costs_per_step}.

We do not make use of several recent techniques that could lead to improvements in the asymptotic limit, but that have large constant factors and are thus not helpful for the system sizes we consider. For instance, the work of \cite{low2018hamiltonian} introduced a technique for computing the diagonal part of the plane wave basis electronic structure Hamiltonian with only $\widetilde {\cal O}(N)$ gates; however, the need to compute the discrete Fourier transform of the potential introduces very large constant factors and increases the spatial complexity to ${\cal O}(N \log N)$.
Haah \textit{et al.}\ have shown an improved algorithm based on locality of lattice models by dividing simulation into layers of disjoint blocks of evolution (blocks in different layers of the algorithm do overlap), though the number of blocks becomes large in two and especially in three dimensions \cite{haah2018quantum}. 
Tran \textit{et al.}\ \cite{tran2018locality} determined the gate count of the Haah \textit{et al.}\ algorithm when the interactions decay as a power law, though there is no asymptotic speedup for the Coulomb interaction. Likewise, the work of \cite{childs2019nearly} achieves nearly $N^{o(1)}$ scaling for Hubbard models but by using arbitrarily high-order formulas which increases gate complexity by a factor of $5^k$ for order $k$. For the extensive error case we achieve the same scaling using low-order formulas. 
While randomized Trotter orderings can yield significant improvements for simulation \cite{childs2018faster,campbell2018random}, these would come at the cost of the improvements we gain from Hamming weight phasing, which requires specific Trotter orderings for full utility.

Finally, we note that while the Hamiltonian of \eq{hamiltonian} exactly describes any electronic structure system (including molecules) with basis set discretization error asymptotically equivalent to Gaussian-based molecular orbitals, there are other discretization schemes (such as finite difference discretization) which also take the form of \eq{hamiltonian}, but with different coefficients from those in \eq{dualbasishamiltonian}. Continuing research on basis functions may yield further examples compatible with these algorithms while also being better suited to molecules. For instance, the basis sets described in \cite{white2017hybrid} are compatible with our algorithms while also being significantly more accurate for molecules than plane waves.

\subsection{Gate costs per Trotter step}

\label{sec:gatecostsperTrotterstep}

We now summarize how we compute the costs,  in terms of the number of arbitrary rotations, Toffoli, and T gates, of the different Trotter steps. We include a more complete description in \app{costs_per_step}, including upper bounds on the number of the different costly gates, as well as more complete descriptions of the two Trotter steps following \app{fermionicswaptrotternetwork} and \app{FFTtrotterstep}. For each Trotter step, we compute these costs numerically. We must calculate this numerically so that we can do it for a range of numbers of ancilla qubits used for Hamming weight phasing. The core idea behind Hamming weight phasing \cite{gidney2018halving} is that the total phasing operation from a group of equiangular rotations can instead be applied to the Hamming weight, using ancilla qubits. The Hamming weight can be computed using only a small number of Toffoli/T gates. We review Hamming weight phasing in greater detail in \app{combining}.

Next, we compute the gate costs per Trotter step with fixed numbers of ancilla qubits assigned for Hamming weight phasing. We then determine the numbers of arbitrary rotations that must be synthesized, T, and Toffoli gates, by iterating through the terms in the Hamiltonian in the order specified by the particular Trotter step, combining arbitrary rotations where possible. The number of arbitrary rotations and Toffoli/T gates in each Trotter step are then totally determined by the simulation order, which we describe next. For all systems and for both Trotter steps, we generate the Hamiltonians and determine simulation order using OpenFermion \cite{openfermion2017}.

For the Hubbard model, we simulate all the on-site interactions in a single layer in which we apply Hamming weight phasing. 
Running through the fermionic swap network (see \app{fermionicswaptrotternetwork}), we continue until we reverse the initial spin-orbital ordering, applying rotations corresponding to hopping terms as we go. 
The small number of terms in the Hamiltonian ($O(N^2)$) allows us to defer evolution until we have many equiangular rotations which can be merged using Hamming weight phasing.
For jellium (the uniform electron gas), there are too many terms and we must apply these rotations immediately, and can only apply Hamming weight phasing within particular layers of fermionic simulation gates.
Finally, for the various materials, we generate the Hamiltonians and run the same simulation procedures as for jellium, but with all the single-qubit rotations repeated in the middle of the Trotter step. 

The gate counting for the split-operator Trotter step is slightly easier to describe. For jellium and the materials, we develop a new scheme using the translation-invariance of the interaction part of the Hamiltonian to significantly reduce the Trotter step cost using Hamming weight phasing (\app{HWP_potential}).  
For the Hubbard model, because the interaction is exclusively on-site, the split-operator Trotter step simulates it exactly as the fermionic swap network Trotter step does, but without the swaps. 
In all cases, we group the single-qubit rotations for the kinetic energy operator together as much as possible for Hamming weight phasing given the number of ancilla qubits assigned for that task. 
Finally, the split-operator step changes from the position to the momentum basis and vice versa: using Hamming weight phasing, arbitrary rotations which might appear in the FFFT are catalyzed rather than synthesized (see \app{catalyzing}) using only a handful of Toffoli gates, and when the Givens rotation procedure is used we apply Hamming weight phasing within it as much as possible given the number of ancilla qubits.

\section{Resource Analysis for Fault-Tolerant Phase Estimation}
\label{sec:resourceanalysis}

\subsection{Trotter-Suzuki errors}
\label{sec:phaseestimation}

Consistent with past work in molecular simulation \cite{poulin2015trotter,babbush2015chemical}, we focus on the second-order Trotter formula given by
\be
\label{eq:2o_trotter}
e^{-iHt} \approx \left( \prod_{\ell=1}^L e^{-iH_\ell t/2} \prod_{\ell=L}^1 e^{-iH_\ell t/2} \right) \equiv e^{-i H_{\rm eff} t},
\qquad \qquad
H = \sum_{\ell=1}^L H_\ell
\ee
which holds for sufficiently small $t$. The algorithms of \sec{trottersteps} naturally implement the second-order Trotter formula if run forward for time $t/2$ and then again in reverse. 
In \app{costs_per_step} as well as the previous subsection, we describe how to compute the T-costs of optimized circuits for second-order Trotter steps with both the split-operator and fermionic swap network algorithms. Low-order Trotter decompositions may be useful in variational algorithms to approximate the ground state energy \cite{wecker2015progress,babbush2018low}. 
However, for non-heuristic algorithms using Trotterization to simulate time dynamics or to prepare eigenstates via phase estimation, we must ensure that the discretization errors incurred by the Trotter-Suzuki decomposition can be controlled. 

In \app{new_bounds} we show that for a second-order Trotter step, the difference between the exact and effective unitary evaluations is bounded by
\be
\label{eq:error_operator}
\left\| e^{-i H t} - e^{-i H_\text{eff} t} \right\| \leq \frac{t^3}{12} \sum_{b=1}^{L-1} \left(\left\|\sum_{c>b} \sum_{a>b} \left[\left[H_b,H_c\right],H_a\right]\right\| + \frac 12 \left\|\sum_{c>b} \left[\left[H_b,H_c\right],H_b\right]\right\| \right)  \equiv  W t^3
\ee
where we call $W$ the ``Trotter error norm'', and the sums run over the terms in the Hamiltonian. Unlike a similar bound \cite[(Eq.\ 11)]{poulin2015trotter}, the result above is tighter, and additionally is non-perturbative and holds for all values of $t$. 

Computing the Trotter error norm $W$ allows us to bound the number of applications of the Trotter step circuit, and hence the gate count, required for phase estimation to a desired precision.
Of particular interest to us is the problem of sampling eigenvalues using phase estimation. As we show in \app{new_bounds},
the maximum shift in the unitary eigenphases (which encode the energies) from Trotterization is
\begin{equation}
\label{eq:ts_shift}
\left |E_n - E_n^{\rm eff} \right | t \leq  \arctan \left(\left\|  e^{-i H t} - e^{-i H_\text{eff} t} \right\| \frac{\sqrt{4-\left\|  e^{-i H t} - e^{-i H_\text{eff} t} \right\|^2}}{2-\left\|  e^{-i H t} - e^{-i H_\text{eff} t} \right\|^2}\right)
=W  t^3 + \frac{W^3 t^9}{24} + {\cal O}\left( W^5 t^{15}\right)
\end{equation}
for all $n$, where 
$E_n$ and $E_n^{\rm eff}$ are corresponding eigenvalues of $H$ and $H_{\rm eff}$, respectively. Thus, the error in the eigenphases from Trotterization is roughly $W t^3$ whenever $W t^3 \gg W^3 t^9 / 24$. For simplicity, we will assume we can approximate the errors by $W t^3$ so long as $W t^3 \leq 1$, implying the first-order term is at least 24 times larger than the next order correction. As we shall see, the Trotterized evolution time is $t = \sqrt{\Delta E_{TS} / W}$, where $\Delta E_{TS}$ accounts for energy errors from the Trotter-Suzuki approximation, so for the condition to hold we need $(\Delta E_{TS})^3 \leq W$.

Rather than approximating the error as $W t^3$, we could instead use the exact expression with the arctan, which holds for $W t^3\le \sqrt{2}$; however, here we focus on the simpler form from the series expansion. The first reason for this is that the condition $W t^3 \leq 1$ is strongly satisfied for all the numerics in this paper. The second reason is that it is much easier to perform the optimizations of \app{trotter_error_numerics} if we can assume the error is given by $W t^3$. Unlike the work of~\cite{reiher2017elucidating} which uses Monte Carlo sampling to estimate Trotter errors, we exactly numerically evaluate $W$ for all systems we consider, and are not subject to sampling errors. While the work of \cite{babbush2015chemical} performed similar numerics, that work focused on arbitrary basis chemistry (which involved much more complicated error operators), and as a result, was not able to make calculations for larger than $N=30$, whereas here we are able to go as high as $N=512$. Still, the bounds in \app{new_bounds} can dramatically overestimate the error in the limit of small $t$. For this reason, it is important to bear in mind that the estimates we provide for the cost are almost certainly quite pessimistic.

As discussed in \sec{trottersteps}, up to a single layer of rotation gates implementing the local external potential terms, the same circuit simulates Trotter steps of any molecule in the plane wave basis. For simplicity, much of our study focuses on the case when these gates are dropped ($U_p=0$), corresponding to simulation of the uniform electron gas (jellium) discussed in detail in \cite{babbush2018low}. There, it is argued that the scientific importance of jellium, the classical difficulty of its simulation, and its history as a benchmark for classical electronic structure methods, position it as an intriguing system through which to contrast quantum and classical simulations. There are open questions about the physics of jellium (especially pertaining to the nature of the biased errors introduced to control the sign problem in quantum Monte Carlo) which one could begin to study on a quantum computer with fewer than one hundred logical qubits. Additionally, we compute the Trotter errors after re-introducing the external potential, so as to determine the costs of simulating several different periodic materials.

Jellium has a single phase parameter, the density given by $\eta/\Omega$ in \eq{dualbasishamiltonian}, which scales the system between the limits of strong and weak correlation. Classically, jellium is particularly challenging to study near half-filling for densities where the average electron radius (the Wigner-Seitz radius) is approximately $r_s=10$ Bohr radii. 
We conduct most of our numerics for jellium in this regime, computing the Trotter errors for spinful jellium with and without spin in two and three dimensions at $r_s=10$.
Additionally, we compute the Trotter error for jellium in 3D, varying the Wigner-Seitz radius in logarithmically spaced steps from $r_s=0.5$ to $r_s=32$ Bohr radii.

For the Hubbard model, we determine the costs of simulation in the intermediate and strongly coupled regimes, $U/\tau=4$ and $U/\tau=8$, respectively, with the hopping integral $\tau=1$. 
We base our energy precision requirements on classical state-of-the-art ground state energy estimates for the Hubbard model at these parameter values for weakly doped systems (filling fraction $0.875$) \cite{leblanc2015solutions}.

For all systems studied, our numerics were performed using code which we have contributed to the open-source package OpenFermion \cite{openfermion2017}. 
However, the larger system sizes studied here may be difficult to access without distributed calculations. We discuss full details of these calculations in \app{trotter_error_numerics}.

\subsection{T gate requirements for Trotterized phase estimation}
\label{sec:t_gates}

In order to give concrete T gate counts for the cost of phase estimation, we need to discuss how errors propagate throughout the algorithm.  We focus here on quantum simulation within a fault-tolerant architecture, similar to \cite{reiher2017elucidating}, but provide further details about the optimal balance between these errors. Neglecting individual gate errors, the sources of error for estimating the energy in the simulation are: 
\begin{enumerate}
\item Trotter-Suzuki errors $\Delta E_{TS}$, from the Trotter approximation to $e^{-iHt}$ due to Hamiltonian terms not commuting. From \eq{error_operator} and \eq{ts_shift}, we know that the error in the eigenvalue of the simulated Hamiltonian obeys 
\begin{equation}
\label{eq:eTS}
\Delta E_{TS} \le W  t^2 + {\cal O}(W^3 t^8).
\end{equation}
We numerically computed the Trotter error norm $W$ as shown in \sec{phaseestimation} for a variety of systems for the fermionic swap and split-operator algorithms outlined in \sec{trottersteps}.
\item Phase estimation errors $\Delta E_{PE}$, due to uncertainty in the value returned by phase estimation; i.e.\ errors due to not computing enough bits of the phase. We use the adaptive phase estimation techniques of \cite{berry2009how,higgins2007entanglement} to reach a root mean squared error of $\Delta E_{PE} t$ using
\begin{equation}
N_{\rm PE} \approx \frac{0.76 \pi}{\Delta E_{PE} t}=\frac{0.76 \pi\sqrt{W}}{\Delta E_{PE} \sqrt{\Delta E_{TS}}}
\end{equation}
applications of the fundamental simulation circuit. This relies on using directionally-controlled evolution \cite{wecker2014gate} to reduce prior estimates on the cost of phase estimation \cite{berry2009how} by a factor 2. This approach uses a single control qubit.
It is possible to match the ultimate lower bound of $\pi / (2\Delta E_{PE} t)$ using multiple control qubits, as in \cite{babbush2018encoding}.
The number of control qubits needed would be the log of the dynamic range needed.
The difference in performance can be accounted for by dividing the gate counts we present by a factor $1.52$.
Alternatively, the approach of \cite{wiebe2015efficient} allows median error performance of $\Delta E_{PE} t$ using $1.65 / (\Delta E_{PE} t)$ applications of the fundamental simulation circuit and just a single control qubit; this could similarly be accounted for by dividing our gate counts by $1.45$.
\item Circuit synthesis errors $\Delta E_{HT}$, due to the fact that there is typically some approximation in compiling arbitrary rotations $R_z(\theta)$ into single-qubit Clifford and T gates. Relative to the other two sources of error, the impact of circuit synthesis error can be reduced at substantially lower cost. If one wishes to synthesize a single-qubit rotation, $R_z(\theta)$ within error $\epsilon$ then the number of T gates required using repeat-until-success synthesis is on average $T_\text{synth} \approx 1.15\log_2(1/\epsilon)+9.2$~\cite{roe2015efficient}.  Following the arguments in~\cite{reiher2017elucidating}, such errors in the individual rotations add at most linearly to the error in the phase estimated.  The phase estimated is the energy eigenvalue multiplied by the time step used for the Trotter decomposition.  Thus, if the error desired in the estimate of the energy eigenstate is $\Delta E_{HT}$, the cost of circuit synthesis is approximately 
\begin{equation}
N_{HT} \approx 1.15\log_2\left(\frac{N_r}{\Delta E_{HT} t}\right)+9.2\label{eq:NHT1}
\end{equation} 
T gates per arbitrary rotation, where $N_r$ is the number of rotations used in a single Trotter step.
\end{enumerate}
In the worst case, these errors add linearly \cite{reiher2017elucidating}. Thus, to guarantee that the total error is at most $\Delta E$, we assume
\begin{equation}
\label{eq:eq_constraint}
\Delta E \ge \Delta E_{TS} + \Delta E_{PE} + \Delta E_{HT}.
\end{equation}

The total T-cost of phase estimation is the product of these, plus the number of ``direct'' T and Toffoli gates $N_d$ (those T/Toffoli gates which appear in the circuit before synthesis) multiplied by $N_{PE}$. Again, we convert Toffoli to T gates in computing $N_d$ at a cost of 2 T gates each \cite{gidney2018efficient}. Of the different variables described so far, $N_r$ and $N_d$ are the only fixed values: irrespective of the desired total precision $\Delta E$, $N_r$ and $N_d$ are set by the Trotter step circuit and number of ancillae used for Hamming weight phasing. $N_{PE}$ depends on $\Delta E_{PE}$ and $\Delta E_{TS}$, and $N_{HT}$ depends directly on $\Delta E_{HT}$ and through the time $t$ on $\Delta E_{TS}$. The T count for phase estimation is thus
\begin{equation}
\label{eq:full_t_cost}
(N_r N_{HT} + N_d) N_{PE} \approx  \frac{0.76 \pi \sqrt{W} }{\Delta E_{PE}\sqrt{\Delta E_{TS}}} \left( N_r \left[1.15\log_2\left(\frac{N_r \sqrt{W}}{\Delta E_{HT} \sqrt{\Delta E_{TS}}}\right)+9.2\right] + N_d \right) .
\end{equation}
We numerically minimize this cost for each system, subject to the constraint $0 < \Delta E_{TS} + \Delta E_{PE} + \Delta E_{HT} \le \Delta E$, where each $\Delta E$ is positive. Note that these generally follow the order $\Delta E_{TS} > \Delta E_{PE} \gg \Delta E_{HT}$. 

We perform this minimization for two different precision targets: relative precision, where $\Delta E$ is set as a fraction of a proxy for the ground state energy $\tilde E_0$, and thus scales with system size, and absolute precision, where $\Delta E$ is set as some fixed constant irrespective of the system size. 
We discuss these precision targets, as well as some subtleties of the minimization and other numerical results, in more detail in \app{t_count}. In both cases, in minimizing and plotting the cost, we convert Toffoli to T gates at a cost of 2 T gates per Toffoli. This is because using the techniques of \cite{gidney2018efficient}, one can distill a Toffoli state at twice the cost of a magic state.

\begin{figure}[ht]
\centering
\subfloat[][]{
\includegraphics[width=0.5\textwidth]{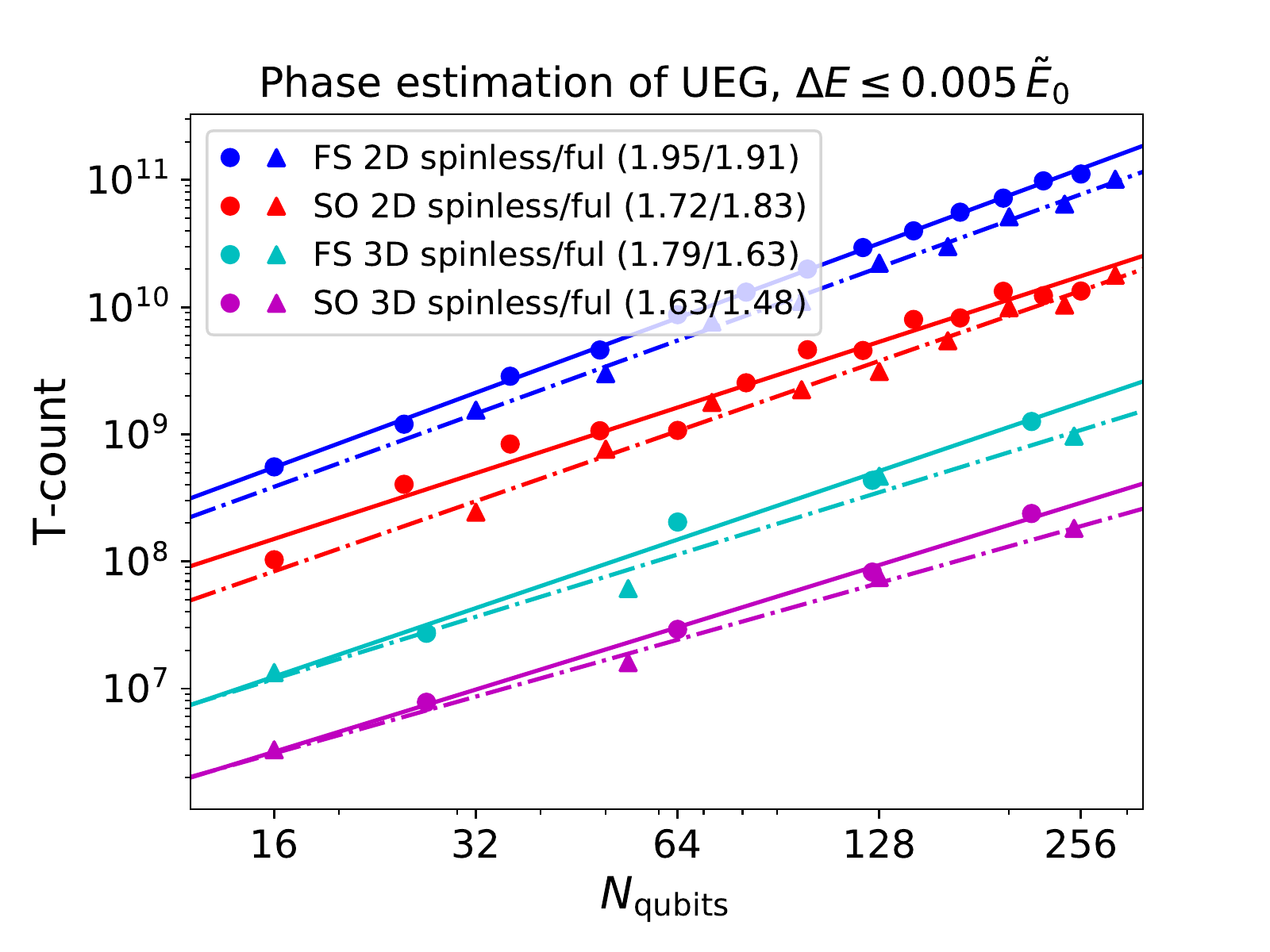}
\label{fig:TcountUEGrel}
}
\subfloat[][]{
\includegraphics[width=0.5\textwidth]{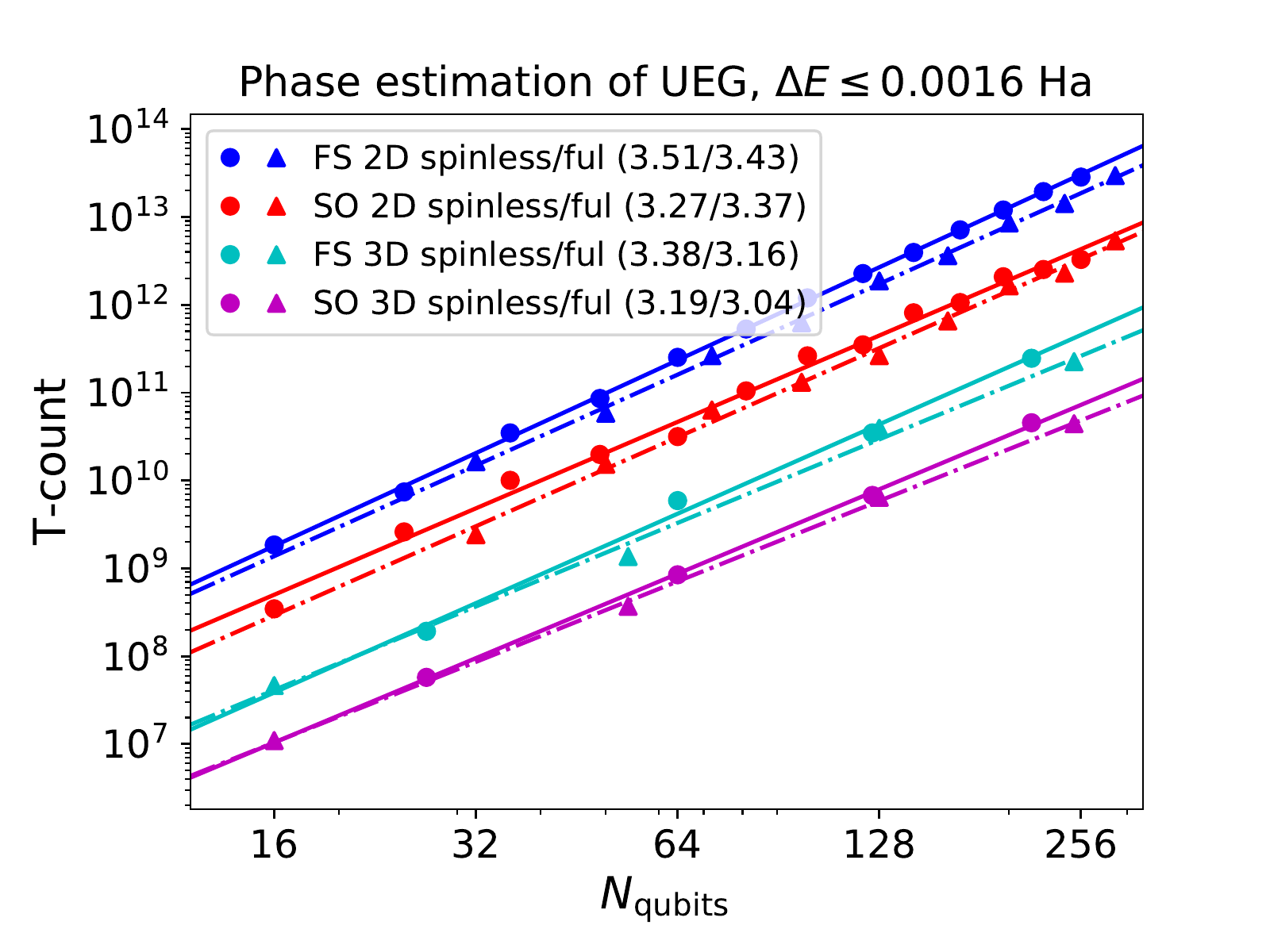}
\label{fig:TcountUEGabs}
}
\caption{\label{fig:TcountUEG} Loose but rigorous upper bound on the number of T gates in the circuit which performs our Trotterized phase estimation simulation of the uniform electron gas (UEG, or jellium) at Wigner-Seitz radius of 10 Bohr radii at half filling $\eta= \lfloor N/2 \rfloor$ based on \eq{full_t_cost}. \protect\subref{fig:TcountUEGrel} shows data for the relative precision target (to within half a percent of the total system energy) as a function of the number of spin-orbitals (qubits); \protect\subref{fig:TcountUEGabs} shows data for the absolute precision target of 0.0016 Hartree (``chemical accuracy''). Bounds were computed for jellium both in 2D and in 3D, using both the fermionic swap network (FS) and split-operator (SO) Trotter steps. Exponents of a power law fit are shown in the caption for each system. 14 ancilla qubits are used for Hamming weight phasing for all systems. We can see that for the extensive error target the T cost scales between ${\cal O}(N^{3/2})$ and ${\cal O}(N^2)$ and for the intensive error target the T cost scales between ${\cal O}(N^3)$ and ${\cal O}(N^{7/2})$.}
\end{figure}

\begin{figure}[htb]
\centering
\subfloat[][]{
\includegraphics[width=0.5\textwidth]{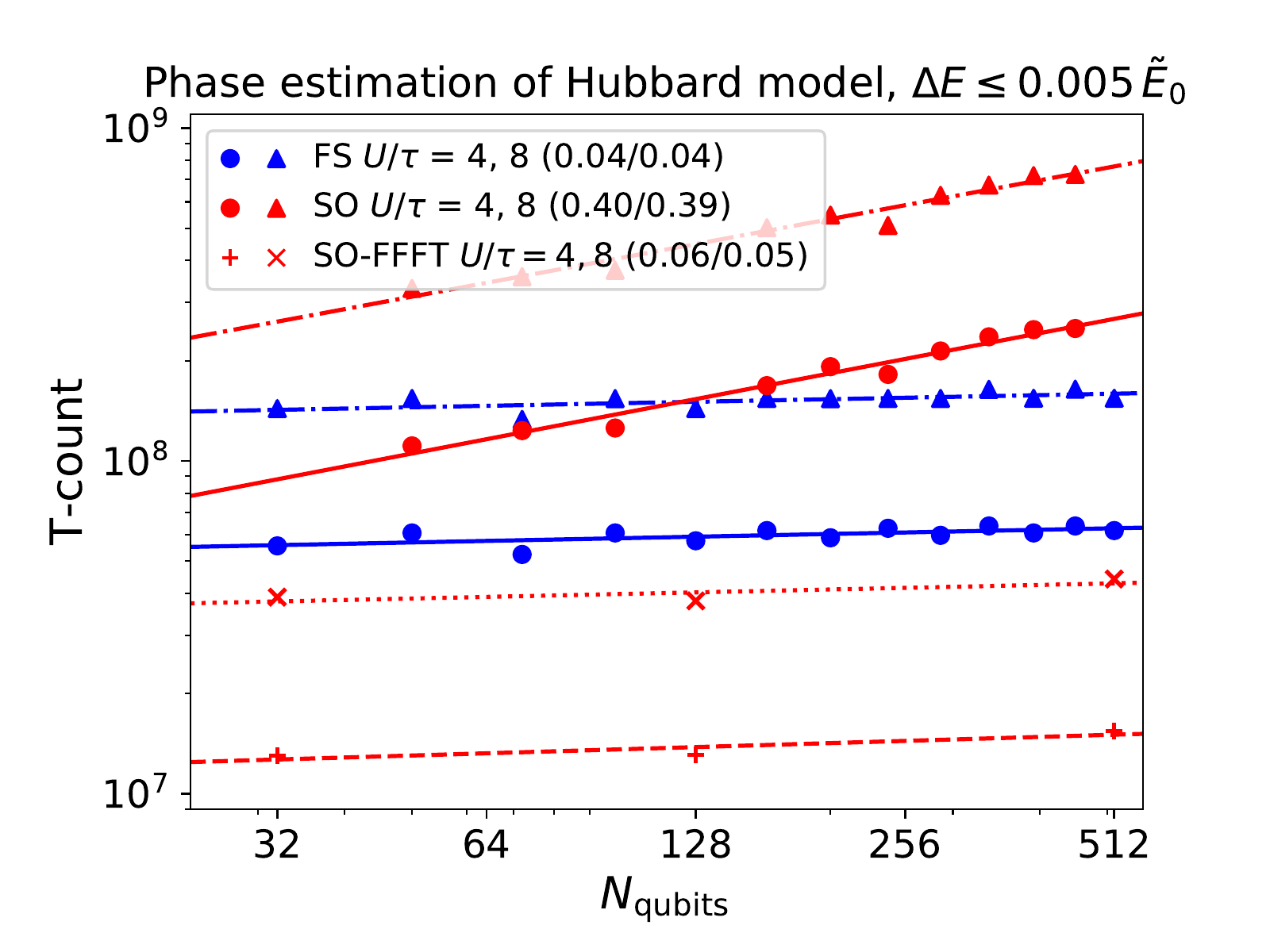}
\label{fig:TcountFHrel}
}
\subfloat[][]{
\includegraphics[width=0.5\textwidth]{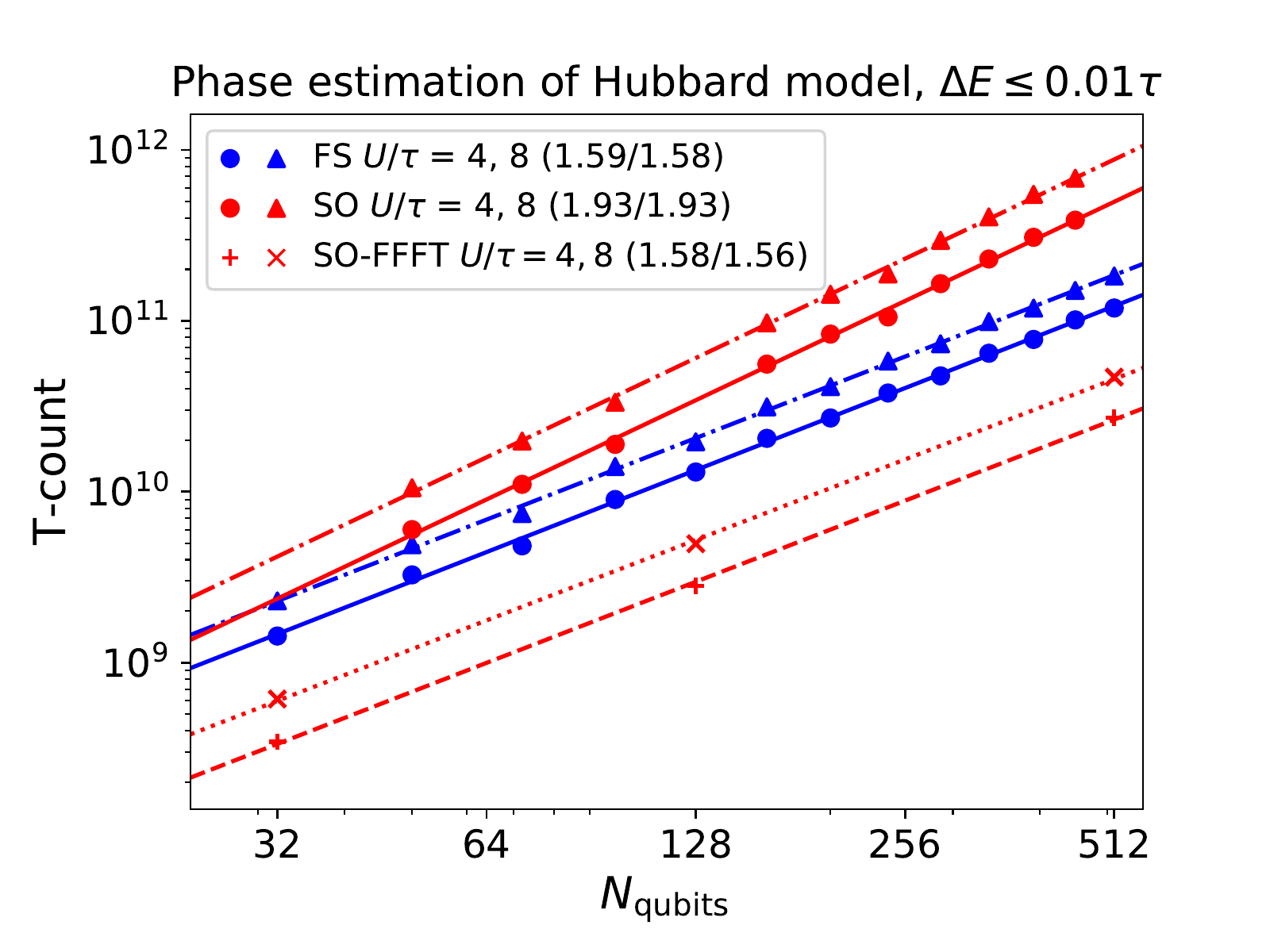}
\label{fig:TcountFHabs}
}
\caption{Loose but rigorous upper bound on the number of T gates in the circuit which performs our Trotterized phase estimation simulation of the Hubbard model in the strongly correlated regime ($U/\tau=$ 4 and 8) \protect\subref{fig:TcountFHrel} to within half a percent of the total system energy (based on data from \cite{leblanc2015solutions}), and \protect\subref{fig:TcountFHabs} to absolute precision $\Delta E=\tau/100$. SO-FFFT (red $+$'s (x's) for $U/\tau=4$ (8)) denotes the split-operator Trotter step using the fast fermionic Fourier transform to change bases (only for side lengths which are binary powers), SO (red circles (triangles) for $U/\tau=4$ (8)) denotes the split-operator Trotter step using Givens rotations to change bases, and FS (blue circles (triangles) for $U/\tau=4$ (8)) denotes the fermionic swap Trotter step. The fermionic swap Trotter step more efficiently combines repeated rotations for the Hubbard model than for jellium. Lines show power law fits to the data, with the exponents of the fits shown in the caption for each system. 14 ancilla qubits are used for Hamming weight phasing for all systems. We see scaling between ${\cal O}(1)$ and ${\cal O}(N^{1/2})$ for the extensive error simulations and scaling between ${\cal O}(N^{3/2})$ and ${\cal O}(N^2)$ for the intensive error simulations. \label{fig:TcountFH}}
\end{figure}

\begin{figure}[ht]
\centering
\subfloat[][]{
\includegraphics[width=0.5\textwidth]{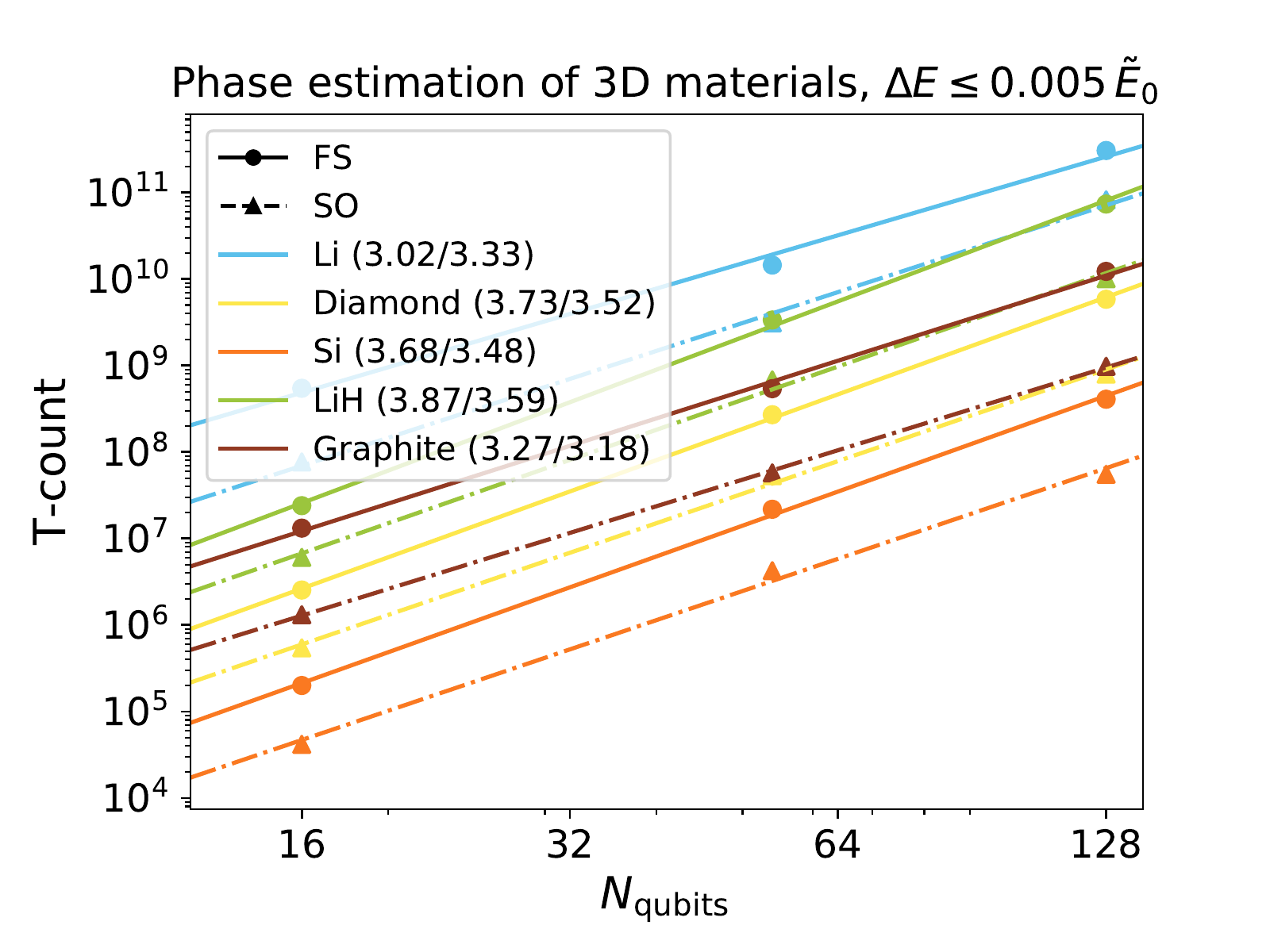}
\label{fig:Tcountmaterials_rel}
}
\subfloat[][]{
\includegraphics[width=0.5\textwidth]{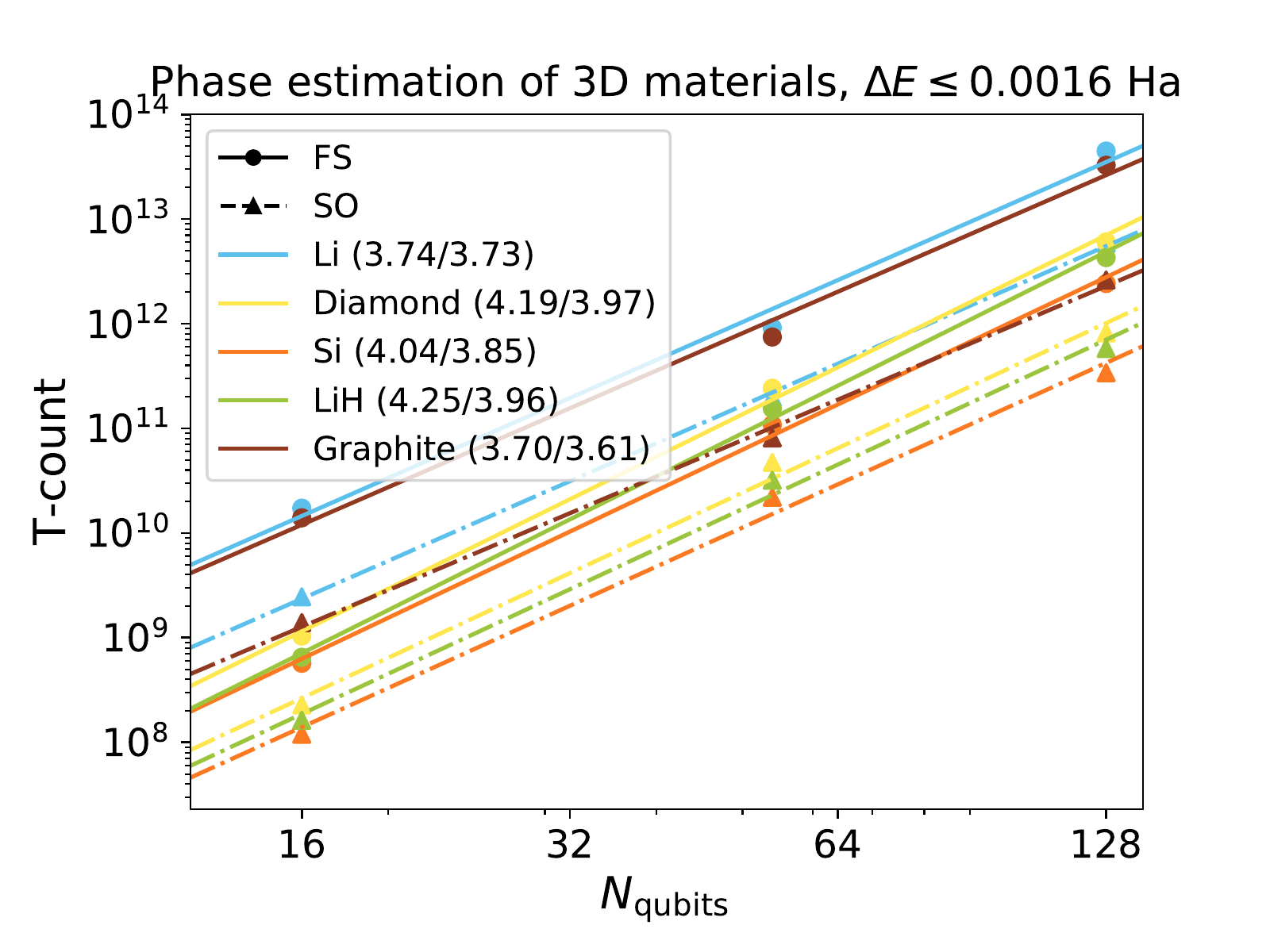}
\label{fig:Tcountmaterials_abs}
}
\caption{Loose but rigorous upper bound on the number of T gates in the circuit which performs our Trotterized phase estimation simulation for a range of materials \protect\subref{fig:Tcountmaterials_rel} to within half a percent of the total system energy, and \protect\subref{fig:Tcountmaterials_abs} to absolute precision $\Delta E = 0.0016$ Hartree. Exponents of a power law fit are shown in the legend for each system in the form ($x_{FS}$/$x_{SO}$), where $x_{FS}$ is the exponent for the fermionic swap network Trotter step and $x_{SO}$ is the exponent for the split-operator Trotter step. Lines show power law fits to the data, with the exponents of the fits shown in the caption for each system. 14 ancilla qubits are used for Hamming weight phasing for all systems. \label{fig:Tcountmaterials}}
\end{figure}

We plot the minimized T-count in \fig{TcountUEGrel} for the uniform electron gas in two and three dimensions to relative precision $\Delta E=0.005 \, \tilde E_0$ (comparable to errors associated with the sign problem in quantum Monte Carlo \cite{tanatar1989ground,shepherd2012full}) and absolute precision $\Delta E=0.0016$ Hartree (i.e.\ ``chemical accuracy''), which shows that we can realistically expect to simulate classically intractable instances of two- and three-dimensional jellium (up to $\sim$300 qubits) with Trotter methods using fewer than one billion T gates. Counts for the Hubbard model to the same relative precision and to absolute precision $\Delta E = \tau/100$ are shown in \fig{TcountFH}. 
In \fig{Tcountmaterials} we plot the minimized T-counts for phase estimation of several different periodic materials.

We separately plot data for the fermionic swap network and split-operator Trotter steps; within the split-operator Trotter step, we separate data at system side lengths which are a power of two from the rest of the data. When the side length is a power of two, the fermionic fast Fourier transform (FFFT) can be applied allowing significantly more efficient basis changes than the Givens rotation procedure. Because this is not a significant component of the cost for the uniform electron gas, we do not separate these data points for that algorithm. We discuss this difference in costs further in \app{FFTtrotterstep}.
As observed in \cite{babbush2015chemical,reiher2017elucidating,poulin2015trotter,berry2019qubitization}, estimates based on bounds on the Trotter error operator norm are often extremely loose, often by several orders of magnitude \cite{babbush2015chemical}, so we expect that these values significantly overestimate the true costs. But even if the bounds were tight for the unitary, we are really interested in how the error accumulates for a particular state \cite{poulin2015trotter,babbush2015chemical}: this difference is yet another potential source of looseness in our cost estimates.

An interesting observation from \fig{TcountFHrel} is that the required number of T gates scales sublinearly in the number of Hamiltonian terms. This is because the number of Trotter steps required to achieve the extensive error target actually decreases with $N$, implying that the allowable error $\Delta E = \Theta(N)$ grows at a rate comparable to the Trotter error norm. As discussed in \sec{intro}, fixed relative error is often a sensible target. Despite this, for sufficiently large system sizes, the number of circuit repetitions in phase estimation will eventually reach one (it cannot go lower), at which point the number of gates will return to linear scaling in the number of terms in the Hamiltonian. We analyze this further in \app{t_count} and show that the number of phase estimation repetitions will not reach this limit until system sizes of hundreds of thousands of spin-orbitals.

The full T-cost (\eq{full_t_cost}) scales as $\widetilde{\cal O}(N_r \sqrt{W} / \Delta E^{3/2})$ in terms of the Trotter error norm $W$, the number of arbitrary rotations per Trotter step $N_r$, and the target precision $\Delta E$. In \app{trotter_error_numerics} we show that the Trotter error norm $W$ scales cubically with $N$ (the number of spin-orbitals) for the uniform electron gas and linearly with $N$ for the Hubbard model. 
Our relative precision target is $\Delta E = \Theta(N)$ for both the uniform electron gas and the Hubbard model and our absolute precision target is $\Delta E = \Theta(1)$. The number of arbitrary rotations in a Trotter step $N_r$ scales linearly for the Hubbard model with the fermionic swap network Trotter step and either quasilinearly or quadratically for the uniform electron gas and the Hubbard model with the split-operator Trotter step (due to the gates required for changing bases). 
From these scalings, we expect the T-costs in the absolute precision case to be ${\cal O}(N^{7/2})$, ${\cal O}(N^2)$, and ${\cal O}(N^{3/2})$ for the uniform electron gas, the Hubbard model with fermionic swap network Trotter steps, and the Hubbard model with the split-operator step, respectively. Changing to relative precision reduces these scalings by $N^{3/2}$ in all cases. This is consistent with what we see in \fig{TcountUEG} and \fig{TcountFH}.

We can understand the condition from the previous subsection $Wt^3 = \sqrt{(\Delta E_{TS})^3 / W} \le 1$, required for the error to be well estimated by $W t^3$ as in \eq{ts_shift}, in terms of these scalings. In the absolute precision case the condition becomes more strongly satisfied with growing system size; however, in the relative precision case the quantity $W t^3$ will grow. In particular, for the Hubbard model $W$ is only ${\cal O}(N)$, so $\sqrt{(\Delta E_{TS})^3 / W} = {\cal O}(N)$ and we must be careful to work with systems where the value remains small.  For the Hubbard model with $U/\tau=4$, $(\Delta E_{TS})^3 / W$ will not reach unity until hundreds of thousands of spin-orbitals, far larger than any system we study.

\subsection{Surface code resource estimate analysis}

\label{sec:surfacecoderesourceanalysis}

Any resource estimate depends sensitively on the assumed hardware. We shall assume the availability of a large planar square array of qubits with nearest-neighbor interactions capable of implementing all physical gates with non-uniform error rates sufficiently low that the overall performance is approximated by a uniform gate error rate of $p=10^{-3}$ or $p=10^{-4}$. We will assume the gates are sufficiently fast to implement a full round of surface code error detection in 1 $\mu$s. Finally, we will assume that a decoder with error suppression performance comparable to \cite{fowler2013optimal} is available that is capable of extracting the logical measurement value from potentially many physical measurements associated with a given logical qubit in no more than 10 $\mu$s. This is 100 times slower than has been previously assumed \cite{fowler2012surface}, as recent (unpublished) research suggests that it will be challenging to achieve even this level of performance.

We will use lattice surgery \cite{horsman2012surface,litinski2018game,fowler2018low,gidney2018efficient} to protect our computations, and will make use of distillation to produce $\ket{{\rm CCZ}}$ states and catalyzation to efficiently produce $\ket{\rm T}$ states. These states enable Toffoli and T gates, respectively. Only a single factory producing states serially will be assumed, to minimize qubit overhead. We use the approach described in \cite{gidney2018efficient}, and fall back to the methods of \cite{fowler2018low} if a total logical error probability below 0.3 is not attainable during the execution of the phase estimation algorithm. For both techniques, we iterate through level-1 and level-2 distillation code distances of 15 through 51 in steps of 2 to find the scheme and code distances that minimize the total number of physical qubits.

\begin{table*}[tb]
\centering
\small
\begin{tabular}{| c | c | c | c | c | c | c | c | c |}
\hline
\multicolumn{5}{|c|}{} & \multicolumn{2}{c|}{Physical qubits} & \multicolumn{2}{c|}{Execution time (hours)} \\ \hline
System & Anc & Log & Toffoli gates & T gates & $p=10^{-3}$ & $p=10^{-4}$ & $p=10^{-3}$ & $p=10^{-4}$ \\ \hline \hline
FH $8\times8$ & 4 & 132 & 2.5e+05 & 2.3e+07 & 3.2e+05$^b$ & 6.2e+04$^a$ & 6.0e-01 & 5.7e-01 \\ \hline
FH $8\times8$ & 32 & 160 & 6.4e+05 & 9.9e+06 & 3.3e+05$^b$ & 7.2e+04$^a$ & 2.8e-01 & 3.0e-01 \\ \hline
FH $10\times10$ & 4 & 204 & 1.3e+06 & 9.2e+07 & 4.4e+05$^b$ & 8.8e+04$^a$ & 2.4e+00 & 2.3e+00 \\ \hline
FH $10\times10$ & 128 & 328 & 3.1e+06 & 4.8e+07 & 6.4e+05$^b$ & 1.3e+05$^a$ & 1.4e+00 & 1.4e+00 \\ \hline
FH $16\times16$ & 6 & 518 & 4.3e+05 & 2.5e+07 & 9.4e+05$^b$ & 2.0e+05$^a$ & 6.6e-01 & 6.4e-01 \\ \hline
FH $16\times16$ & 130 & 642 & 9.0e+05 & 9.2e+06 & 1.1e+06$^b$ & 2.5e+05$^a$ & 2.8e-01 & 3.1e-01 \\ \hline
UEG $3\times3\times3$ & 4 & 58 & 4.4e+05 & 3.1e+07 & 1.9e+05$^b$ & 3.5e+04$^a$ & 8.2e-01 & 7.8e-01 \\ \hline
UEG $3\times3\times3$ & 16 & 70 & 1.0e+06 & 1.4e+07 & 2.0e+05$^b$ & 4.0e+04$^a$ & 4.0e-01 & 4.3e-01 \\ \hline
UEG $4\times4\times4$ & 4 & 132 & 2.0e+06 & 1.5e+08 & 3.3e+05$^b$ & 6.2e+04$^a$ & 4.0e+00 & 3.8e+00 \\ \hline
UEG $4\times4\times4$ & 128 & 256 & 5.7e+06 & 2.5e+07 & 5.2e+05$^b$ & 1.1e+05$^a$ & 8.9e-01 & 1.1e+00 \\ \hline
UEG $5\times5\times5$ & 4 & 254 & 5.0e+06 & 4.0e+08 & 6.0e+05$^b$ & 1.4e+05$^a$ & 1.1e+01 & 1.0e+01 \\ \hline
UEG $5\times5\times5$ & 128 & 378 & 1.4e+07 & 3.3e+07 & 7.3e+05$^b$ & 2.1e+05$^a$ & 1.5e+00 & 2.1e+00 \\ \hline
\end{tabular}
\caption{Fault-tolerant resource estimates (number of logical ancillae (Anc), logical qubits for encoding the system (Log), Toffoli gates, T gates, physical qubits, and execution time) for the relative precision target, assuming physical operation error rates of $p=10^{-3}$ and $p=10^{-4}$. We include data varying the number of ancilla qubits used for Hamming weight phasing. FH indicates Hubbard model data with $U/\tau=4$, and UEG indicates data for the spinful uniform electron gas with a Wigner-Seitz radius of 10 Bohr radii. For all systems, the total number of logical qubits for the system (Log) is double the number of grid points plus the number of ancillae (Anc). When generating T states, superscript \emph{a} indicates that a single round of 15-1 distillation was used, superscript \emph{b} indicates 15-1 followed by CCZ state generation and catalyzation was used \cite{gidney2018efficient}. Scheme \emph{a} requires fewer qubits, but can be slower than scheme \emph{b}.
\label{tab:finaldata_rel}
}
\end{table*}

\begin{table*}[tb]
\centering
\small
\begin{tabular}{| c | c | c | c | c | c | c | c | c |}
\hline
\multicolumn{5}{|c|}{} & \multicolumn{2}{c|}{Physical qubits} & \multicolumn{2}{c|}{Execution time (hours)} \\ \hline
System & Anc & Log & Toffoli gates & T gates & $p=10^{-3}$ & $p=10^{-4}$ & $p=10^{-3}$ & $p=10^{-4}$ \\ \hline \hline
FH $8\times8$ & 4 & 132 & 4.6e+07 & 5.0e+09 & 4.2e+05$^b$ & 8.1e+04$^a$ & 1.3e+02 & 1.2e+02 \\ \hline
FH $8\times8$ & 32 & 160 & 1.2e+08 & 2.1e+09 & 4.7e+05$^b$ & 9.6e+04$^a$ & 5.9e+01 & 6.2e+01 \\ \hline
FH $10\times10$ & 4 & 204 & 4.7e+08 & 4.2e+10 & 6.7e+05$^b$ & 2.4e+05$^b$ & 1.2e+03 & 1.1e+03 \\ \hline
FH $10\times10$ & 128 & 328 & 1.1e+09 & 2.2e+10 & 9.8e+05$^b$ & 3.2e+05$^b$ & 7.0e+02 & 6.2e+02 \\ \hline
FH $16\times16$ & 6 & 518 & 6.4e+08 & 4.7e+10 & 1.7e+06$^b$ & 4.5e+05$^b$ & 1.4e+03 & 1.2e+03 \\ \hline
FH $16\times16$ & 130 & 642 & 1.3e+09 & 1.6e+10 & 1.8e+06$^b$ & 5.3e+05$^b$ & 5.3e+02 & 4.7e+02 \\ \hline
UEG $3\times3\times3$ & 4 & 58 & 8.9e+06 & 7.2e+08 & 2.3e+05$^b$ & 4.4e+04$^a$ & 1.9e+01 & 1.8e+01 \\ \hline
UEG $3\times3\times3$ & 16 & 70 & 2.0e+07 & 3.2e+08 & 2.5e+05$^b$ & 5.0e+04$^a$ & 9.1e+00 & 9.6e+00 \\ \hline
UEG $4\times4\times4$ & 4 & 132 & 1.4e+08 & 1.3e+10 &  4.8e+05$^b$ & 1.7e+05$^b$ & 3.9e+02 & 3.4e+02 \\ \hline
UEG $4\times4\times4$ & 128 & 256 & 4.2e+08 & 2.2e+09 & 6.9e+05$^b$ & 1.4e+05$^a$ & 7.5e+01 & 9.3e+01 \\ \hline
UEG $5\times5\times5$ & 4 & 254 & 1.0e+09 & 9.9e+10 & 8.9e+05$^c$ & 2.7e+05$^b$ & 5.1e+03 & 2.6e+03 \\ \hline
UEG $5\times5\times5$ & 128 & 378 & 2.8e+09 & 8.4e+09 & 1.1e+06$^b$ & 3.6e+05$^b$ & 3.8e+02 & 3.4e+02 \\ \hline
\end{tabular}
\caption{Fault-tolerant resource estimates (number of logical ancillae (Anc), logical qubits for encoding the system (Log), Toffoli gates, T gates, physical qubits, and execution time) for the absolute precision target, assuming physical operation error rates of $p=10^{-3}$ and $p=10^{-4}$. We include data varying the number of ancilla qubits used for Hamming weight phasing. FH indicates Hubbard model data with $U/\tau=4$, and UEG indicates data for the spinful uniform electron gas with a Wigner-Seitz radius of 10 Bohr radii. For all systems, the total number of logical qubits for the system (Log) is double the number of grid points in it plus the number of ancillae (Anc). When generating T states, superscript \emph{a} indicates a single round of 15-1 distillation was used, superscript \emph{b} indicates 15-1 followed by CCZ state generation and catalyzation was used, superscript \emph{c} indicates two rounds of 15-1 distillation were used \cite{gidney2018efficient}. Note that scheme \emph{a}, while requiring fewer qubits, can be slower than scheme \emph{b}.
\label{tab:finaldata_abs}
}
\end{table*}

Our resource estimates for phase estimation with the relative precision target (0.5\% of the approximate system energy) are in \tab{finaldata_rel}. We include the total number of logical qubits (including system qubits, and ancilla qubits used for Hamming weight phasing and phase estimation), numbers of Toffoli and T gates, and the computational volume in physical qubit-hours for error rates $p=10^{-3}$ and $p=10^{-4}$. We only give results for whichever Trotter step algorithm gives better performance -- the fermionic swap network is superior for the Hubbard model when the side length is not a power of two, and the split-operator step superior in the other cases. \tab{finaldata_abs} shows the same data with absolute precision targets of 0.0016 Hartree for the uniform electron gas (jellium) and $\tau/100$ for the Hubbard model. The error-corrected data clarify the time-space tradeoff in Hamming weight phasing: by increasing the number of logical ancilla qubits, we can reduce the execution time by as much as an order of magnitude. Because the number of physical qubits used for state distillation is generally comparable to the total number of physical qubits, the increase in the number of physical qubits to support the additional ancillae is typically small, only $\sim\!25\%$ of the original requirement.

\section{Discussion}
\label{sec:conclusions}

Prior to this work, the most practical techniques for performing fault-tolerant quantum simulations of condensed-phase correlated electron models were those introduced in \cite{babbush2018encoding}. That work deployed methods based on linear combinations of unitaries which scaled roughly as ${\cal O}(\lambda N / \Delta E)$ where $\lambda$ is the sum of the absolute values of the Hamiltonian terms; $\lambda = {\cal O}(N)$ for the Hubbard model and $\lambda = {\cal O}(N^2)$ for the plane wave basis electronic structure Hamiltonian. In this paper we demonstrate low-order Trotter methods for these problems scaling as roughly ${\cal O}(N^{3/2}/\Delta E^{3/2})$ for the Hubbard model and ${\cal O}(N^{7/2} / \Delta E^{3/2})$ for jellium. While the methods of \cite{babbush2018encoding} are more practical when targeting an intensive error, $\Delta E = \Theta(1)$, the methods here are more practical when targeting an extensive error, $\Delta E = \Theta(N)$. For example, for Hubbard model simulations with extensive error the methods of \cite{babbush2018encoding} have ${\cal O}(N)$ T complexity whereas the methods here have ${\cal O}(1)$ T complexity (at least until $N > 10^5$ at which point the scaling increases).

It is also interesting to compare these asymptotics to other recent work on Trotter based simulations. For instance, the work of \cite{childs2019nearly} will also have effective ${\cal O}(1)$ scaling for the Hubbard model in the extensive error target regime, but does not provide a concrete circuit implementation and their analysis is not directly applicable to electronic structure Hamiltonians. 
Another comparison is the method of \cite{campbell2018random} which essentially scales as ${\cal O}(\lambda^2 / \Delta E^2)$ for the purposes of phase estimation. For extensive error simulations that method will have T complexities of roughly ${\cal O}(N^2)$ and ${\cal O}(1)$ for jellium and the Hubbard model, respectively, which matches our scaling here.

Our work has also made significant progress towards reducing the resources required to perform the first classically intractable electronic structure calculation on a fault-tolerant device that could be realized with the error rates and connectivities accessible to superconducting qubits. In addition to the low scaling of our methods, as discussed above, we put significant effort into optimizing constant factors through strategies such as the Hamming weight phasing techniques introduced here. We also further reduced surface code compilation overheads by using lattice surgery \cite{fowler2018low,gidney2018efficient,litinski2018game} rather than topological braiding (as in \cite{babbush2018encoding}). Ultimately, we found that even at $10^{-3}$ error rates one can error-correct classically intractable instances of these problems with only a few hundred thousand physical qubits.

We conclude with a brief discussion of open research questions relevant to this work.
First, one could investigate the T-costs of higher-order Trotter formulas. Other work has shown these to be advantageous even for small system sizes \cite{childs2017toward,nam2018low}. Numerically evaluating the Trotter error norm would be challenging for higher-order formulas, but even analytic bounds would be interesting to compare with those presented here. 
Second, rather than looking at upper bounds, one could study the true Trotter error in circuit simulations (using inefficient exact classical calculations and extrapolating, or with efficient approximate classical calculations): this could reduce the number of Trotter steps, and hence our T-count estimates, by orders of magnitude. Finally, one could attempt to extrapolate these quantum simulations to zero Trotter error with even fewer steps than we use here by interpolating the trend between Trotter errors and Trotter number, as is common in quantum Monte Carlo electronic structure simulations.

\subsection*{Acknowledgments}
We thank Garnet Kin-Lic Chan, Matthew Foulkes, Jessica Lemieux, Matthias Degroote, and Kostyantyn Kechedzhi for discussions about when it is appropriate to target extensive error in the energies of Hubbard model and jellium simulations, and for what values of extensive and intensive errors the simulations would be classically challenging. We thank Vasil Denchev for assistance parallelizing some calculations required for our numerics, as well as contributors to OpenFermion~\cite{openfermion2017}, which we used extensively for numerics. We thank Yuan Su for feedback on an earlier draft.
I.~D.~K.~acknowledges partial support from the National Sciences and Engineering Research Council of Canada.
D.~W.~B. is funded by Australian Research Council Discovery projects (Grant Nos.\ DP160102426 and DP190102633) and by a grant from Google Quantum.

\bibliographystyle{apsrev4-1}
\bibliography{main}

\appendix

\section{Combining repeated rotations using Hamming weight phasing}
\label{app:combining}

In this appendix, we discuss the technique of \cite{gidney2018halving} for efficiently applying repeated arbitrary rotations by the same angle. 
First, we review how this may be done when the repeated rotations by the same angle are parallelizable, that is, multiple rotations $R_z(\theta)$ can be performed simultaneously in the circuit. For $n$ parallel equiangular rotations, the technique reduces the number of arbitrary rotations which must be synthesized to $\lfloor \log_2 n + 1\rfloor $, at the cost of at most $4n-4$ additional T gates (or $n-1$ Toffoli gates) and $n-1$ ancilla qubits. Because synthesizing arbitrary rotations is very costly on a fault-tolerant quantum computer, this can lead to large savings. 
Following that, we briefly discuss how the method works with a limited number of ancillae $r < n - 1$ as well as some possible extensions.

\subsection{Combining arbitrary parallelizable rotations by Hamming weight phasing}
\label{app:hammingphasingbasic}

Consider a circuit where the same single-qubit rotation $R_z(\theta) = \exp(-i \theta Z/2)$ is simultaneously applied to three different qubits. The action on the logical states is a different phase, depending only on the Hamming weight of that logical state. To be explicit,
\begin{enumerate}
\item the all-zero state $\ket{000}$ picks up the phase $-3\theta/2$,
\item the three states of Hamming weight one $\ket{001}$, $\ket{010}$, and $\ket{100}$ each pick up a phase $-\theta/2$ 
\item the three states of Hamming weight two $\ket{011}$, $\ket{101}$, and $\ket{110}$ are all phased by $\theta/2$, and finally
\item the all-one state $\ket{111}$ picks up the phase $3\theta/2$.
\end{enumerate}
Rather than applying three rotations by the same angle $R_z(\theta)$, we can instead compute the Hamming weight of the input states, and apply two distinct rotations to the Hamming weight: $R_z(\theta)$ to the 1s bit, and $R_z(2\theta)$ to the 2s bit. In this case,
\begin{enumerate}
\item the phase $-3\theta/2$ is applied to the all-zero state $\ket{000}$ (Hamming weight 0),
\item $-\theta/2 = - \theta + \theta/2 $ is applied to $\ket{001}$, $\ket{010}$, and $\ket{100}$ (Hamming weight $1 = 01_b$ in binary),
\item $\theta/2 = + \theta -\theta/2 $ is applied to $\ket{011}$, $\ket{101}$, and $\ket{110}$ (Hamming weight $2 = 10_b$), and finally
\item $3\theta/2 = + \theta + \theta/2 $ is applied to the all-one state $\ket{111}$ (Hamming weight $3 = 11_b$).
\end{enumerate}
The phases on each logical state are identical for the two procedures. However, because arbitrary rotations must be synthesized using costly T gates, reducing the number of rotation gates in the circuit reduces its fault-tolerant cost.

This idea readily generalizes to the case of $n$ repeated equiangular rotations $R_z(\theta)$ appearing in parallel in a circuit: rather than applying the $n$ original arbitrary rotations, we can compute the Hamming weight of the relevant qubits, and instead apply $\lfloor \log_2 n + 1\rfloor $ arbitrary rotations $R_z(\theta)$, $R_z(2\theta)$, $R_z(4\theta)$, $\ldots$, to the Hamming weight \cite{gidney2018halving}. We call this technique Hamming weight phasing.

At this point, Hamming weight phasing would appear to give an improvement for free. However, there are two sources of cost we have not mentioned.
These costs are an additional number of T gates and ancilla qubit requirement, both arising from the use of adder circuits to compute the Hamming weight. 
Using the adder circuits of \cite{gidney2018halving}, we can compute and uncompute the Hamming weight using at most $n-1$ Toffoli gates (or equivalently, $4n-4$ additional T gates) and $n-1$ ancilla qubits, corresponding to $n-1$ single-bit adders. (Each single-bit adder uses one ancilla qubit, and costs one Toffoli gate or 4 T gates \cite{gidney2018halving}.) The strategy for this summation is as follows.

We begin with $n$ qubits to which a rotation $R_z(\theta)$ must be applied. 
We call these initial qubits ``weight-1'' qubits: our goal is to combine them such that we are left with one qubit which captures each digit of the Hamming weight, i.e., one qubit each of weight 1, 2, 4, 8, $\ldots$, $2^{\lfloor \log_2 (n)\rfloor}$. 
Then, we apply the rotation $R_z(w \theta)$ to each qubit depending on its weight $w$, and uncompute the Hamming weight, returning each qubit to its initial state.

We combine the qubits to form the Hamming weight in stages. First, we pick a group of 3 weight-1 qubits: call these three qubits $a$, $b$, and $c$. We initialize one ancilla qubit in the $\ket0$ state, and apply the adder building block of \cite{gidney2018halving} to the three qubits and the ancilla. 
The first two qubits are unaffected (their final state remains $a$ and $b$), but the third qubit and the ancilla qubit are changed to the 1s and 2s bits of $(a+b+c)$: $(a + b + c)_0$ and $(a + b + c)_1$, respectively. 
After the adder, we have introduced one ancilla qubit (now a weight-2 qubit, storing $(a + b + c)_1$), and accounted for the weight-1 qubits $a$ and $b$, thereby decreasing the number of weight-1 qubits unaccounted for in the Hamming weight by two. 
We repeat this process of choosing groups of 3 weight-1 qubits and combining them until either two or one weight-1 qubits remain. At least one weight-1 qubit must remain because the adder primitive always leaves a qubit of weight 1.

At this point, we have applied $\lfloor (n-1)/2\rfloor$ temporary adders, thereby introducing $\lfloor (n-1)/2\rfloor$ ancilla qubits (now of weight 2) and incurring a cost of $\lfloor (n-1)/2\rfloor$ Toffoli gates. $n - 2\lfloor (n-1)/2\rfloor$ weight-1 qubits remain. If $n$ is odd, only one weight-1 qubit remains, and we are done: this is the 1s digit qubit in the Hamming weight. 
If $n$ is even, two weight-1 qubits (call them $a$ and $b$) still have to be accounted for. We introduce an ancilla and apply the temporary adder to the two remaining weight-1 qubits. We are left with the input $a$, $(a+b)_0$---now the weight-1 bit of the Hamming weight---and the ancilla in the state $(a+b)_1$, which now has weight 2. In the worst case (the case where $n$ is even), we applied $\blu \lfloor n/2 \rfloor$ adders, introduced $\blu \lfloor n/2 \rfloor$ ancilla qubits, and used $\blu \lfloor n/2 \rfloor$ Toffoli gates to reduce to a single weight-1 qubit. We repeat this procedure for the weight-2 qubits. 
There are at most $\blu \lfloor n/2 \rfloor$ weight-2 qubits, so the cost of reducing to a single weight-2 qubit for the Hamming weight register is at most $\blu n/4$ temporary adders. 

In general, when converting from weight $w$ to weight $2w$, the odd case is simple. In the even case, two weight-$w$ qubits (again, $a$ and $b$) remain, which are converted to $a$, $(a+b)_{\log w}$, and $(a+b)_{\log w + 1}$; these are the unmodified input, the weight-$w$ bit of the Hamming weight, and the ancilla qubit now in the weight $2w$ state $(a+b)_{\log w + 1}$, respectively.
We continue to repeat this procedure for the higher-weight qubits, determining the Hamming weight bit-by-bit, until we reach a stage where there are exactly two qubits of weight $2^{\lfloor \log_2 n\rfloor - 1}$. At this point, we apply the final adder, and introduce a final ancilla qubit for the most significant bit of the Hamming weight. The worst case for the entire procedure is when $n$ is a power of 2. When $n$ is a power of 2, the number of weight-$k$ qubits is always even up to the final weight to be combined, weight $k=2^{\lfloor \log_2 n\rfloor - 1} = n/2$, for which there are 2 qubits. In general, the number of qubits of weight $k$ is $n / k$ for each $k$, except for the single $k=n$ qubit. So in the worst case the number of adders required is half this expression summed over $k$, 
\begin{equation}
\sum_{i=0}^{\lfloor \log_2 n \rfloor -1} \frac{n}{2^{i+1}} = n - 1.
\end{equation}
The total cost to form the Hamming weight register is thus at most $n-1$ Toffoli gates (equivalently, $4n-4$ T gates) and $n-1$ ancilla. This allows us to reduce the number of parallel equiangular rotations from $n$ to $\lfloor \log_2 n + 1\rfloor$. Uncomputing the Hamming weight requires no further T gates or ancillae \cite{gidney2018halving}.

Finally, note if $\theta = m\pi/2^k$ for integers $k$ and $m$ that there is no reason to continue the addition past weight $2^{k-2}$: at that point, each qubit only needs a T gate applied to it, and further adders eliminate only one T gate, versus the 4 T gates used by the adder. We discuss a specific application of this idea later in \app{catalyzing}.

The total tradeoff for Hamming weight phasing depends on the cost of synthesizing arbitrary rotations. With $T_\text{synth}$ the number of T gates required to synthesize a single-qubit arbitrary rotation $R_z(\theta)$, Hamming weight phasing reduces the number of T gates required by
\begin{equation}
T_\text{synth} \left(n - \lfloor \log_2 n + 1\rfloor\right) - 4(n - 1).
\end{equation}
Even for small $n$, the break-even point for Hamming weight phasing to yield an improvement is $T_\text{synth} \gtrsim 10$. For comparison, the number of T gates required to synthesize the single-qubit rotation $R_z(\theta)$ within error $\epsilon_\text{synth}$ using repeat-until-success circuits is approximately $T_\text{synth} \approx 1.15\log_2(1/\epsilon_\text{synth})+9.2$ \cite{roe2015efficient}.

The additional cost that must be considered is that of allocating $n-1$ ancilla qubits for use in Hamming weight phasing. If $n$ is sufficiently large, these qubits could alternatively be put to use distilling T gates. Together with the additional space requirement, this must be balanced against the savings of Hamming weight phasing.

In the following subsection, we consider the problem of Hamming weight phasing with stronger restrictions on the number of ancilla qubits we can introduce. We present two schemes: one with a constant number of ancillae $r$, and one using $\lfloor \sqrt n\rfloor + \lfloor \log_2 n\rfloor$ ancilla qubits. In both cases, the T gate requirement increases while the ancilla qubit requirement becomes more favourable.

\subsection{Hamming weight phasing with limited ancilla}

The ancilla requirement of Hamming weight phasing may be prohibitive. We discuss two possible modifications with lower ancilla requirements in this subsection. 
First, we consider Hamming weight phasing when limiting to a constant number of ancilla qubits $r < n-1$. 
Second, we discuss a method where we compute the Hamming weight of size-$\lfloor \sqrt n\rfloor$ subsets of the qubits at a time, and then sum those subset Hamming weights, rather than directly computing the full Hamming weight. This second method reduces the number of ancilla qubits required to $\lfloor \sqrt n\rfloor + \lfloor \log_2 n\rfloor$. 
For both modifications, the number of ancilla is reduced at the cost of requiring more Toffoli or T gates.

First, we consider limiting to a constant number of ancilla qubits $r < n-1$. Because the number of parallel equiangular rotations $n$ is greater than $r + 1$, it is no longer possible to combine all the rotations at once. 
The problem is that we do not have enough ancilla qubits to compute the Hamming weight. However, we can still combine $r+1$ rotations at a time into $\lfloor \log_2 (r+1) + 1 \rfloor $ arbitrary rotations, and repeat the process until we have combined all $n$ arbitrary rotations. 
Within each repetition, there are at most $r$ adders using all $r$ ancilla qubits, reducing $r+1$ arbitrary rotations to $\lfloor \log_2 (r+1) + 1 \rfloor $ at a cost of $r$ Toffoli ($4r$ T) gates. The number of repetitions required to include every rotation is $\lceil n / (r+1) \rceil$. 
The total cost is at most $4r \lceil n / (r+1) \rceil$ T gates and $\lceil n / (r+1) \rceil \lfloor \log_2 (r+1) + 1 \rfloor$ arbitrary rotations. 
So even with only a constant $r$ ancilla qubits, we reduce the number of T gates by at least
\begin{equation}
T_\text{synth} \left(n - \left\lceil \frac{n}{r+1} \right\rceil \lfloor \log_2 (r+1) + 1 \rfloor \right) - 4r \left\lceil \frac{n}{r+1} \right\rceil,
\end{equation}
where $T_\text{synth}$ is the number of T gates per arbitrary rotation required for synthesis. 
This is the simplest approach to reducing the number of rotations with a limited number of ancillae.

Second, we present a method for reducing the number of arbitrary rotations by computing the Hamming weight of groups of qubits, and adding it to a second register storing the total Hamming weight. We divide the original $n$ rotations into groups of size $\lfloor \sqrt{n}\rfloor$. 
For each group, we compute the Hamming weight using $\lfloor \sqrt{n} - 1\rfloor $ ancillae, add it to a second register of $\lfloor \log_2 n + 1\rfloor $ qubits, and then uncompute the Hamming weight of the group. 
The same $\lfloor \sqrt{n} - 1\rfloor $ ancillae are used to compute the Hamming weight of each group. 
After the Hamming weight of all $\left\lceil n / \lfloor \sqrt{n}\rfloor \right\rceil$ groups has been computed and added to the total, we apply phases to the total Hamming weight as before.
Then, we recompute and subtract each of the group Hamming weights to uncompute the total Hamming weight register. 
This reduces the number of ancillae required from $n-1$ as in the previous section, where the Hamming weight of all $n$ initial rotation qubits was computed at once, to $\lfloor \sqrt{n}\rfloor + \lfloor \log_2 n\rfloor$.

The number of T gates required when we divide into groups of size $\lfloor \sqrt{n}\rfloor$, and add to a total Hamming weight register, is as follows. Computing the Hamming weight within each group can be done at a cost of at most $4\lfloor \sqrt{n} - 1\rfloor$ T gates. Adding that group Hamming weight to the total Hamming weight register costs at most $4\lfloor \log_2 n\rfloor$ T gates.
Uncomputing the group Hamming weight after adding it to the total register (to continue on to the next group) costs no T gates; however, to uncompute the total Hamming weight we must subtract the Hamming weight of each group. Because we uncompute the group Hamming weights to free up the $\lfloor \sqrt n\rfloor$ ancillae, this subtraction requires us to sequentially recompute and subtract each group Hamming weight. For the final group added to the total, we do not need to recompute the group Hamming weight because we can subtract immediately after phasing the total Hamming weight, without uncomputation. 

Thus, we compute the Hamming weight of $\left\lceil n / \lfloor \sqrt{n}\rfloor - 1 \right\rceil $ groups twice. The total cost of this is $8 \left\lceil n / \lfloor \sqrt{n}\rfloor - 1 \right\rceil  \lfloor \sqrt{n} - 1\rfloor $ T gates, and we compute the Hamming weight of the final group once at a cost of $4\lfloor \sqrt{n} - 1\rfloor $ T gates. 
We add and subtract the Hamming weight of each group to the total Hamming weight register once, at a total cost of $8\lfloor \log_2 n\rfloor \left\lceil n / \lfloor \sqrt{n}\rfloor \right\rceil$ T gates. All of these values are upper bounds on the cost. The full T-cost of the $\lfloor \sqrt{n}\rfloor$-grouping method is then
\begin{equation}
8 \left( \left\lceil n / \lfloor \sqrt{n}\rfloor \right\rceil - \frac12\right) \lfloor \sqrt{n} - 1\rfloor + 8\lfloor \log_2 n\rfloor \left\lceil n / \lfloor \sqrt{n}\rfloor \right\rceil.
\end{equation}
We can get a better understanding of this by simplifying using the inequality $\left\lceil n / \lfloor \sqrt{n}\rfloor \right\rceil < \sqrt n + 3/2$ for $n \in \mathbb{N}^{>0}$. This shows that the number of T gates required to reduce the original $n$ rotations to $\lfloor \log_2 n + 1\rfloor $ rotations is (loosely) upper bounded by
\begin{equation}
8n + 8\sqrt{n} \log_2 n + 12 \log_2 n - 8.
\end{equation}
The grouping method thus requires slightly more than double the $4n-4$ T gates of the method described in the previous section. On the other hand, it only requires $\lfloor \sqrt{n}\rfloor + \lfloor \log_2 n\rfloor$ ancilla qubits, compared with the $n$ required for that method.

We can use the different methods presented in this section within the same circuit, depending on how many rotations we wish to combine at a given point. Say we have at most $n=500$ arbitrary repeated rotations to combine in a circuit. Then we assign $\lfloor \sqrt{n}\rfloor + \lfloor \log_2 n\rfloor = 30$ ancilla qubits for combining those 500 rotations via Hamming weight phasing. If at some other point in the circuit we have 50 rotations to combine, we can do so using the existing 30 ancillae and $4r \lceil n/(r+1)\rceil = 4\times24\times2 = 192$ T gates, reducing the number of arbitrary rotations to be synthesized from 50 to 10. Further improvements in the number of ancilla qubits required are possible using other grouping strategies, but are beyond the scope of this work. We refer the interested reader to work on pebble games (see, e.g., Chapter 10 of \cite{savage1998models}, or \cite{nordstrom2013pebble}).

\subsection{Catalyzing two \texorpdfstring{$\sqrt{\mathrm{T}}$}{sqrt T} gates using 5 T gates}
\label{app:catalyzing}

In this subsection, we present an application of Hamming weight phasing discussed in \cite{gidney2018efficient}. We mentioned in \app{hammingphasingbasic} that when the rotation angles are $\theta = m\pi/2^k$ for integers $k$, $m$, it is most efficient to only combine until the rotations are T gates. These rotations appear in some of the circuits in the paper (specifically, in the FFFT for side lengths which are a power of two), and we present a circuit tailored to $\sqrt{\mathrm{T}}$ and $\mathrm{T} \sqrt{\mathrm{T}}$ gates in this section.

\begin{figure}[ht]
\begin{center}
\(\Qcircuit @R=1em @C=0.7em {
\lstick{\ket\psi} & \qw & \ctrl{2} & \qw & \ctrl{3} & \qw & \ctrl{3} & \qw & \qw & \ctrl{2} & \qw & \rstick{\sqrt{\mathrm{T}} \ket\psi} \qw 
\\
\lstick{\ket\phi}  & \qw & \targ & \ctrl{1} & \qw & \qw & \qw & \ctrl{1} & \ctrl{1} & \targ & \qw & \rstick{\sqrt{\mathrm{T}} \ket\phi} \qw \\
\lstick{\sqrt{\mathrm{T}} \ket+} & \targ & \targ & \ctrl{1} & \qw & \qw & \qw & \ctrl{1} & \targ & \targ & \targ & \rstick{\sqrt{\mathrm{T}} \ket+} \qw \\
& & & \qwx & \targ & \gate{\mathrm{T}} & \targ & \qw
} \rule{1.4cm}{-0.2cm} \raisebox{-2.0em}{~=~} \rule{1cm}{-0.2cm} \raisebox{0.5em}{
\Qcircuit @R=1em @C=0.7em {
\lstick{\ket\psi} & \ctrl{2} & \qw & \ctrl{3} & \qw & \qw & \qw & \ctrl{3} & \qw & \ctrl{2} & \rstick{\sqrt{\mathrm{T}} \ket\psi} \qw \\
\lstick{\ket\phi}  & \targ & \ctrl{1} & \qw & \ctrl{1} & \qw & \ctrl{1} & \qw & \ctrl{1} & \targ & \rstick{\sqrt{\mathrm{T}} \ket\phi} \qw \\
\lstick{\ket+} & \targ & \ctrl{1} & \targ & \targ & \gate{\sqrt{\mathrm{T}}} & \targ & \targ & \ctrl{1} & \targ & \rstick{\sqrt{\mathrm{T}} \ket+} \qw \\
& & \qwx & \targ & \qw & \gate{\mathrm{T}} & \qw & \targ & \qw 
}
}\)
\caption{\label{fig:catalyze_sqrtT} Circuit to catalyze two $\sqrt {\mathrm{T}}$ gates using 5 T gates (left), and equivalent adder-based circuit (right). The left circuit is constructed from the right by moving the $\sqrt{\mathrm{T}}$ gate through to the beginning using circuit identities. While the right circuit applies three $\sqrt{\mathrm{T}}$ gates using a $\sqrt{\mathrm{T}}$ and 5 T gates, the left circuit takes one of the $\sqrt{\mathrm{T}}$ gates as input in the form of a seed state, and applies the other two $\sqrt{\mathrm{T}}$ gates using 5 T gates. We use the ``cornering'' notation for temporary logical \textsc{and} of \cite{gidney2018halving} (see \fig{temporary_and_notation}). The same circuit catalyzes two $\sqrt{\mathrm{T}}^3 = \mathrm{T}\sqrt{\mathrm{T}}$ gates if we replace the seed state with $\sqrt{\mathrm{T}}^3\ket+$ and add an $S$ gate on the ancilla qubit after the T gate.}
\end{center}
\end{figure}
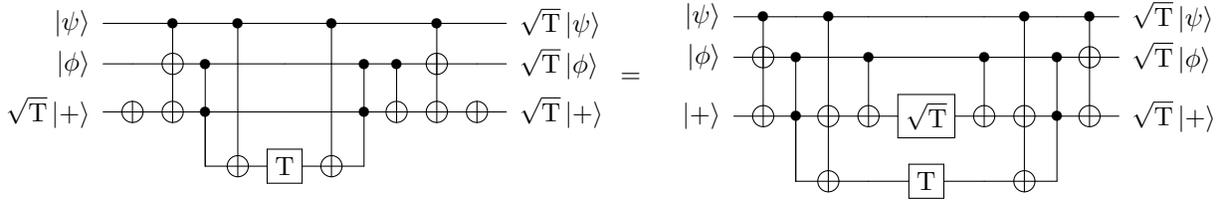

This circuit uses Hamming weight phasing with an ancilla qubit to catalyze $\sqrt {\mathrm{T}}$ gates on two other qubits using only one Toffoli and one T gate, or 5 T gates, given the seed state $\sqrt{\mathrm{T}} \ket+$. The same circuit can catalyze two $\sqrt{\mathrm{T}}^3 = \mathrm{T}\sqrt{\mathrm{T}}$ gates for the same fault-tolerant cost by changing the seed state and applying an $S$ gate on the ancilla qubit in addition to the T gate. The circuit is constructed by moving the $\sqrt{\mathrm{T}}$ gate of the standard adder-based Hamming weight phasing circuit through to the beginning to create the seed state $\sqrt{\mathrm{T}}\ket+$. The seed state must be synthesized at the beginning of the entire computation at full cost, but can be used anywhere after that point---it is not consumed as we generate $\sqrt{\mathrm{T}}$ gates. We show the catalysis circuit and its adder-based counterpart in \fig{catalyze_sqrtT}.

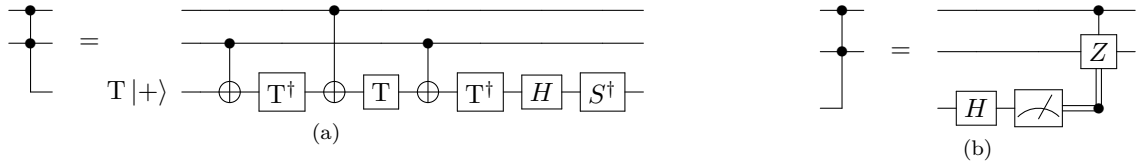
\begin{figure}[hb]
\begin{center}
\subfloat[][]{
\(\Qcircuit @R=1em @C=0.7em {
&\ctrl{1}  &\qw           && &&&&&&\qw                &\qw      &\qw              &\ctrl{2} &\qw      &\qw      &\qw              &\qw      &\qw              &\qw \\
&\ctrl{1}  &\qw          &&=&&&&&&\qw                &\ctrl{1} &\qw              &\qw      &\qw      &\ctrl{1} &\qw              &\qw      &\qw              &\qw \\
&          &\qw           && &&&&&\lstick{\mathrm{T}\ket+} & \qw &\targ    &\gate{\mathrm{T}^\dagger} &\targ    &\gate{\mathrm{T}} &\targ    &\gate{\mathrm{T}^\dagger} &\gate{H} &\gate{S^\dagger} &\qw
}\)
\label{fig:computing_and}
}\hspace{2cm}
\subfloat[][]{
\(\Qcircuit @R=0.8em @C=0.7em {
&\ctrl{1}  &\qw       && &&&\qw                &\qw      &\ctrl{1}          &\qw\\
&\ctrl{1}  &\qw        &&=&&&\qw                &\qw      &\gate{Z}          &\qw\\
&\qw       && &&&&\gate{H}           &\meter   &\control \cw \cwx &
}\)
\label{fig:uncomputing_and}
}
\caption{\label{fig:temporary_and_notation}
Circuits with cornering notation for computing and uncomputing logical \textsc{and} operations, together with the equivalent circuits in the Clifford+T basis \cite{gidney2018halving}. The cornering notation indicates that the qubit is only needed during that section of the circuit. In \fig{catalyze_sqrtT}, the ancilla qubit is available for use before computing and after uncomputing the logical \textsc{and}. \protect\subref{fig:computing_and} Computing the logical \textsc{and} requires one Toffoli or 4 T gates; \protect\subref{fig:uncomputing_and} uncomputing it requires none. }
\end{center}
\end{figure}

We use the ``cornering'' notation of \cite{gidney2018halving,gidney2018efficient} for computing and uncomputing the logical \textsc{and} to indicate that the ancilla qubit is free for use elsewhere in the circuit outside that part of the computation. We review this notation in \fig{temporary_and_notation} and present equivalent Clifford+T circuits.

\section{Trotter steps by fermionic swap network}
\label{app:fermionicswaptrotternetwork}

We review in this appendix the fermionic swap network Trotter step developed in \cite{kivlichan2018quantum}. 
The idea for this Trotter step is that the Hamiltonian \eq{hamiltonian} can be simulated in linear depth in a network which changes the Jordan-Wigner ordering as it runs, such that the qubits corresponding to each spin-orbital are adjacent to the qubits corresponding to each other spin-orbital in the system at some point in the network.

The fermionic swap network does this by reversing the initial Jordan-Wigner ordering of the spin-orbitals with fermionic swaps using odd-even transposition sort (parallel bubble sort), simulating terms in the Hamiltonian that act on neighboring qubits as it goes. 
The sorting algorithm alternates between two different layers of swaps.
In the first layer, each qubit in an odd-numbered position is fermionically swapped with the even-numbered qubit to its right. 
In the second layer, each qubit in an even-numbered position is fermionically swapped with the odd-numbered qubit to its right. 
This is shown for 5 qubits in the leftmost column of \tab{operatorswaps}. After the final set of swaps, the original Jordan-Wigner ordering has been reversed: because this reversal was performed using only nearest-neighbor operations, for each term in the Hamiltonian there must have been a point in the network where the spin-orbitals the term acts on were mapped to adjacent qubits.

Now, suppose in a particular layer of the swap network that spin-orbital $p$ (encoded by qubit $i_p$ in the Jordan-Wigner ordering) undergoes a fermionic swap with spin-orbital $q$ (encoded by qubit $i_q$). 
Since all fermionic swaps occur between adjacent qubits, either $i_q = i_p + 1$ or $i_p = i_q + 1$. 
Thus, the fermionic operator $V_{pq} n_p n_q$ can be simulated by the qubit operators $V_{pq} (\openone - Z_{i_p} - Z_{i_p+1} + Z_{i_p} Z_{i_p+1}) / 4$, and the fermionic operators $T_{pq} (a^\dagger_{p} a_{q} + a^\dagger_{q} a_{p})$ can be simulated by the qubit operators $T_{pq} (X_{i_p} X_{i_p+1} + Y_{i_p} Y_{i_p+1}) / 2$.

This shows how to simulate all the terms in the Hamiltonian involving spin-orbitals $p$ and $q$. We integrate these evolutions with the fermionic swap, and refer to the combined unitary as the ``fermionic simulation gate'' following \cite{kivlichan2018quantum}. 
The fermionic simulation gate performs three operations: it simulates evolution for time $t$ under $T_{pq} (X_{i_p} X_{i_p+1} + Y_{i_p} Y_{i_p+1}) / 2$, simulates evolution for time $t$ under $V_{pq} (Z_{i_p} Z_{i_p+1} - Z_{i_p} - Z_{i_p+1} + \openone) / 4$, and finally fermionically swaps the two spin-orbitals. In matrix form, this gate is
\begin{align}
\label{eq:specialF}
{\cal F}_t\left(i_p, i_q \right) & = \begin{pmatrix}
1 & 0 & 0 & 0 \\
0 & -i \sin \left( T_{pq} t\right) & \cos \left( T_{pq} t\right) & 0 \\
0 & \cos \left( T_{pq} t\right) & -i \sin \left( T_{pq} t\right) & 0 \\
0 & 0 & 0 & -e^{-i V_{pq} t}
\end{pmatrix}.
\end{align}

We can see then that \tab{operatorswaps} actually depicts an entire first-order Trotter step if we interpret the boxes in the canonical ordering as the unitary ${\cal F}_t\left(i_p, i_q \right)$. If they are nonzero, the external potential terms $U_p n_p$ can be simulated by applying single-qubit rotations after any layer of fermionic simulation gates. 

\begin{table*}[tb]
\centering
\begin{tabular}{| c | c | c | c | c |}
\hline
\raisebox{-0.4em}{Canonical Ordering} & \raisebox{-0.4em}{$ \frac12\left(X_1 X_2 + Y_1 Y_2\right)$} & \raisebox{-0.4em}{$ \frac12\left(X_2 X_3 + Y_2 Y_3\right)$} & \raisebox{-0.4em}{$ \frac12\left(X_3 X_4 + Y_3 Y_4\right)$} & \raisebox{-0.4em}{$ \frac12\left(X_4 X_5 + Y_4 Y_5\right)$} \\[0.8em] \hline \hline
\raisebox{-0.7ex}{\framebox(21,10){$\varphi_1 \varphi_2$}} \raisebox{-0.7ex}{\framebox(22,10){$\varphi_3 \varphi_4$}} \raisebox{0.3ex}{$\varphi_5$} & $a^\dagger_1 a_2 + a^\dagger_2 a_1$ & \diagbox[width=2.0cm, height=0.62cm] & $a^\dagger_3 a_4 + a^\dagger_4 a_3$ & \diagbox[width=2.0cm, height=0.62cm]{}{} \\ \hline
\raisebox{0.3ex}{$\varphi_2$} \raisebox{-0.7ex}{\framebox(21,10){$\varphi_1 \varphi_4$}} \raisebox{-0.7ex}{\framebox(22,10){$\varphi_3 \varphi_5$}} & \diagbox[width=2.0cm, height=0.62cm] & $a^\dagger_1 a_4 + a^\dagger_4 a_1$ & \diagbox[width=2.0cm, height=0.62cm] & $a^\dagger_3 a_5 + a^\dagger_5 a_3$ \\ \hline
\raisebox{-0.7ex}{\framebox(21,10){$\varphi_2 \varphi_4$}} \raisebox{-0.7ex}{\framebox(22,10){$\varphi_1 \varphi_5$}} \raisebox{0.3ex}{$\varphi_3$} & $a^\dagger_2 a_4 + a^\dagger_4 a_2$ & \diagbox[width=2.0cm, height=0.62cm] & $a^\dagger_1 a_5 + a^\dagger_5 a_1$ & \diagbox[width=2.0cm, height=0.62cm]{}{} \\ \hline
\raisebox{0.3ex}{$\varphi_4$} \raisebox{-0.7ex}{\framebox(21,10){$\varphi_2 \varphi_5$}} \raisebox{-0.7ex}{\framebox(22,10){$\varphi_1 \varphi_3$}} & \diagbox[width=2.0cm, height=0.62cm] & $a^\dagger_2 a_5 + a^\dagger_5 a_2$ & \diagbox[width=2.0cm, height=0.62cm] & $a^\dagger_1 a_3 + a^\dagger_3 a_1$ \\ \hline
\raisebox{-0.7ex}{\framebox(21,10){$\varphi_4 \varphi_5$}} \raisebox{-0.7ex}{\framebox(22,10){$\varphi_2 \varphi_3$}} \raisebox{0.3ex}{$\varphi_1$} & $a^\dagger_4 a_5 + a^\dagger_5 a_4$ & \diagbox[width=2.0cm, height=0.62cm] & $a^\dagger_2 a_3 + a^\dagger_3 a_2$ & \diagbox[width=2.0cm, height=0.62cm]{}{} \\ \hline
\raisebox{0.3ex}{$\varphi_5$} \raisebox{0.3ex}{$\varphi_4$} \raisebox{0.3ex}{$\varphi_3$} \raisebox{0.3ex}{$\varphi_2$} \raisebox{0.3ex}{$\varphi_1$} & \diagbox[width=2.0cm, height=0.62cm] & \diagbox[width=2.0cm, height=0.62cm] & \diagbox[width=2.0cm, height=0.62cm] & \diagbox[width=2.0cm, height=0.62cm]{}{} \\ \hline
\end{tabular}
\caption{The five layers of swaps for the five spin-orbital fermionic swap network. The boxes in the leftmost column of each row indicate the pairs of qubits involved in a fermionic simulation gate in a given layer. The right columns show which kinetic energy fermion operators the different nearest-neighbor qubit operators correspond to at that layer. The final ordering is shown in the last row of the table for completeness. All $\binom{5}{2} = 10$ possible kinetic energy interactions appear at some point in the table, and are thus applied during the steps shown. The fermionic simulation gate also applies the potential energy interactions $V_{pq} n_p n_q$ at that time. Thus, these five layers implement an entire Trotter step of the Hamiltonian. A slash through a cell indicates that no interaction is performed between that pair of qubits in the layer.
\label{tab:operatorswaps}
}
\end{table*}

The second-order (symmetric) Trotter step can be constructed from the Trotter step described so far by running it once forward, and symmetrizing by running it again in reverse. The spin-orbitals are restored to their original ordering by the second-order Trotter step. 

This Trotter step can simulate any Hamiltonian of the form \eq{hamiltonian}, including the Fermi-Hubbard model, the uniform electron gas, and the electronic structure problem. 
For the Fermi-Hubbard model on an $n\times n$ lattice without periodic boundary conditions, we only need to run through the first ${\cal O}(\sqrt{n})$ layers of fermionic simulation gates \cite{kivlichan2018quantum}. With periodic boundary conditions, the full network must be run to ensure that interactions between the top and bottom rows of the lattice are simulated. 
For the electronic structure problem, the external potential terms $U_p n_p$ arise from the charges of the nuclei $\zeta_j$. These nuclear charges are all that distinguish various molecules and materials from the uniform electron gas. However, because they contribute only to the terms $U_p n_p$, they enter the simulation only as a layer of single-qubit rotations appended to the Trotter step circuit.

\section{Trotter steps of the split-operator algorithm}
\label{app:FFTtrotterstep}

In this appendix, we review how to simulate Trotter steps of the Hamiltonian \eq{hamiltonian} using a variation of the split-operator algorithm introduced in \cite{babbush2018low}. The general principle of a split-operator algorithm is to divide (``split'') the Hamiltonian into parts which can be more easily simulated separately.
In this case, the algorithm separately simulates the kinetic energy operator in the momentum basis, and the potential energy operator in the position basis. We use an additional circuit to change back and forth between the position and momentum bases.
In the momentum basis, the kinetic energy operator is diagonal, and moreover can be simulated using only single-qubit rotations.
In the position basis, the potential energy operator is diagonal, containing only terms of the form $Z_p Z_q$. While this can be simulated using a swap network as in \cite{babbush2018low} or \cite{kivlichan2018quantum}, we will instead develop a specialized network tailored to take full advantage of Hamming weight phasing.

We begin this appendix by describing how to construct the split-operator Trotter step. We then introduce our specialized network for taking full advantage of Hamming weight phasing when simulating the potential energy terms.
Following that, we present the circuits used in the split-operator algorithm to change between the position and momentum bases. When the system side length is a power of 2, we use the fermionic fast Fourier transform (FFFT) as in \cite{verstraete2009quantum,babbush2018low}. We compute the T-cost of the FFFT using rotation catalysis as discussed in \app{catalyzing}. For other system side lengths, we use an alternative approach to diagonalizing the kinetic energy operator using Givens rotations \cite{wecker2015solving,kivlichan2018quantum}. While this second approach has fewer restrictions on the kinetic energy operators it can diagonalize than the FFFT, in the fault-tolerant setting, it is more costly because it uses more arbitrary rotations.

\subsection{Constructing the split-operator Trotter step}

We discuss here how to construct the split-operator Trotter step. Specifically, we will describe the circuit for the second-order (first-order symmetric) split-operator Trotter step.

Recall from \eq{hamiltonian} that we wish to simulate evolution under a fermionic Hamiltonian of the form 
\begin{equation}
H = \sum_{p q} T_{pq} a^\dagger_{p} a_{q}
+ \sum_p U_p n_p
+ \sum_{p\neq q} V_{pq} n_{p} n_{q}.
\end{equation}
In the context of the electronic structure problem, $T_{pq}$, $U_p$, and $V_{pq}$ are real numbers defined by integrals over molecular orbitals, arising from the kinetic energy, external potential, and electron-electron Coulomb repulsion, respectively. The algorithms we discuss can simulate arbitrary Hamiltonians of this form, not necessarily arising from any physical system.
The creation and annihilation operators index the spin-orbitals through vectors $p$ and $q$. This second-quantized Hamiltonian can be mapped by the Jordan-Wigner transform to a qubit Hamiltonian of the form
\begin{equation}
H = \sum_{ p q} \tilde T_{p q} \left(X_{ p} Z_{ r_{i}} Z_{ r_{i+1}} \cdots Z_{ r_{i+k}} X_{ q} + Y_{ p} Z_{ r_{i}} Z_{ r_{i+1}} \cdots Z_{ r_{i+k-1}} Z_{ r_{i+k}} Y_{ q}\right)
+ \sum_{ p} \tilde U_{ p} Z_{ p}
+ \sum_{ p q} \tilde V_{ p q} Z_{ p} Z_{ q},
\end{equation}
where we have omitted the identity term because it is trivial to simulate.
Here, $X_{ p}$, $Y_{ p}$, and $Z_{ p}$ are the Pauli $X$, $Y$, and $Z$ gates, respectively, acting on the qubit indexed by $ p$ in the Jordan-Wigner ordering. $Z_{ r_{i}} \cdots Z_{ r_{i+k}}$ is the string of Pauli $Z$ gates between qubit $ p$ and qubit $ q$ in the Jordan-Wigner order, needed to preserve the appropriate commutation relations for the transformed fermionic creation and annihilation operators.  The coefficients $\tilde T_{ p q}$, $\tilde U_{ p}$, and $\tilde V_{ p q}$ in the qubit Hamiltonian can be directly computed from the coefficients without tildes in the second-quantized Hamiltonian.

The split-operator Trotter step splits evolution $e^{-iHt}$ under the Hamiltonian $H$ into separate evolutions under the kinetic and potential energy parts of the Hamiltonian. The reason this is useful is that the terms in the kinetic energy part of the Hamiltonian (those with coefficients $\tilde T_{p q}$) can be diagonalized by an efficient circuit transformation $C$, leading to a sum of single-qubit terms, that is,
\begin{equation}
\sum_{ p q} \tilde T_{ p q} \left(X_{ p} Z_{ r_{i}} Z_{ r_{i+1}} \cdots Z_{ r_{i+k}} X_{ q} + Y_{ p} Z_{ r_{i}} Z_{ r_{i+1}} \cdots Z_{ r_{i+k-1}} Z_{ r_{i+k}} Y_{ q}\right) = C \left(\sum_{ p} T_{ p} Z_{ p} \right) C^\dagger.
\end{equation}
The diagonalizing circuits $C$ and $C^\dagger$ change between the position and momentum bases using either the FFFT or Givens rotations.
The idea of the split-operator method is that the Trotter-Suzuki approximation can be applied to divide evolution under the total Hamiltonian $H$ into two evolutions: one under the kinetic energy terms
\begin{equation}
T = C \left(\sum_{ p} T_{ p} Z_{ p} \right) C^\dagger,
\end{equation}
that is $e^{-iTt}$, and one under the potential energy terms
\begin{equation}
U+V = \sum_{ p} \tilde U_{ p} Z_{ p} + \sum_{ p q} \tilde V_{ p q} Z_{ p} Z_{ q},
\end{equation}
that is $e^{-i (U+V) t}$. Symmetrizing these evolutions to give the second-order Trotter step, we approximate time evolution under the full Hamiltonian by blocks of evolutions separated by the circuit $C$ and its inverse $C^\dagger$,
\begin{equation}
\label{eq:TV_step}
e^{-iHt} \approx C \left[  \prod_{ p} e^{-i T_{ p} Z_{ p} t / 2} \right] C^{\dagger} \left[ \prod_{ p} e^{-i \tilde U_{ p} Z_{ p} t}
\prod_{ p q} e^{-i \tilde V_{ p q} Z_{ p} Z_{ q} t} \right] C \left[  \prod_{ p} e^{-i T_{ p} Z_{ p} t / 2} \right] C^{\dagger}.
\end{equation}
The equation above describes a single second-order Trotter step: the approximation can be refined by dividing into $\blu r$ such steps, each of time $t/r$.

In fact, we have two choices for how to symmetrize the split-operator Trotter step: as opposed to \eq{TV_step}, in which we approximate evolution for $r$ Trotter steps as $e^{-iHt} \approx (e^{-iTt/2r} e^{-i (U+V) t/r} e^{-iTt/2r})^r$, we can instead approximate it as $e^{-iHt} \approx (e^{-i (U+V) t/2r} e^{-i T t/r} e^{-i (U+V) t/2r})^r$. Including the circuits for diagonalizing the kinetic energy operator, a single Trotter step of this form is
\begin{equation}
\label{eq:VT_step}
e^{-iHt} \approx \left[ \prod_{ p} e^{-i \tilde U_{ p} Z_{ p} t/2}
\prod_{ p q} e^{-i \tilde V_{ p q} Z_{ p} Z_{ q} t} \right] C \left[  \prod_{ p} e^{-i T_{ p} Z_{ p} t} \right] C^{\dagger} \left[ \prod_{ p} e^{-i \tilde U_{ p} Z_{ p} t}
\prod_{ p q} e^{-i \tilde V_{ p q} Z_{ p} Z_{ q} t/2} \right] .
\end{equation}
At first glance, the number of gates required for the two Trotter steps might appear to be different. The number of terms in $T$ is linear in the number of spin-orbitals, whereas the number of terms in $U+V$ is quadratic in the number of spin-orbitals, which would lead us to expect that \eq{VT_step}---in which $e^{-i (U+V) t}$ appears twice---might require many more gates than \eq{TV_step}, in which $e^{-i (U+V) t}$ only appears once. On the other hand, $C$ or $C^\dagger$ appear four times in \eq{TV_step} but only twice in \eq{VT_step}, which might lead us to believe that \eq{TV_step} could use more gates.

However, this intuition is deceptive: so long as the number of Trotter steps $r$ is large, since the beginning and end of both split-operator steps are the same and hence can be merged together, the difference in the costs is negligible.
The result of this merging is that the cost of a simulation with $r$ Trotter steps is nearly identical for the two competing orderings of terms. Only in the final Trotter step do the two orderings differ in cost: \eq{TV_step} applies $C$ one more time and $e^{-i (U+V) t}$ one fewer time than \eq{VT_step}. Both variants apply $C$, $C^\dagger$, $e^{-i T t}$, and $e^{-i (U+V) t}$ once in each of the remaining $r-1$ Trotter steps, far outweighing this final difference.

Rather, it is differences in the number of Trotter steps $r$ required for accurate simulation---dictated by the Trotter error of the two steps---that dominate the difference in cost. For all systems considered, we construct the Trotter step depending on which split-operator step yields the smaller Trotter error.
For the system parameters chosen, simulating $V$ followed by $T$ yields a smaller Trotter error norm for the uniform electron gas with Wigner-Seitz radius $r_s \gtrsim 1$, whereas $T$ followed by $V$ yields a smaller Trotter error for the Hubbard model and the uniform electron gas with Wigner-Seitz radius $r_s \lesssim 1$. 
There is little significance to this beyond the relative sizes of the norms $\|T\|$ and $\|V\|$ (see \eq{TV_order}, \eq{VT_order}, and the following discussion).

Implementing the different circuits in the split-operator steps is straightforward. Each term $e^{-iT_p Z_p t}$, $e^{-i \tilde U_p Z_p t}$, or $e^{-i \tilde V_{pq} Z_p Z_q t}$ requires one arbitrary rotation, whether we implement time evolution, or the evolution directionally controlled on an ancilla $\ket{c}\ket{\psi} \mapsto \ket{c}e^{-i H t (-1)^c} \ket{\psi}$ that we shall use for phase estimation (\app{phase_estimation}). Thus, with $N$ spin-orbitals, at most $N(N+3)/2$ arbitrary rotations are needed to simulate all the terms in the Hamiltonian, and the Trotter step circuit can be completed with one application each of the basis change circuit $C$ and its inverse $C^\dagger$. We address two remaining questions in this appendix: first, how to order the simulation of the terms in $\prod_{pq} e^{-i \tilde V_{pq} Z_p Z_q t}$ so as to take full advantage of Hamming weight phasing, thereby reducing the number of arbitrary rotations required; and second, how to construct the circuit $C$ using either the FFFT or Givens rotations.

\subsection{Reducing the cost of simulating the potential energy operator using Hamming weight phasing}
\label{app:HWP_potential}

In this subsection we present a technique for reducing the number of arbitrary rotations needed to implement evolution under the interaction part of the potential energy operator $\prod_{pq} e^{-i \tilde V_{pq} Z_p Z_q t}$ in the split-operator Trotter step. Evolution under the terms $\prod_{p} e^{-i \tilde U_{p} Z_p t}$ can be done using only a single arbitrary rotation on each qubit. By comparison, $\prod_{pq} e^{-i \tilde V_{pq} Z_p Z_q t}$ in general appears to require ${\cal O}(N^2)$ arbitrary rotations for $N$ qubits. We will show here how to use Hamming weight phasing to reduce this cost for the electronic structure problem to ${\cal O}(N \log N)$ arbitrary rotations, at a cost of $(N-1)(N-2)$ additional Toffoli or $4(N-1)(N-2)$ additional T gates. Because the number of T gates needed to synthesize each arbitrary rotation is much larger than 4, this greatly reduces the fault-tolerant cost of simulating these terms.

The idea behind our technique is that in the basis of \cite{babbush2018low}, the coefficients $\tilde V_{ p,  q}$ are translation-invariant for the electronic structure Hamiltonian, that is, for any index vector $ s$,
\begin{equation}
\label{eq:Vpq_transinv}
\tilde V_{ p,  q} = \tilde V_{ p + s,  q + s}.
\end{equation}
We will harness this property so that we can apply Hamming weight phasing to groups of angles that are as large as possible. Hamming weight phasing allows $M$ parallel rotations by the same angle to be reduced to $\lfloor \log_2 M + 1\rfloor$ rotations by different angles, and $4M-4$ additional T gates (alternatively, $M-1$ additional Toffoli gates), using $M-1$ ancilla qubits. We can fully make use of Hamming weight phasing if we can order the terms $e^{-i \tilde V_{pq} Z_p Z_q t}$ in the simulation such that equiangular rotations are grouped together as much as possible. 
Note that we can compute and uncompute the parity of qubits $p$ and $q$ at no additional cost using a \textsc{CNOT} gate.

We do this by performing the evolution $\prod_{pq} e^{-i \tilde V_{pq} Z_p Z_q t}$ in layers such that the value of every $\tilde V_{ p q}$ simulated within a given layer is the same. In this case, the arbitrary rotations used in each layer are all by the same angle, allowing them to be fully combined using Hamming weight phasing. We accomplish this as follows. We choose the interactions simulated in each layer according to the set of pairs of qubits $(p, p +  s)$ with fixed $ s$, with the index vector $p$ running over all qubits. The translation-invariance of $\tilde V_{ p,  q}$ guarantees, for all qubit indices $p$, that
\begin{equation}
\tilde V_{{ 0},  s} = \tilde V_{{p},  p +  s}.
\end{equation}

\begin{figure}[htb]
\centering
\subfloat[][]{
\includegraphics[width=0.3\textwidth]{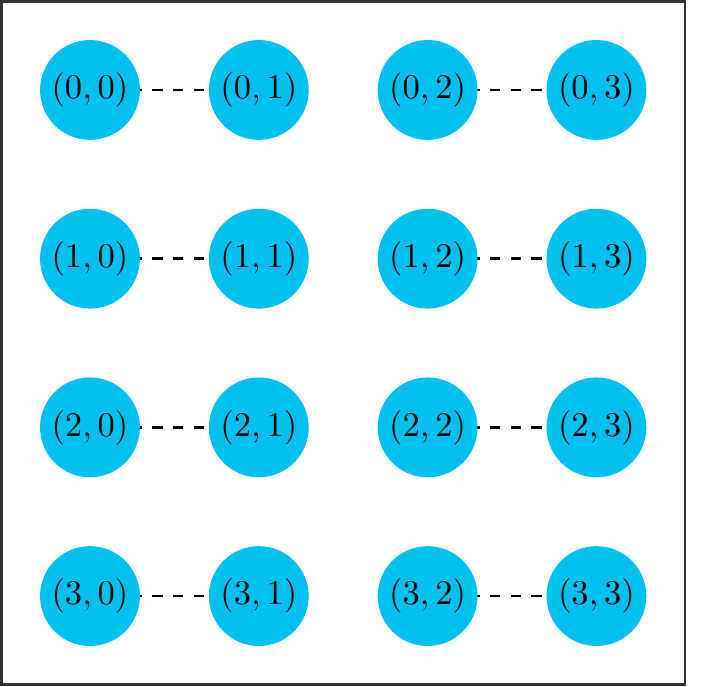}
\label{fig:matrix}
}\hspace{0.5cm}
\subfloat[][]{
\includegraphics[width=0.3\textwidth]{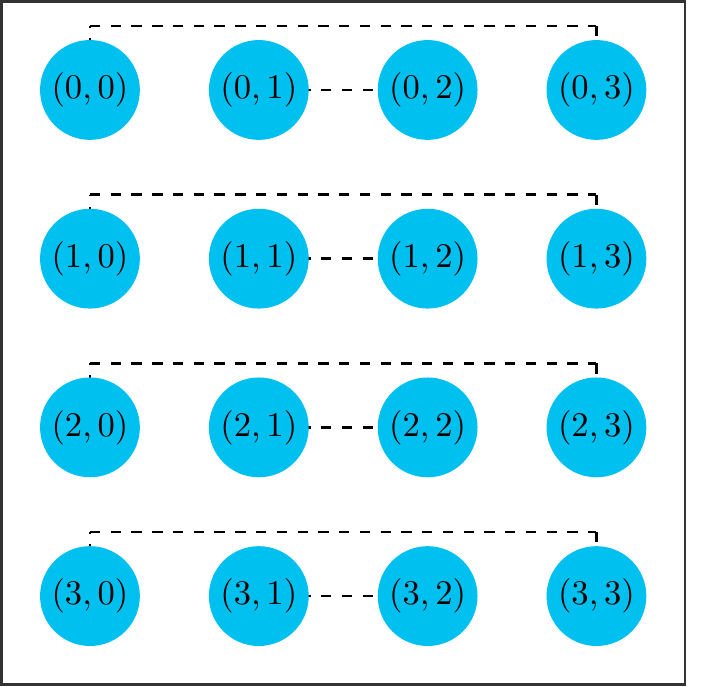}
\label{fig:matrix_offset}
}
\caption{Two-qubit evolution terms (dotted lines) simulated in the two layers of the swap network with the index vector $ s = (0, 1)$. In both layers, qubit $p$ interacts with qubit $p +  s$, where the elements of the index vectors $p$ and $p+s$ are evaluated modulo the grid length. \protect\subref{fig:matrix} Qubit $ p$ interacts with qubit $ p +  s$, starting with $ p = (0, y)$ and $ p = (2, y)$. \protect\subref{fig:matrix_offset} Qubit $ p$ interacts with qubit $ p +  s$, starting with $ p = (1, y)$ and $ p = (3, y)$.
\label{fig:interactions}
}
\end{figure}

We first consider the case where the number of qubits $N$ is even; when $N$ is odd the layers differ only slightly from the even $N$ case. For $N$ qubits (with $N$ even), there are $N$ such pairs of qubits $( p,  p +  s)$, given by the $N-1$ different vectors $s$ on the grid. However, because each of these interactions is between two qubits, at most $N/2$ of them can be performed simultaneously in a layer. We show a diagram of the two layers used for $ s = (0, 1)$ in \fig{interactions} for the case of 16 qubits on a $4\times4$ grid. Note that the vector indices on the qubits are evaluated modulo 4 for the $4\times4$ grid. In each layer, we are able to simulate $N/2$ interactions of equal strength, yielding $N/2$ equiangular rotations combinable using Hamming weight phasing.

In general, there are $N-1$ different index vectors $s$ that must be iterated over. (The final vector, $s = 0$, does not appear because it corresponds to a qubit interacting with itself.) Each index vector is applied in two layers, leading to $2N-2$ layers each simulating $N/2$ interactions. In each of the $2N-2$ layers, Hamming weight phasing can be applied over every simulated term $e^{-i \tilde V_{ p,  p +  s} Z_{ p +  s} Z_{ s} t}$, because each coefficient $\tilde V_{ p,  p +  s}$ is the same, and hence each of the $N/2$ rotation angles is the same. 

This allows Hamming weight phasing to be applied over every rotation in each layer, reducing the number of arbitrary rotations required per layer from $N/2$ to $\lfloor \log_2 (N/2) + 1\rfloor$, at a cost of an additional $2N-4$ T gates or $N/2 - 1$ Toffoli gates. Over all the $2N-2$ layers, the T-cost to simulate all the two-qubit terms in a step of Trotter evolution is
\begin{equation}
4(N-1)(N-2) + T_\text{synth} (2N-2) \lfloor \log_2 N \rfloor ,
\end{equation}
where $T_\text{synth} = {\cal O}(\log(1/\epsilon))$ is the number of T gates to synthesize each arbitrary rotation, with $\epsilon$ the required synthesis precision.

When the number of qubits $N$ is odd, we instead must simulate each of the $N-1$ index vectors using three rather than two layers. 
Because the number of qubits is odd, one qubit is left out in each layer as compared to the even case---its partner is already matched with another qubit. As a result there are $2N-2$ layers of $(N-1)/2$ equiangular rotations and $N-1$ layers with a lone rotation in the simulation. 
We use Hamming weight phasing to reduce the $(N-1)/2$ equiangular rotations per layer to $\lfloor \log_2 (N-1)/2 + 1\rfloor$ at a cost of an additional $(N-1)/2$ Toffoli or $2N-2$ T gates. Over all the layers, the T-cost to simulate all the two-qubit terms in a step of Trotter evolution is
\begin{equation}
4(N-1)^2 + T_\text{synth} (N-1) \left( 2 \lfloor \log_2 (N-1) \rfloor + 1\right) ,
\end{equation}
for the case of odd $N$.

For both even and odd $N$, we reduce the cost of the arbitrary rotations to ${\cal O}(N \log N \log(1/\epsilon))$ using at most $4(N-1)^2$ additional T gates or $(N-1)^2$ Toffoli gates. This is to be compared with the ${\cal O}(N^2 \log(1/\epsilon))$ cost were we to not apply Hamming weight phasing, or were the interaction terms not translation-invariant. This improvement applies to any quantum simulation of electronic structure using the basis of \cite{babbush2018low}, or more generally to any simulation of a two-qubit interaction with the translation-invariance property \eq{Vpq_transinv}.

\subsection{The fast fermionic Fourier transform}

The fermionic fast Fourier transform (FFFT) is one way of implementing the circuit $C$ which diagonalizes the kinetic energy operator. In this appendix, we define the FFFT, present circuits for it, and compute the T-cost of using it to change from the position to momentum bases using rotation catalysis.

The effect of the fermionic Fourier transform is a single-particle rotation
\begin{equation}
c^\dagger_\nu = \frac{1}{\sqrt{N}} \sum_p a^\dagger_p e^{-ik_\nu \cdot r_p} ~\qquad~~ c_\nu = \frac{1}{\sqrt{N}} \sum_p a_p e^{ik_\nu \cdot r_p}
\end{equation}
with $r_p = p(2\Omega / N)^{1/d}$ and $k_\nu = 2\pi\nu / \Omega^{1/d}$ the centroid of the orbital given by the index vector $p$, and $k_\nu$ a momentum mode as in \sec{trottersteps}.

The circuit for the FFFT is composed of two types of local gates. The first is the fermionic swap gate, which has the effect of exchanging two orbitals, while maintaining proper anti-symmetrization. Under the Jordan-Wigner transform, the fermionic swap gate $f_\text{swap}$ has the matrix representation
\be
f_\text{swap} = \begin{pmatrix}
1 & 0 & 0 & 0 \\
0 & 0 & 1 & 0 \\
0 & 1 & 0 & 0 \\
0 & 0 & 0 & -1
\end{pmatrix},
\ee
i.e.\ it swaps two qubits and applies a sign $-1$ to the $\ket{11}$ state. The second gate used for the FFFT has the matrix representation
\be
F_{k,n} = \begin{pmatrix}
1 & 0 & 0 & 0 \\
0 & \frac{1}{\sqrt2} & \frac{\exp(2\pi i k/n)}{\sqrt2} & 0 \\
0 & \frac{1}{\sqrt2} & -\frac{\exp(2\pi i k/n)}{\sqrt2} & 0 \\
0 & 0 & 0 & -\exp(2\pi i k/n)
\end{pmatrix}
\ee
where $n$ is the number of qubits the FFFT acts on, and $k$ is an integer. This has the effect of a rotation on the first qubit followed by a Hadamard transform in the single-particle subspace. Our notation for this gate is slightly different from other works discussing the FFFT \cite{verstraete2009quantum,babbush2018low} which denote the same gate by $F_k$ (with the number of qubits $n$ left implicit). In this sense our presentation is closer to that in \cite{ferris2014fourier}.

The FFFT on $n$ qubits can be implemented recursively in depth ${\cal O}(n \log n)$ using only the fermionic swap gate $f_\text{swap}$ and the gates $F_{k,n}$ for integer $k$ \cite{babbush2018low}. This recursive structure constrains the FFFT to operate on systems whose side lengths are powers of 2. 
The strategy for this recursive circuit parallels the recursive structure of the classical fast Fourier transform, and can be broken into four stages. 
\begin{enumerate}
\item First, the $n/2$-qubit FFFT is applied separately to the first and second halves of the qubit array. The base case is the two-qubit FFFT, which is the gate $F_{0,2}$.
\item Next, an interleaving step of fermionic swaps puts the qubits in bit-reversed order. (The qubits are re-ordered based on their index, viewed as binary digits, in reverse. 
For example $[0, 1, 2, 3]$ written in binary digits is $[00, 01, 10, 11]_b$ and hence would be re-ordered as $[00, 10, 01, 11]_b = [0, 2, 1, 3]$.) The purpose of this reordering step is to separate the groups of qubits by parity of the index---they are now separated by their final digit. The recursive substructures of the FFFT operate differently depending on whether the final digit is 0 or 1. 
\item The reordered qubits are paired, and $F_{i, n}$ applied to phase the $i$th pair for $i \in [0, n/2-1]$. 
\item Finally, a de-interleaving step restores the qubits to their initial order.
\end{enumerate}

\begin{figure}[ht]
\begin{center}
\scalebox{0.9}{
\Qcircuit @R=1.0em @C=0.7em {
& & & & & \raisebox{1.4em}{\hspace{-6em} \mbox{$2\times$ FFFT$_4$}} & & & & & & \raisebox{1.4em}{\hspace{+1.5em} \mbox{interleave + phases on bits}} & & & & & & \raisebox{1.4em}{\mbox{de-interleave}} & & \\
& \qw & \multigate{1}{F_{0,2}} & \qw & \qw & \multigate{1}{F_{0,4}} & \qw & \qw & \qw & \qw & \qw & \qw & \qw & \multigate{1}{F_{0,8}} & \qw & \qw & \qw & \qw & \qw & \qw & \qw \\
& \qw & \ghost{F_{0,2}} & \qw & \multigate{1}{f_\text{swap}} & \ghost{F_{0,4}} & \multigate{1}{f_\text{swap}} & \qw & \qw & \qw & \qw& \qw  & \multigate{1}{f_\text{swap}} & \ghost{F_{0,8}} & \qw & \qw & \multigate{1}{f_\text{swap}} & \qw & \qw & \qw & \qw \\
& \qw & \multigate{1}{F_{0,2}} & \qw & \ghost{f_\text{swap}} & \multigate{1}{F_{1,4}} & \ghost{f_\text{swap}} & \qw & \qw & \qw& \qw  & \multigate{1}{f_\text{swap}} & \ghost{f_\text{swap}} & \multigate{1}{F_{1,8}} & \qw & \qw & \ghost{f_\text{swap}} & \multigate{1}{f_\text{swap}} & \qw & \qw & \qw \\
& \qw & \ghost{F_{0,2}} & \qw & \qw & \ghost{F_{1,4}} & \qw & \qw & \qw & \qw & \multigate{1}{f_\text{swap}} & \ghost{f_\text{swap}} & \multigate{1}{f_\text{swap}} & \ghost{F_{1,8}} & \qw & \qw & \multigate{1}{f_\text{swap}} & \ghost{f_\text{swap}} & \multigate{1}{f_\text{swap}} & \qw & \qw \\
& \qw & \multigate{1}{F_{0,2}} & \qw & \qw & \multigate{1}{F_{0,4}} & \qw & \qw & \qw & \qw & \ghost{f_\text{swap}} & \multigate{1}{f_\text{swap}} & \ghost{f_\text{swap}} & \multigate{1}{F_{2,8}} & \qw & \qw & \ghost{f_\text{swap}} & \multigate{1}{f_\text{swap}} & \ghost{f_\text{swap}} & \qw & \qw \\
& \qw & \ghost{F_{0,2}} & \qw & \multigate{1}{f_\text{swap}} & \ghost{F_{0,4}} & \multigate{1}{f_\text{swap}} & \qw & \qw & \qw & \qw & \ghost{f_\text{swap}} & \multigate{1}{f_\text{swap}} & \ghost{F_{2,8}} & \qw & \qw & \multigate{1}{f_\text{swap}} & \ghost{f_\text{swap}} & \qw & \qw & \qw \\
& \qw & \multigate{1}{F_{0,2}} & \qw & \ghost{f_\text{swap}} & \multigate{1}{F_{1,4}} & \ghost{f_\text{swap}} & \qw & \qw & \qw & \qw & \qw & \ghost{f_\text{swap}}  & \multigate{1}{F_{3,8}} & \qw & \qw & \ghost{f_\text{swap}} & \qw & \qw & \qw & \qw \\
& \qw & \ghost{F_{0,2}} & \qw & \qw & \ghost{F_{1,4}} & \qw & \qw & \qw & \qw & \qw & \qw & \qw & \ghost{F_{3,8}} & \qw & \qw & \qw & \qw & \qw & \qw & \qw \gategroup{2}{7}{6}{3}{1.4em}{^\}} \gategroup{2}{8}{5}{2}{1.5em}{--} \gategroup{6}{8}{9}{2}{1.5em}{--} \gategroup{2}{10}{9}{14}{0.6em}{--} \gategroup{2}{10}{8}{14}{1.3em}{^\}} \gategroup{2}{16}{9}{20}{1.2em}{--} \gategroup{2}{17}{9}{19}{2.2em}{^\}} \\
}
}
\caption{\label{fig:ffft8} Circuit for the 8-qubit fermionic fast Fourier transform (FFFT$_8$) using only two-qubit nearest-neighbor gates. The dotted boxes denote the different stages of the decomposition of the FFFT circuit. Furthest left are the two 4-qubit FFFTs (FFFT$_4$) that appear in the recursive structure of the 8-qubit FFFT. Within the 4-qubit FFFT, the two-qubit gate $F_{0,2}$ is the recursive base case: the 2-qubit FFFT is exactly the gate $F_{0,2}$.
The triangular structures of fermionic swap gates interleave the qubits (putting them in bit-reversed order, as described in the text) before they are phased by the gates $F_{i,8}$. Finally, the de-interleaving step returns the qubits to their original ordering.
Note that several of the phasing gates in the circuit are the same: $F_{k,n} = F_{j,m}$ if $k/n=j/m$. For example, $F_{1,4}$ in the 4-qubit gate is equivalent to $F_{2,8}$ in the 8-qubit FFFT.}
\end{center}
\end{figure}
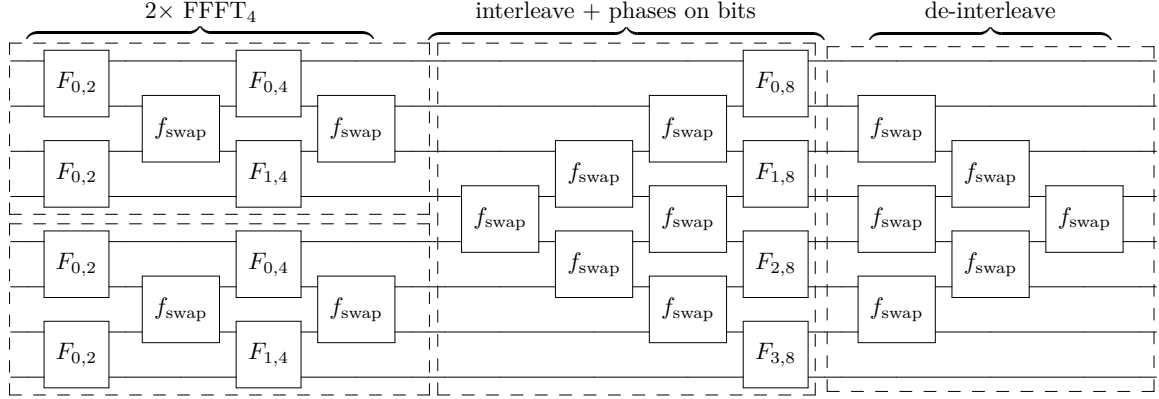

We present the circuit for the 8-qubit FFFT on a nearest neighbor architecture in \fig{ffft8}, emphasizing the different stages in the circuit. Stage 3 is only a single layer of gates and we merge it into stage 2.

\newcommand\coolover[2]{\mathrlap{\smash{\overbrace{\phantom{%
    \begin{matrix} #2 \end{matrix}}}^{\mbox{$#1$}}}}#2} 

So far, we have discussed the structure of the FFFT circuit. Next we will compute the number of T gates required to implement it to determine the cost of the FFFT on a fault-tolerant quantum computer. We do this by presenting explicit circuits for the gate $F_{k,n}$, and counting how many of them appear in the circuit.

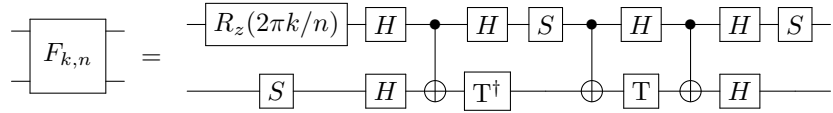
\begin{figure}[ht]
\begin{center}
\[\Qcircuit @R=1em @C=0.7em {
& \multigate{1}{F_{k,n}} & \qw \\
& \ghost{F_{k,n}} & \qw
} \rule{.3em}{-0.2cm} \raisebox{-1.0em}{~=~} \rule{.3em}{-0.2cm} \raisebox{0.5em}{
\Qcircuit @R=1em @C=0.7em {
& \gate{R_z(2\pi k/n)} & \gate{H} & \ctrl{1} & \gate{H} & \gate{S} & \ctrl{1} & \gate{H} & \ctrl{1} & \gate{H} & \gate{S} & \qw \\
& \gate{S} & \gate{H} & \targ & \gate{\mathrm{T}^\dagger} & \qw & \targ & \gate{\mathrm{T}} & \targ & \gate{H} & \qw & \qw
}}\]
\caption{\label{fig:betterFk} Fault-tolerant circuit for the two-qubit fermionic Fourier transform gate $F_{k,n}$ in the $n$-mode FFFT, up to global phase. The T and S gates are $\mathrm{T}=R_z(\pi/4)$ and $S=\mathrm{T}^2$. If $4k/n$ is an integer, the rotation at the beginning is at most an S gate; if instead $8k/n$ is an integer, it is an additional (third) T gate. The largest case we consider is applying the FFFT to a system with side length $n=16$: in that case, $\sqrt{\mathrm{T}}$ gates can appear in the rotation. These can be synthesized in pairs using only 5 T gates per pair, rather than the cost of a full arbitrary rotation.}
\end{center}
\end{figure}

The recursive structure of the $n$-qubit FFFT (where $n$ is necessarily a power of two) means that it uses $n/2$ $F_{k,n}$ gates in $\log_2 n$ recursive layers, yielding $(n \log_2 n)/2$ $F_{k,n}$ gates in total. The second gate used in the circuit, the fermionic swap gate, appears $n(n - 2)/4$ times directly in the interleave and de-interleave steps of the $n$-qubit FFFT, as well as in the interleave/de-interleave steps of the two recursive $n/2$-qubit FFFTs. In total, this is
\be
\sum_{i=1}^{\log_2 n - 1} n (n - 2^{i-1}) / 2^i = n(n - \log_2(8n) / 2)
\ee
$f_\text{swap}$ gates. In fact, $f_\text{swap}$ gates are free in this model, requiring only Clifford gates, and we will not discuss them further here. By comparison, any fault-tolerant circuit construction for $F_{k,n}$ will require T gates. We present one such construction in \fig{betterFk}---note that $T^\dagger = Z S T$. $F_{k,n}$ requires two T gates when $4k/n$ is an integer, three T gates if instead $8k/n$ is an integer, and two T gates and an arbitrary rotation otherwise. In the particular case that $16k/n$ is an integer, the arbitrary rotation is a $\sqrt{\mathrm{T}}$ gate. Recall from \app{catalyzing} that provided that $\sqrt{\mathrm{T}}$ gates appear two at a time in the circuit, we can implement them using only 2.5 T gates each.

This means that the T-cost of the FFFT circuit is 8 for the $4$-qubit case, 26 for the $8$-qubit case, and 71 T gates + 4 $\sqrt{\mathrm{T}}$ gates $=$ 81 T gates for the $16$-qubit case, not accounting for the cost of preparing the seed $\sqrt{\mathrm{T}} \ket+$ state. Next, we will shift our attention to the FFFT in multiple dimensions. Following that, we will show how to minimize the T-cost of the 2D FFFT using rotation catalysis, when the system being Fourier transformed is large enough that the FFFT has non-trivial rotations (rotations which must be synthesized).

\subsubsection{Changing basis in multiple dimensions}
\label{app:multidim_basis_change}

We use the fermionic Fourier transform or Givens rotations to implement the circuit $C$ to diagonalize the kinetic energy terms in the Hamiltonian in the split-operator algorithm. Earlier in this section, we discussed the number of arbitrary rotations required for the fast fermionic Fourier transform on $n$ qubits in one dimension. For a $d$-dimensional system, however, the 1D FFFT circuit must be applied along each of the system dimensions. Though we have not yet discussed it, this also holds if we implement $C$ using Givens rotations, in the sense that whether we use the FFFT or Givens rotations, the $d$-dimensional basis change is composed of the same pattern of 1D basis changes. The discussion in this section applies identically to both approaches.

In each dimension, the 1D basis change must be applied once for each value of the other coordinates. For example, for a hypercube with side length $n$ in $d$ dimensions, transforming along one dimension requires $n^{d-1}$ applications of the 1D basis change. Repeating this once for each dimension, the 1D basis changing circuit must be applied $d n^{d-1}$ times, in $d$ layers of $n^{d-1}$ basis changing circuits. 

More concretely, for a $4\times4\times4$ grid, we apply the 4-qubit FFFT 16 times to Fourier transform the $x$ dimension. The 1D 4-qubit FFFT is applied to the 4 qubits with fixed $(y, z)$ and varying $x$ coordinate. There are 16 pairs $(y, z)$ so the 4-qubit FFFT is applied 16 times. This applies the fermionic Fourier transform in the $x$ dimension: the procedure is repeated for the $y$ dimension (to each of the 16 groups of 4 qubits with the same $x$ and $z$ coordinates, with the $y$ coordinate varying), and again for the $z$ dimension (to groups with the $x$ and $y$ coordinates now fixed, and the $z$ coordinate varying). In total, the 4-qubit FFFT is applied $d n^{d-1} = 48$ times for this system.

For a system with spin, there are still $d$ parallel stages in which the 1D basis changing circuit is applied multiple times. However, in each stage, the 1D circuit is applied $2n^{d-1}$ times, twice as many as in the corresponding spinless system. This doubling occurs because of the two spin sectors.

In general, this means that the 1D basis transformation circuits are applied many times in parallel, whether we use the FFFT or Givens rotations. Because the Givens rotation circuits contain arbitrary rotations, this allows us to apply Hamming weight phasing (\app{combining}) to reduce the number of these arbitrary rotations which must be synthesized. For the FFFT, because the rotations are always of the form $R_z(2\pi k/n)$ with $n$ a power of two, Hamming weight phasing is less efficient. Instead, as we discuss in the next section, we can use rotation catalysis as in \app{catalyzing} to apply e.g.\ $\sqrt{\mathrm{T}}$ and $\mathrm{T} \sqrt{\mathrm{T}}$ gates using only T gates, without the usual costs of rotation synthesis. We discuss this in the next subsection for the case of the $16\times16$ grid, both with and without spin.

\subsubsection{Rotation catalysis and the 2D FFFT}

Of the systems we study, there is only one grid size for which rotation catalysis (see \app{catalyzing}) is beneficial for the FFFT. This is because the FFFT only has a non-trivial rotation---a rotation which is not entirely composed of T, $S$, and $Z$ gates---for systems with side lengths greater than 8. Given that the FFFT can only be applied when the side length is a power of 2, the first such side length is 16. At this size, the FFFT circuit has several $\sqrt{\mathrm{T}}$ and $\sqrt{\mathrm{T}}^3 = \mathrm{T}\sqrt{\mathrm{T}}$ gates. The $16\times16$ grid appears for the largest spinless uniform electron gas system we consider, and with spin for the Fermi-Hubbard model.

Fourier transforming with side length $n=16$ in $d=2$ dimensions applies the $16$-qubit (1D) FFFT 32 times without spin, and 64 times with spin. This represents two layers each of 16 (32) parallel applications of the 1D circuit without (with) spin. For the 16-qubit FFFT, the rotation $R_z(2\pi k/n)$ in the gate $F_{k,16}$ is a $\{\sqrt{\mathrm{T}}, ~ \mathrm{T}\sqrt{\mathrm{T}}, ~ S\sqrt{\mathrm{T}}, ~ S\mathrm{T}\sqrt{\mathrm{T}}\}$ gate for $k \in \{1, ~ 3, ~ 5, ~ 7\}$, respectively. Ignoring the Clifford gates, the 16-qubit FFFT uses two $\sqrt{\mathrm{T}}$ and two $\mathrm{T}\sqrt{\mathrm{T}}$ gates. Recall from \fig{betterFk} that every $F_{k,n}$ gate costs two T gates in addition to these, and that there are $(n \log_2 n) / 2 = 32$ $F_{k,n}$ gates in the $16$-qubit FFFT. Further, for $k = 2$ and $k=6$, the additional rotation in $F_{k,16}$ contains a third T gate, and $F_{1,8}$ and $F_{3,8}$ contribute another 4 T gates (these gates appear twice, since the 8-qubit FFFT appears twice in the recursive circuit for the 16-qubit FFFT). There are thus 70 T gates in the 16-qubit FFFT.

Because the 16-qubit FFFT is applied 32 (64) times without (with) spin, changing between the position and momentum basis using it would appear to require us to distill 2240 (4480) T gates and additionally synthesize 128 (256) $\sqrt{\mathrm{T}}$ gates. Instead, by using rotation catalysis, we can reduce this to two arbitrary rotations to create the seed states at the beginning of the computation, regardless of how many split-operator Trotter steps we apply. We then use these seed states to catalyze all the $\sqrt{\mathrm{T}}$ and $\sqrt{\mathrm{T}}^3 = \mathrm{T}\sqrt{\mathrm{T}}$ gates at a cost of half a Toffoli and half a T gate each, rather than the $\sim\!40$--$50$ T gates per rotation required for synthesis. The only downside is the need for three ancilla qubits: two to store the seed states $\sqrt{\mathrm{T}}\ket+$ and $\mathrm{T}\sqrt{\mathrm{T}}\ket+$, and one to temporarily compute logical \textsc{and}s as in \fig{catalyze_sqrtT}.

In total, converting between the position and momentum bases using the 2D fermionic Fourier transform on a $16\times16$ grid requires 2304 T and 64 Toffoli gates without spin and 4608 T and 128 Toffoli gates with spin. Because the other cases for which we use the FFFT (systems with side length 4 and 8) do not require arbitrary rotations in the FFFT, there is no advantage to these catalysis techniques. Finally, for side lengths that are not a power of two, we cannot construct the diagonalizing circuit $C$ using the above FFFT construction. For these cases, we instead diagonalize using Givens rotations.  We elaborate upon this diagonalization process in the next subsection.

\subsection{Diagonalizing the kinetic energy operator using Givens rotations}
\label{app:Givens_Tcost}

Whereas the FFFT allows transformations of the fermionic mode operators based on the Fourier transform, Givens rotations allow arbitrary unitary transformations to be performed on the mode operators \cite{kivlichan2018quantum}
\begin{equation}
\tilde a^\dagger_i = \sum_j U_{ij} a^\dagger_j, \qquad \tilde a_i = \sum_j U^*_{ij} a_j
\end{equation}
for unitary $U$, with $a^\dagger_i$ and $a_i$ the initial creation and annihilation operators, respectively. This means we can construct a basis change circuit $C$ capable of diagonalizing any number-conserving quadratic Hamiltonian $\sum_{pq} T_{pq} a^\dagger_p a_q$. By comparison, implementations of the diagonalizing circuit $C$ using the FFFT are limited to the case where $\tilde a^\dagger_i$ is the Fourier transform of $a^\dagger_i$; the recursive structure (radix-2 decimation) further restricts the FFFT to act on a number of qubits $n$ which is a power of 2. 
Other radices in the fermionic Fourier transform are possible in principle (and in fact their recursive base cases could be implemented using Givens rotations), but the corresponding circuits require more rotation gates than the radix-2 FFFT. 
Whereas the most challenging gates in the FFFT are rotation gates $R_z(2\pi k/n)$ with $n$ a power of 2, as we shall see the majority of the rotation gates used when we construct $C$ with Givens rotations are by arbitrary angles.

The reasoning behind the diagonalization of the kinetic energy operator using Givens rotations is as follows. By the Thouless theorem, applying the single-particle ($n\times n$) unitary transformation $U$ is equivalent to applying the $2^n\times 2^n$ unitary
\begin{equation}
\exp \left( \sum_{ij} [\log U]_{ij} (a^\dagger_i a_j - a^\dagger_j a_i) \right).
\end{equation}
This unitary can be implemented using a sequence of rotations of the form \cite{kivlichan2018quantum}
\begin{equation}
R_{ij}(\theta) = \exp[\theta_{ij} (a^\dagger_i a_j - a^\dagger_j a_i)].
\end{equation} 
The action of $R_{ij}(\theta)$ is a Givens rotation by $\theta$ in the single-particle subspace. 
Appropriate choices of the angles $\theta$ allows us to diagonalize the matrix $U$, leaving $n$ phases along the diagonal. The unitary $U$ can be applied by applying these $n$ phases to the qubit corresponding to that spin-orbital, followed by the generated sequence of Givens rotations in reverse. For $n$ orbitals, $\binom{n}{2}$ Givens rotations are sufficient for each spin sector. The spinless case thus requires exactly $n(n-1)/2$ Givens rotations, whereas for the case with spin $n(n-1)$ Givens rotations are required \cite{kivlichan2018quantum}. 

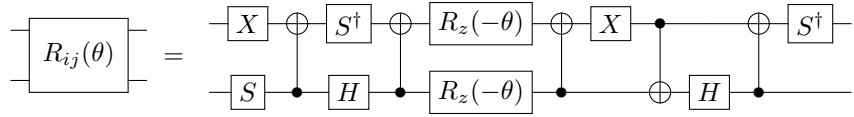
\begin{figure}[ht]
\begin{center}
\[\Qcircuit @R=1em @C=0.7em {
& \multigate{1}{R_{ij}(\theta)} & \qw \\
& \ghost{R_{ij}(\theta)} & \qw
} \rule{.3em}{-0.2cm} \raisebox{-1.0em}{~=~} \rule{.3em}{-0.2cm} \raisebox{0.5em}{
\Qcircuit @R=1em @C=0.7em {
& \gate{X} & \targ & \gate{S^\dagger} & \targ & \gate{R_z(-\theta)} & \targ & \gate{X} & \ctrl{1} & \qw & \targ & \gate{S^\dagger} & \qw \\
& \gate{S} & \ctrl{-1} & \gate{H} & \ctrl{-1} & \gate{R_z(-\theta)} & \ctrl{-1} & \qw & \targ & \gate{H} & \ctrl{-1} & \qw & \qw
}}\]
\caption{\label{fig:givens} Fault-tolerant circuit for $R_{ij}(\theta) = \exp[\theta_{ij} (a^\dagger_i a_j - a^\dagger_j a_i)]$ on neighboring qubits, corresponding to a Givens rotation in the single-particle subspace. Two arbitrary rotations $R_z(-\theta)$ are required. Because these rotations are in parallel, when the same Givens rotation is applied simultaneously many times (i.e.\ when changing bases in multiple dimensions) they can all be combined using Hamming weight phasing. For $n$ orbitals, the circuit must be applied $n(n-1)/2$ times in each spin sector.}
\end{center}
\end{figure}

We show a fault-tolerant circuit implementing $R_{ij}(\theta)$ between neighboring qubits in \fig{givens}. Each Givens rotation $R_{ij}(\theta)$ requires $2$ arbitrary rotations, so the total number of arbitrary rotations required per spin sector is $n(n-1)$.

In total, without spin, the Givens rotations require $n(n-1)$ arbitrary rotations for any single-particle basis change using this procedure, while with spin $2n(n-1)$ are needed. The diagonal phases require $n$ arbitrary rotations without spin and $2n$ arbitrary rotations with spin. 

For diagonalizing the kinetic energy operator of a system with side length $n$ in $d$ dimensions, the discussion of \app{multidim_basis_change} exactly carries over. The 1D ($n$-qubit) basis change circuit is applied several times in parallel, in $d$ separate stages. Within each of the $d$ stages, the 1D circuit is applied $n^{d-1}$ times in parallel for the spinless system and $2n^{d-1}$ times in parallel for the spinful system. Hamming weight phasing can be applied across all these parallel circuits to reduce the number of arbitrary rotations.

\section{Costs per Trotter step}
\label{app:costs_per_step}

The number of T gates and arbitrary rotations per Trotter step figure significantly in the total costs of the two algorithms. For the fermionic swap network-based Trotter step, no T gates are directly required: instead, each application of the fermionic simulation gate requires up to 4 arbitrary rotations which must be compiled into T gates. On the other hand, for the split-operator algorithm, changing basis using either the FFFT or Givens rotations may itself require T gates and arbitrary rotations. While the split-operator algorithm is in the momentum basis, evolution under the entire kinetic energy part of the Hamiltonian can be implemented by an arbitrary rotation on each qubit. While the algorithm is in the position basis, each interaction between spin-orbitals requires an arbitrary rotation, as well as an additional arbitrary rotation on each qubit to implement evolution under the single-qubit $Z$ terms.

In this section, we discuss how to count the number of T gates and arbitrary rotations required per Trotter step for both the fermionic swap-based Trotter step and the split-operator Trotter step for all systems discussed in the paper.

\subsection{The fermionic swap network}

For the fermionic swap network simulating the uniform electron gas (jellium), the second-order (symmetric) Trotter step requires $4 (N-1)^2$ arbitrary rotations for $N$ qubits in the spinless case. Each application of the fermionic swap gate requires 4 arbitrary rotations (see \fig{directional_fsim}), and the fermionic swap gate is applied $2\binom{N}{2} = N(N-1)$ times, twice for each pair of orbitals because of the symmetric step. This would imply a cost at most $4 N(N-1)$ arbitrary rotations. However, we can slightly reduce this cost by noting that (i) the arbitrary rotations in the first and final layers of the fermionic swap network can be merged in the symmetric step, and (ii) for the case with spin, many interactions do not occur between electrons in different spin sectors. If there are an equal number of interactions between same-spin and opposite-spin sectors at the start and end of the Trotter step, there are $3N-4$ arbitrary rotation gates total in the first and last layer of the step which may be merged.

The first of these observations allows us to eliminate $4(N-1)$ arbitrary rotations from the fermionic swap network Trotter step. This reduces the number of arbitrary rotations to $4 (N-1)^2$ for the case of the uniform electron gas without spin. In the case of jellium with spin, the key for the second observation is that the kinetic energy terms $T_{pq}$ are zero if $\sigma_p \neq \sigma_q$, i.e.\ if $p$ and $q$ belong to different spin sectors. In this case, ${\cal F}_t(i_p, i_q)$ only requires 2 arbitrary rotations. Only considering the potential energy terms, the fermionic swap network requires $2N(N-1)$ arbitrary rotations. The kinetic energy only acts within the same spin sector. There are $\binom{N/2}{2} = N^2/8 - N/4$ pairs of orbitals within each spin sector requiring 2 arbitrary rotations for the kinetic energy terms, two spin sectors, and the symmetric Trotter step applies the corresponding rotations twice. In total then, the kinetic energy operator requires at most $N^2 - 2N$ arbitrary rotations. Combining this with the $2N^2-2N$ arbitrary rotations for the potential energy operator, the fermionic swap network requires at most $3N^2-4N$ arbitrary rotations for $N$ qubits for the system with spin. Using the improvement of the first observation, there are at most $(N-1)(3N-4)$ arbitrary rotations to perform. Materials are nearly identical to jellium, but add up to $N$ arbitrary rotations (one on each qubit) to the Trotter step.

For simulating the Hubbard model with spin, there are $N/2$ on-site interactions in the Hamiltonian (one between the two spin sectors for each of the $N/2$ spatial orbitals), and $2N$ hopping terms (two for each of the $N$ spin-orbitals in the system).  Each on-site interaction and each hopping term requires two arbitrary rotations to simulate.  Symmetrizing for the second-order Trotter step adds an additional $4N$ arbitrary rotations for the hopping terms; however, the interaction terms in the symmetric step can be combined.  Totaling these costs, at most $9N$ arbitrary rotations are required to simulate the Hubbard model.

\subsection{The split-operator Trotter step}

The costs for the split-operator Trotter step for the uniform electron gas and Hubbard model are as follows.  For both systems, applying the kinetic energy operator in the momentum basis requires $2N$ arbitrary rotations, two for each qubit for the symmetric Trotter step. Because all terms in the potential energy operator commute, each rotation can be performed just once even for the symmetric step: $\binom{N}{2} + N$ arbitrary rotations are required at most. While the uniform electron gas saturates this requirement, the Hubbard model only needs $3N/2$ arbitrary rotations for the potential.  As discussed in \app{FFTtrotterstep}, because we work with a large number of Trotter steps, the cost of simulating terms in the symmetric split-operator Trotter step is approximately the sum of the cost of simulating the kinetic terms and the cost of simulating the potential terms.  (For a single symmetric split-operator step, either the kinetic or the potential terms would be simulated twice, i.e.\ $\exp(-iHt) \approx \exp(-iTt/2) \exp(-iVt) \exp(-iTt/2)$ or $\exp(-iHt) \approx \exp(-iVt/2) \exp(-iTt) \exp(-iVt/2)$: when the number of steps is large, we can merge the halved term with the same term in the next Trotter step. We are thus neglecting in our counts the cost of the final simulation term, which cannot be merged in this way, but this is negligible when the number of Trotter steps is large.)  In total, then, implementing the terms in the uniform electron gas Hamiltonian requires $\binom{N}{2} + 3N$ arbitrary rotations while the Hubbard model requires $7N/2$ arbitrary rotations. 

We have not yet discussed the cost of changing bases. Changing between the position and momentum bases represents an additional cost for the split-operator Trotter step. Recall from \app{FFTtrotterstep} that the Fourier transform must be applied $dM^{d-1}$ times to transform a spinless system of side length $M$ in $d$ dimensions, and $2dM^{d-1}$ times for the same system with spin. When the side length is a power of two, the FFFT can be used, with a cost of 8 T gates for the 4-qubit case, 26 for the 8-qubit case, and 54 T gates and 4 arbitrary rotations for the 16-qubit case. Otherwise, the Givens rotation-based basis change algorithm must be used \cite{kivlichan2018quantum}. The basis change algorithm on $M$ qubits requires $2\binom{M}{2}$ arbitrary rotations. In either case, the rotation from the position to the momentum basis and its reverse must be applied twice for the symmetric Trotter step.

Recall from \app{combining} that Hamming weight phasing may be used to reduce the number of arbitrary rotations required for changing the basis, whether the FFFT or Givens rotation algorithm is used. The basis must be applied twice in the symmetric split-operator Trotter step, i.e.\ each of the following costs will be doubled in our final accounting.

For the $16\times16$ FFFT without spin, we can catalyze all non-Clifford rotations at a total cost of 2304 T and 64 Toffoli gates. The $16\times16$ FFFT with spin uses 4608 T and 128 Toffoli gates.

To diagonalize the kinetic energy operator using Givens rotations with Hamming weight phasing on a $d$-dimensional grid of side length $M$, the costs are as follows.  Without spin, $d \binom{M}{2} \left( \lfloor (d-1) \log_2 M\rfloor + 2 \right)$ arbitrary rotations must be synthesized; with spin, $d \binom{M}{2} \left( \lfloor (d-1) \log_2 M\rfloor + 3 \right)$ need to be synthesized.  This introduces an additional cost in terms of T and Toffoli gates.  Without spin, $d \binom{M}{2} (2M^{d-1} + 2 \lfloor (d-1) \log_2 M\rfloor)$ additional Toffoli gates are introduced by this procedure whereas with spin the number of additional Toffoli gates is $d \binom{M}{2} (4M^{d-1} + 2 \lfloor (d-1) \log_2 M\rfloor + 2)$.

\subsection{Gate counts per Trotter step with Hamming weight phasing}
\label{app:final_costs}

The Trotter step gate counts we use are actually slightly smaller than those discussed in the previous two subsections. There are three reasons for this, which we describe in order of increasing complexity. First, we do not include zero-rotations. Second, for both the fermionic swap and split-operator Trotter steps, we neglect the cost of the final term in the simulation (the final layer, for the fermionic swap Trotter step), since those arbitrary rotations can be merged with the first term or layer of the next Trotter step. Third, we use Hamming weight phasing with a fixed number of ancilla qubits to combine equiangular rotations where possible. We discuss these aspects further in \sec{trottersteps}.

\section{Improved second-order Trotter-Suzuki error bounds}
\label{app:new_bounds}

Bounds on the error in Trotterized evolution typically assume that the evolution time $t$ is small. Here, we present a tighter bound on the second-order Trotter error than that in Appendix B of \cite{wecker2014gate}, which, similar to the proof in that paper, does not require $t$ to be small. Broadly, our proof is quite similar overall to that in \cite{wecker2014gate}, and we adopt similar notation so as to make it easier to identify where they differ. 
However, our result is tighter by constant factors, matching the factor of $1/12$ from the Baker-Campbell-Hausdorff expansion discussed in \cite{poulin2015trotter}. Additionally, we are able to keep the norm on the outside of some of the sums, which is another reason it is tighter than the bound in \cite{wecker2014gate}. 
It has been pointed out that the bound described here can in principle be reached from the work in \cite{childs2019nearly} using slightly different methods. 

The second-order Trotter formula for $e^{-iHt}$ is given by \eq{2o_trotter},
\begin{equation}
U = e^{-iHt} \approx \left( \prod_{\ell=1}^L e^{-iH_\ell t/2} \prod_{\ell=L}^1 e^{-iH_\ell t/2} \right) \equiv U^\text{TS} = e^{-i H_{\rm eff} t}
\end{equation}
for the Hamiltonian $H = \sum_{\ell=1}^L H_\ell$. We first consider the error when $H$ is a sum of only two terms. 
We write $Ht=A+B$, and let $H(x)=B+(1-x)A$. Then the error in the Trotter formula for $A$ and $B$ can be written as
\begin{equation}
\begin{aligned}
&\left\| \exp(-iA/2) \exp(-iB) \exp(-iA/2) - \exp(-iH(0)) \right\| \\
= &\left\| \int_0^1 \partial_x \left(\exp(-ixA/2) \exp(-iH(x)) \exp(-ixA/2)\right) \mathrm dx \right\|.
\end{aligned}
\end{equation}
Now, define $A(\tvar,x)=\exp(iH(x)\tvar) A \exp(-iH(x)\tvar)$.
Then
\begin{align}
&\partial_x \left[\exp(-ixA/2) \exp(-iH(x)) \exp(-ixA/2)\right] \nn
&= \exp(-ixA/2) \left(-\frac i2 A \exp(-iH(x)) + \partial_x \exp(-iH(x)) -\frac i2 \exp(-iH(x)) A \right) \exp(-ixA/2)\, .
\end{align}
In bounding this, \cite{wecker2014gate} use 
\begin{align}
\partial_x \exp(-iH(x)) &= \int_0^1 \exp(-i(1-\tvar)H(x)) (iA) \exp(-i\tvar H(x)) \, \mathrm d\tvar \\
&= i\exp(-iH(x)) \int_0^1 A(\tvar,x) \, \mathrm d\tvar \, .\nonumber
\end{align}
Additionally, $A\exp(-iH(x))=\exp(-iH(x))A(1,x)$. That then gives the expression for the derivative
\begin{align}
\label{eq:trotter_derivative_bound1}
&\partial_x \left(\exp(-ixA/2) \exp(-iH(x)) \exp(-ixA/2)\right) \\
&= 
\frac{i}2 \exp(-ixA/2)\exp(-iH(x)) \left(-A-A(1,x) +2\int_0^1 A(\tvar,x) \, \mathrm d\tvar \right) \exp(-ixA/2)\, .\nonumber
\end{align}
Next, using $A(0,x)=A$ and
\begin{equation}
A(\tvar,x)=A(0,x)+\tvar A'(0,x)+\int_0^\tvar (\tvar-s) A''(s,x) \, \mathrm ds\, ,
\end{equation}
where $A'$ and $A''$ are the respectively the first and second derivatives of $A(\tvar,x)$ with respect to $\tvar$, we have that the terms in parentheses in \eq{trotter_derivative_bound1} are equivalent to
\begin{align}
-A-A(1,x) +2\int_0^1 A(\tvar,x) \, \mathrm d\tvar &= -A-A-A'(0,x) - \int_0^1 (1-s) A''(s,x) \, \mathrm ds \\
& \quad + 2\int_0^1 A \, \mathrm d\tvar +2 \int_0^1 \tvar A'(0,x) \, \mathrm d\tvar + 2\int_0^1 \mathrm d\tvar \int_0^\tvar \mathrm ds \, (\tvar-s) A''(s,x) \nn
&= -\int_0^1 (1-s) A''(s,x)\, \mathrm ds + 2\int_0^1 \mathrm ds \int_s^1 \mathrm d\tvar \, (\tvar-s) A''(s,x) \nn
&=  -\int_0^1 (1-s) A''(s,x)\, \mathrm ds + \int_0^1 (1-s^2-2s+2s^2) A''(s,x) \, \mathrm ds \nn
&= \int_0^1   (s^2-s) A''(s,x) \, \mathrm ds\nonumber
\end{align}
Evaluating the derivatives gives
\begin{align}
A'(s,x) = i \left( H(x) A(s,x) - A(s,x)  H(x) \right)
\end{align}
and
\begin{align}
\label{eq:appsx}
A''(s,x) &= - \left( (H(x))^2 A(s,x) + A(s,x) (H(x))^2 - 2 H(x) A(s,x) H(x) \right) \\
&= -\exp(iH(x)s) \left( (H(x))^2 A + A (H(x))^2 - 2 H(x) A H(x) \right) \exp(-iH(x)s)\, .\nonumber
\end{align}
Substituting in the expression for $H(x)$ gives
\begin{align}
&(H(x))^2 A + A (H(x))^2 - 2 H(x) A H(x) =
(B^2+(1-x)(AB+BA)+(1-x)^2 A^2) A \\ & \quad+ A(B^2+(1-x)(AB+BA)+(1-x)^2 A^2)
-2(BAB+(1-x)(BAA+AAB)+(1-x)^2 A^3) \nn
&= B^2A+AB^2-2BAB + (1-x)(ABA+BAA+AAB+ABA-2BAA-2AAB) \nn
&= B^2A+AB^2-2BAB + (1-x)(2ABA-BAA-AAB) \nn
&= [[A,B],B] + (1-x)[[A,B],A]\, .\nonumber
\end{align}
Combining these results in \eq{trotter_derivative_bound1}, we therefore have
\begin{align}
&\partial_x \left(\exp(-ixA/2) \exp(-iH(x)) \exp(-ixA/2)\right) \\
&= \frac{i}2 \exp(-ixA/2) \left(\int_0^1 ds \, (s-s^2) \exp(iH(x)(s-1)) \left( [[A,B],B] + (1-x)[[A,B],A] \right) \exp(-iH(x)s) \right) \times\nonumber \\&\quad~~~ \exp(-ixA/2).\nonumber
\end{align}
Therefore,
\begin{align}
\left\|\partial_x \left(\exp(-ixA/2) \exp(-iH(x)) \exp(-ixA/2)\right)\right\|&\le \frac{1}2 \int_0^1 ds \, (s-s^2) \left\| [[A,B],B] + (1-x)[[A,B],A] \right\| \\
&=  \frac{1}{12} \left\| [[A,B],B] + (1-x)[[A,B],A] \right\| .\nonumber
\end{align}
This has given a tighter bound by a factor of $1/12$ compared to that given in \cite{wecker2014gate}, which matches the factor from a BCH expansion. 
Lastly, integrating over $x$ gives
\begin{align}
&\left\| \exp(-iA/2) \exp(-iB) \exp(-iA/2) - \exp(-iH(0)) \right\|\nonumber \\ \le& \int_0^1 \left\|\partial_x \left(\exp(-ixA/2) \exp(-iH(x)) \exp(-ixA/2)\right)\right\| \mathrm dx \\
 \le&  \frac{1}{12}\int_0^1 \left\| [[A,B],B] + (1-x)[[A,B],A] \right\| \mathrm dx \nn
 \le& \frac{1}{12}\left\| [[A,B],B] \right\| + \frac 1{24} \left\|[[A,B],A] \right\| .\nonumber
\end{align}
Next, one can use the iterative approach from \cite{wecker2014gate}.
Let
\begin{align}
U_j &= \exp\left( -i \sum_{k=j}^L H_k t \right) \\
U_j^{TS} &= \exp\left( -i H_j \frac{t}{2} \right)\ldots \exp\left( -i H_{L-1} \frac{t}{2} \right) \exp\left( -i H_L \frac{t}{2} \right)
\exp\left( -i H_{L-1} \frac{t}{2} \right) \ldots \exp\left( -i H_j \frac{t}{2} \right) .
\end{align}
Using the triangle inequality
\begin{align}
&\left\| U_{j-1} - U_{j-1}^{TS} \right\| \\ &\le
\left\| U_{j-1} - \exp\left( -i H_{j-1} \frac{t}{2} \right)U_j\exp\left( -i H_{j-1} \frac{t}{2} \right) \right\|
+\left\| \exp\left( -i H_{j-1} \frac{t}{2} \right)U_j\exp\left( -i H_{j-1} \frac{t}{2} \right) - U_{j-1}^{TS} \right\| \nn
&=\left\| U_{j-1} - \exp\left( -i H_{j-1} \frac{t}{2} \right)U_j\exp\left( -i H_{j-1} \frac{t}{2} \right) \right\|+\left\| U_{j} - U_j^{TS} \right\| .\nonumber
\end{align}
Noting that $U_1=U$ and $U_1^{TS}=U^{TS}$, and $U_L=U_L^{TS}$, we have
\begin{align}
\left\| U - U^{TS} \right\| \le \sum_{b=1}^{L-1} \left\| U_{b} - \exp\left( -i H_{b} \frac{t}{2} \right)U_{b+1}\exp\left( -i H_{b} \frac{t}{2} \right) \right\| .
\end{align}
Now, using the bounds above with
\begin{equation}
A=H_b t, \qquad B=\sum_{c>b} H_c t,
\end{equation}
we have
\begin{align}
[[A,B],B] = t^3 \sum_{c>b} \sum_{a>b} [[H_b,H_c],H_a], \qquad \qquad
[[A,B],A] &= t^3 \sum_{c>b} [[H_b,H_c],H_b]\, .
\end{align}
This gives
\begin{align}
\left\| U - U^{TS} \right\| \le \frac{t^3}{12} \sum_{b=1}^{L-1} \left(\left\|\sum_{c>b} \sum_{a>b} [[H_b,H_c],H_a]\right\| + \frac 12 \left\|\sum_{c>b} [[H_b,H_c],H_b]\right\| \right) .
\end{align}

Lastly, since we are using the evolution to measure the energy, we need to bound that in terms of the difference in the unitaries.
The difference in the eigenvalues of unitaries may be upper bounded as \cite{bhatia2007bound}
\begin{equation}
\left|e^{-iEt}-e^{-iE_{TS}t}\right| \le \left\| U - U^{TS} \right\| .
\end{equation}
Using $\Delta\equiv \left\| U - U^{TS} \right\|$, for $\Delta^2\le 2$
\begin{align}
\label{eq:full_trotter_error_bound}
\left|E-E_{TS}\right| t \le \arctan \left(\Delta \frac{\sqrt{4-\Delta^2}}{2-\Delta^2}\right)= \Delta + \frac{\Delta^3}{24} + {\cal O}(\Delta^5).
\end{align}

\section{Phase estimation circuit primitives from a Trotter step}
\label{app:phase_estimation}

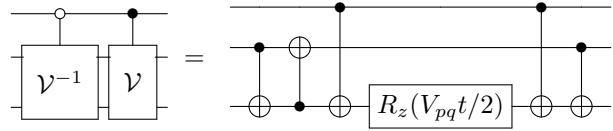
\begin{figure}[ht]
\begin{center}
\[\Qcircuit @R=1em @C=0.4em {
& \ctrlo{1} & \ctrl{1} & \qw \\
& \multigate{1}{{\cal V}^{-1}} & \multigate{1}{{\cal V}} & \qw \\
& \ghost{{\cal F}^{-1}} & \ghost{{\cal F}} & \qw
} \rule{.3em}{-0.2cm} \raisebox{-1.7em}{~=~} \rule{.3em}{-0.2cm} \raisebox{0.5em}{
\Qcircuit @R=1em @C=0.7em {
& \qw & \qw & \ctrl{2} & \qw & \ctrl{2} & \qw & \qw \\
& \ctrl{1} & \targ & \qw & \qw & \qw & \ctrl{1} & \qw 
\\
& \targ & \ctrl{-1} & \targ & \gate{R_z( V_{pq} t / 2)} & \targ & \targ & \qw
}}\]
\caption{\label{fig:directional_Vsim} Directionally controlled fault-tolerant circuit for interaction terms. 
}
\end{center}
\end{figure}

Phase estimation typically is assumed to use a controlled unitary rotation, however, \cite{wecker2015solving} showed that phase estimation can be made more efficient by eschewing controlled quantum evolutions in exchange for circuits that control the direction of the time evolution.  That is to say, we want a circuit such that $\ket{c}\ket{\psi} \mapsto \ket{c}e^{-i H t (-1)^c} \ket{\psi}$.  We build such a circuit by exploiting the structure of the Trotter decomposition.  If the symmetric Trotter formula is used then we can simulate the inverse time evolution by flipping the sign of each evolution in the decomposition.  That is to say, if $U_1\vcentcolon=\prod_{j=1}^M e^{-i H_j t} \prod_{j=M}^1 e^{-iH_j t}$ then $U_1^\dagger = \prod_{j=1}^M e^{i H_j t} \prod_{j=M}^1 e^{iH_j t}$. This does not hold for non-symmetric splittings such as the lowest-order Trotter formula.  Thus, we simply need to find a circuit for controllably reversing the direction of each $H_j$.  Such circuits are simple to produce at no extra cost (compared to the Trotter step circuits) for the potential terms; this allows implementation of the directionally controlled circuit for the split-operator Trotter step.  For that case, the directionally-controlled circuits can be constructed by adding a controlled-\textsc{not} gate (controlled on the phase estimation ancilla, targeting the qubit in question) on either side of each of the arbitrary-angle $R_z$ gates in the circuit.  
We show the directionally-controlled version of ${\cal V}_t$, the circuit for simulating the interaction terms $-V_{pq} Z_p Z_q/4$, in \fig{directional_Vsim}. 

The number of arbitrary rotations and T gates for the directionally-controlled circuit is thus the same as for the bare Trotter step.  However, integrating the kinetic energy terms between electrons in a constant spin manifold into this, and hence constructing the directionally-controlled fermionic swap Trotter step, is more challenging.

\begin{figure}[ht]
\scalebox{0.9}{
\Qcircuit @R=1em @C=0.75em {
&\ctrlo{1}                         &\ctrl{1}                         &\qw && &&&\qw &\qw      &\qw      &\qw      &\ctrlo{2} &\qw           &\ctrlo{2} &\qw       &\qw      &\qw       &\qw      &\ctrl{3} &\gate{R_z(V_{pq} t)} &\qw        &\qw &\gate{Z^{m_1 }}    &\qw&\\
&\multigate{1}{{\cal F}^{-1}}    &\multigate{1}{{\cal F}} &\qw && &&&\qw &\ctrl{1} &\gate{H} &\ctrl{1} &\targ     &\gate{R_z(\delta_{\sigma, \sigma^\prime} T_{pq} t)} &\targ     &\targ     &\targ    &\targ     &\ctrl{1} &\qw      &\qw             &\qw     & \qw   &\ctrl{1} &\qw&\\
&       \ghost{{\cal F}^{-1}}      &       \ghost{{\cal F}} &\qw &&=&&&\qw &\targ    &\targ    &\targ    &\targ     &\gate{R_z(\delta_{\sigma, \sigma^\prime} T_{pq} t)} &\targ     &\ctrl{-1} &\gate{H} &\ctrl{-1} &\ctrl{1} &\qw      &\qw             &\qw        &\qw &       \gate{Z^{m_1}} &\qw&\\
&                               &                            &    &    &                                                  &                                 &         &    &         &         &         &          &              &          &          &         &\lstick{\ket 0}&\targ    &\targ    &\gate{R_z(V_{pq} t)}   &\gate{H} & \meter & \rstick{m_1} \cw            &   &\\
}
}
\caption {
How to implement simultaneous ${\cal F}$ operations followed by fermionic swaps.
Note that the phasing operation on the control has negligible cost.
Although it may appear that there are ${\cal O}(M_k \cdot n^2)$ phasing operations applied to the control in the course of evaluating a Trotter step, all of the control phasing operations can be merged into a single phasing operation. \label{fig:directional_fsim}
}
\end{figure}
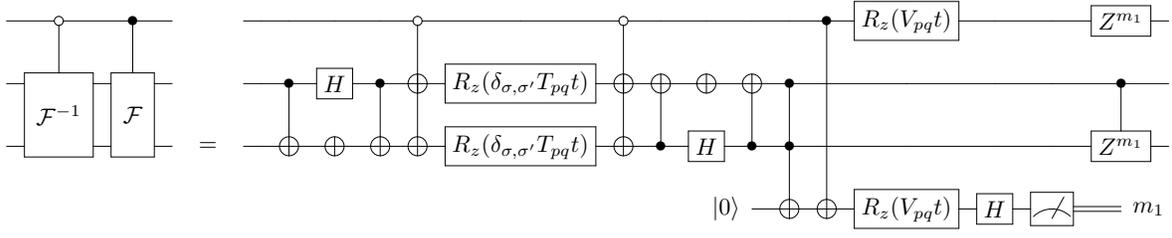

The simulation circuit in \fig{directional_fsim} achieves this directional control for the fermionic swap network. It follows from the circuit for $\mathcal{F}_t(i_p, i_q)$  
after noting that conjugating $R_z$ gates with \textsc{not} gates flips the direction of rotation.  The result given in the circuit follows from this observation and elementary identities for simplifying chains of \textsc{cnot} gates. Importantly, the number of arbitrary rotations required for the circuit is identical to that of the corresponding uncontrolled circuit.

\section{Trotter error numerics}
\label{app:trotter_error_numerics}

Thus far, we have discussed in the appendices the costs of single Trotter steps. We have presented this cost in terms of arbitrary rotations and T gates.  In this appendix, we discuss our computational strategy for determining the errors in Trotterizing evolution, enabling us to compute the total cost of the two algorithms using phase estimation.  We do this by computing the second-order Trotter error norm $W$ as in \eq{error_operator},
\begin{equation}
\left\| e^{-i H t} - e^{-i H_\text{eff} t} \right\|  \leq\frac{t^3}{12} \sum_{b=1}^{L-1} \left(\left\|\sum_{c>b} \sum_{a>b} [[H_b,H_c],H_a]\right\| + \frac 12 \left\|\sum_{c>b} [[H_b,H_c],H_b]\right\| \right) \equiv Wt^3,
\end{equation}
and thereby bounding the shifts in the energy as in \eq{error_operator}. Recall that we call $W$ the Trotter error norm. The key difficulty in bounding the Trotter error is the sheer number of terms in the Hamiltonian: if there are $L$ terms in the Hamiltonian, then $W$ is a sum of $L^3$ double commutators. 
Given that the number of terms in the uniform electron gas Hamiltonian scales quadratically with the number of spin-orbitals $N$, this represents at most ${\cal O}(N^6)$ double commutators that must be evaluated. 
In fact, the number of double commuators that are non-zero in this expansion is significantly smaller than this as each term in the nested commutators must share at least one index for the commutator to be non-zero. 
This results in ${\cal O}(N^4)$ non-trivial terms in the expansion of the error.
By comparison, the locality of the Hubbard model means there are only ${\cal O}(N)$ non-zero double commutators.

The Trotter error norms are computed as follows. For each system, the Trotterization schemes of \sec{trottersteps} (and further explained in \app{FFTtrotterstep} and \app{fermionicswaptrotternetwork})  give the ordered decomposition $\sum_{a} H_a$ in which the terms in the Hamiltonian are simulated. 
Note that the number of terms in the Trotter order is only the same as the number of terms in the Hamiltonian up to constant factors. For the fermionic swap network, the terms $a^\dagger_i a_j$ and $a^\dagger_j a_i$ can be grouped together at each stage, as can $a^\dagger_i a^\dagger_j a_i a_j$, $a^\dagger_i a_i$, and $a^\dagger_j a_j$. For the split-operator algorithm, all the terms in the kinetic energy can be grouped together in the sum, as can all the terms in the potential energy operator. As a result, the Trotter error norm for the split-operator Trotter step has only two terms. When the kinetic energy operator is simulated before the potential, the second-order Trotter error norm for the split-operator step is
\begin{equation}
\label{eq:TV_order}
W_{T+V} = \left \| [V, [V, T]] \right\| + \frac12 \left\| [T, [V, T]]  \right\|.
\end{equation}
On the other hand, when the potential is simulated first, the resulting Trotter error norm is
\begin{equation}
\label{eq:VT_order}
W_{V+T} = \left \| [T, [V, T]] \right\| + \frac12 \left\| [V, [V, T]]  \right\|.
\end{equation}
Depending on the norms of $T$ and $V$ this can yield different bounds on the Trotter error. So long as the number of Trotter steps $r$ is large, the number of gates in the simulation does not depend significantly on whether $T$ or $V$ is simulated first, and the user can freely choose the case yielding the smaller error. The fermionic swap network is less flexible in this regard.

Given these ordered sums, we then compute the double commutators, and total them to find the Trotter error norm. The order in the sums is specified by the Trotter step. Computing the error norms for larger sizes requires significant computational resources. For example, the Hamiltonian for one of the larger system sizes we study for the uniform electron gas ($12\times12$ with spin) has 47,952 terms in the Hamiltonian and 44,496 terms in the Trotter ordering for the fermionic swap network.

This means that computing $W$ requires the evaluation of as many as $10^{14}$ double commutators. We partially ameliorate this by performing preprocessing to determine which double commutators $[[H_b, H_c],H_a]$ are trivially zero and do not need to be evaluated. The Trotter error norm for the larger systems can require significant memory to load into memory ($>\!100$ GB). We work around this issue by parallelizing computation of across orthogonal components as follows. Writing the error norm as a sum of Pauli strings $\blu W = \sum_i a_i \|P_i\|$, let the smallest qubit index $P_i$ acts on be $q_i$. Then we can compute the sum of $|o_i|$ for which $q_i$ is some fixed value (i.e., the triangle inequality bound for a given qubit index), and the sum over $q_i$ of these values is equal to the full Trotter error norm. This allows us to control the maximum memory requirement of any partial computation.

\begin{table*}[tb]
\centering
\begin{tabular}{| c | c | c | c | c | c |}
\hline
UEG System & System Logical Qubits & $W_{FS}$ & $W_{SO}$  \\ \hline \hline
$4\times4$ spinless  &  16  & 2.2e+00 & 9.9e-01 \\ \hline
$5\times5$ spinless  &  25  & 6.5e+00 & 2.3e+00 \\ \hline
$6\times6$ spinless  &  36  & 3.2e+01 & 1.3e+01 \\ \hline
$7\times7$ spinless  &  49  & 5.7e+01 & 1.9e+01 \\ \hline
$8\times8$ spinless  &  64  & 1.9e+02 & 7.0e+01 \\ \hline
$9\times9$ spinless  &  81  & 2.8e+02 & 8.5e+01 \\ \hline
$10\times10$ spinless  &  100  & 6.9e+02 & 2.5e+02 \\ \hline
$11\times11$ spinless  &  121  & 9.8e+02 & 2.6e+02 \\ \hline
$12\times12$ spinless  &  144  & 2.0e+03 & 6.7e+02 \\ \hline
$13\times13$ spinless  &  169  & 2.6e+03 & 6.6e+02 \\ \hline
$14\times14$ spinless  &  196  & 4.9e+03 & 1.6e+03 \\ \hline
$15\times15$ spinless  &  225  & 6.2e+03 & 1.4e+03 \\ \hline
$16\times16$ spinless  &  256  & 1.0e+04 & 3.2e+03 \\ \hline
$4\times4$ spinful  &  32  & 1.4e+01 & 6.9e+00 \\ \hline
$5\times5$ spinful  &  50  & 3.3e+01 & 1.4e+01 \\ \hline
$6\times6$ spinful  &  72  & 1.6e+02 & 6.9e+01 \\ \hline
$7\times7$ spinful  &  98  & 2.6e+02 & 9.8e+01 \\ \hline
$8\times8$ spinful  &  128  & 8.5e+02 & 3.4e+02 \\ \hline
$9\times9$ spinful  &  162  & 1.2e+03 & 4.0e+02 \\ \hline
$10\times10$ spinful  &  200  & 3.0e+03 & 1.1e+03 \\ \hline
$11\times11$ spinful  &  242  & 3.8e+03 & 1.2e+03 \\ \hline
$12\times12$ spinful  &  288  & 8.3e+03 & 3.0e+03 \\ \hline
\end{tabular}
\caption{Trotter error data for simulating the uniform electron gas, for both the fermionic swap (FS) and split-operator (SO) Trotter steps, at Wigner-Seitz radius $r_s=10$ in 2D. $W_X$ denotes the Trotter error norm for algorithm $X$. Hartree-Fock energies $E_\text{HF}$ serve as an upper bound on the true ground state energies for the systems considered. Energies are in units of Hartree (Ha), Trotter error norms are in cubed Hartree (Ha$^3$). 
\label{tab:rawdata_ueg2d}
}
\end{table*}

\begin{table*}[tb]
\centering
\begin{tabular}{| c | c | c | c | c | c |}
\hline
UEG System & System Logical Qubits & $W_{FS}$ & $W_{SO}$  \\ \hline \hline
$3\times3\times3$ spinless  &  27  & 7.1e-03 & 2.4e-03 \\ \hline
$4\times4\times4$ spinless  &  64  & 1.7e-01 & 6.8e-02 \\ \hline
$5\times5\times5$ spinless  &  125  & 4.7e-01 & 1.6e-01 \\ \hline
$6\times6\times6$ spinless  &  216  & 2.6e+00 & 9.7e-01 \\ \hline
$2\times2\times2$ spinful  &  16  & 2.4e-03 & 1.5e-03 \\ \hline
$3\times3\times3$ spinful  &  54  & 2.6e-02 & 1.3e-02 \\ \hline
$4\times4\times4$ spinful  &  128  & 6.0e-01 & 2.5e-01 \\ \hline
$5\times5\times5$ spinful  &  250  & 1.5e+00 & 5.9e-01 \\ \hline
\end{tabular}
\caption{Uniform electron gas (UEG) Trotter error data, for both the fermionic swap (FS) and split-operator (SO) Trotter steps, at Wigner-Seitz radius $r_s=10$ in 3D with varying system side length. $W_X$ denotes the Trotter error norm for algorithm $X$. Hartree-Fock energies $E_\text{HF}$ serve as an upper bound on the true ground state energies for the systems considered. Energies are in units of Hartree (Ha), Trotter error norms are in cubed Hartree (Ha$^3$).
\label{tab:rawdata_ueg3d}
}
\end{table*}

\begin{figure}[ht]
\centering
\subfloat[][]{
\includegraphics[width=0.5\textwidth]{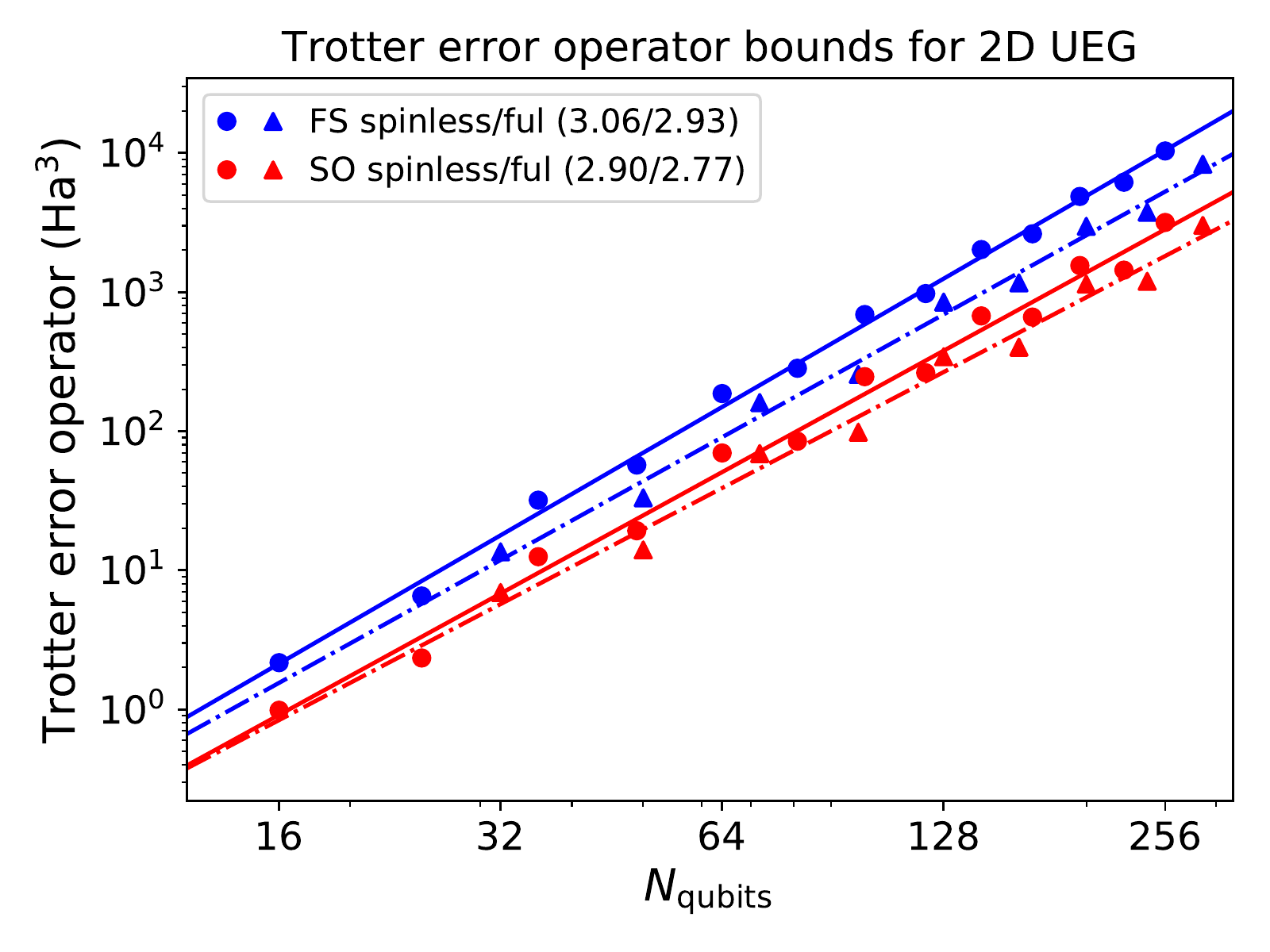}
\label{subfig2D_trotter_ueg}
}
\subfloat[][]{
\includegraphics[width=0.5\textwidth]{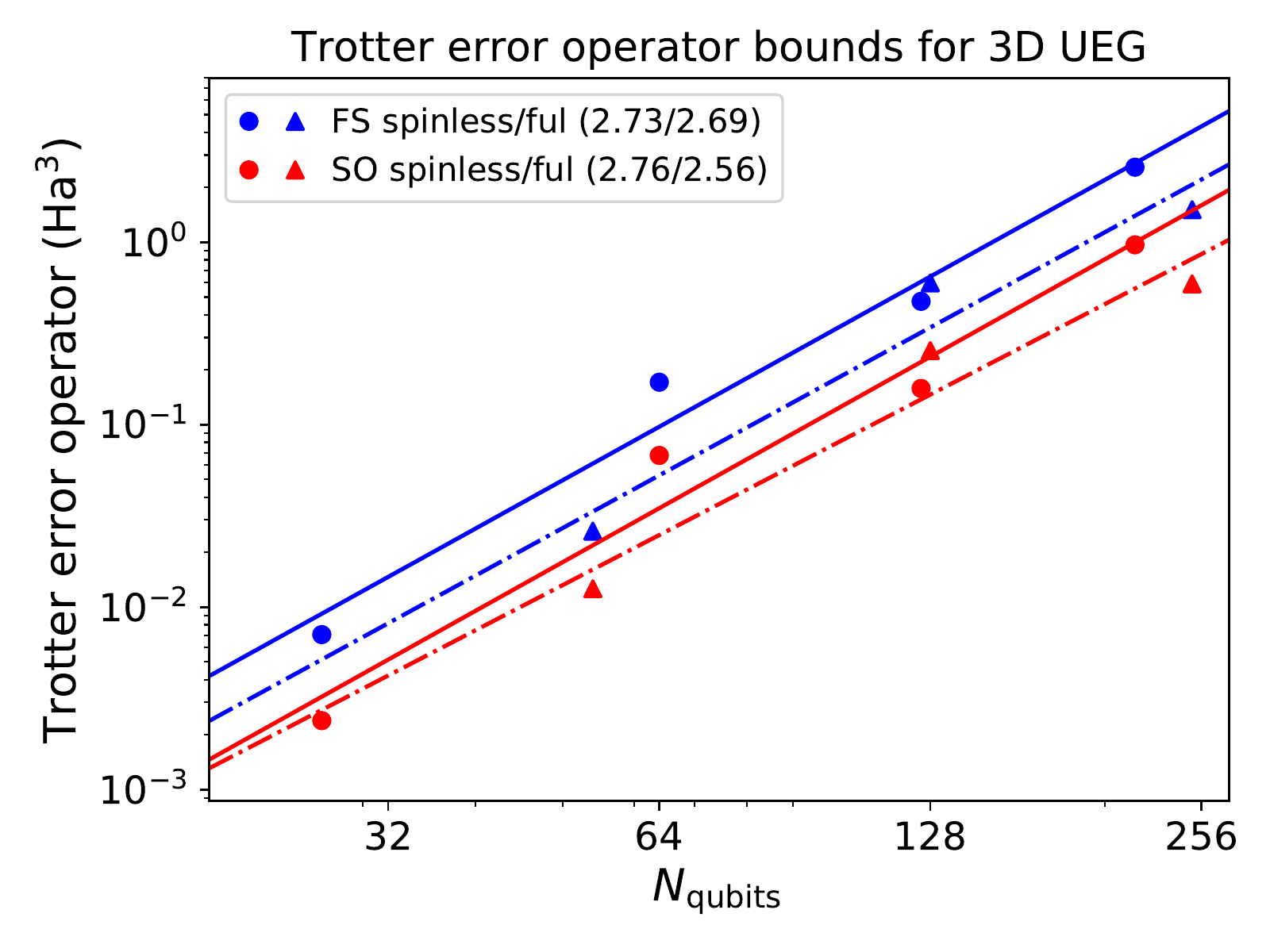}
\label{subfig3D_trotter_ueg}
}
\caption{
Trotter error norms for the uniform electron gas (UEG) Hamiltonian in \protect\subref{subfig2D_trotter_ueg} 2D and \protect\subref{subfig3D_trotter_ueg} 3D, for both spinful and spinless systems, in units of cubed Hartree (Ha$^3$). Lines show power law fits to the data, with the exponents of the fits shown in the caption for each system.
\label{fig:ueg_trotter}}
\end{figure}

\begin{table*}[htb]
\centering
\begin{tabular}{| c | c | c | c | c | c |}
\hline
Hubbard System & System Log.\ Qubits & $W_{FS}$, $U=4$ & $W_{SO}$, $U=4$ & $W_{FS}$, $U=8$ & $W_{SO}$, $U=8$ \\ \hline \hline
$4\times4$   &   32   & 1.9e+03 & 1.4e+03 & 4.3e+03 & 4.1e+03 \\ \hline
$5\times5$   &   50   & 2.9e+03 & 2.1e+03 & 6.7e+03 & 6.4e+03 \\ \hline
$6\times6$   &   72   & 4.2e+03 & 3.1e+03 & 9.6e+03 & 9.2e+03 \\ \hline
$7\times7$   &   98   & 5.7e+03 & 4.2e+03 & 1.3e+04 & 1.3e+04 \\ \hline
$8\times8$   &   128   & 7.5e+03 & 5.5e+03 & 1.7e+04 & 1.6e+04 \\ \hline
$9\times9$   &   162   & 9.4e+03 & 6.9e+03 & 2.2e+04 & 2.1e+04 \\ \hline
$10\times10$   &   200   & 1.2e+04 & 8.5e+03 & 2.7e+04 & 2.6e+04 \\ \hline
$11\times11$   &   242   & 1.4e+04 & 1.0e+04 & 3.2e+04 & 3.1e+04 \\ \hline
$12\times12$   &   288   & 1.7e+04 & 1.2e+04 & 3.8e+04 & 3.7e+04 \\ \hline
$13\times13$   &   338   & 2.0e+04 & 1.4e+04 & 4.5e+04 & 4.3e+04 \\ \hline
$14\times14$   &   392   & 2.3e+04 & 1.7e+04 & 5.2e+04 & 5.0e+04 \\ \hline
$15\times15$   &   450   & 2.6e+04 & 1.9e+04 & 6.0e+04 & 5.8e+04 \\ \hline
$16\times16$   &   512   & 3.0e+04 & 2.2e+04 & 6.8e+04 & 6.6e+04 \\ \hline
\end{tabular}
\caption{Hubbard model Trotter error data for $t=1$, with $U=4$ and $U=8$, for both the fermionic swap (FS) and split-operator (SO) Trotter steps, for a range of system sizes. $ W_X$ denotes the Trotter error norm for algorithm $X$.
\label{tab:hubbardtrotterdata}
}
\end{table*}

\begin{figure}[ht]
\centering
\includegraphics[width=0.5\textwidth]{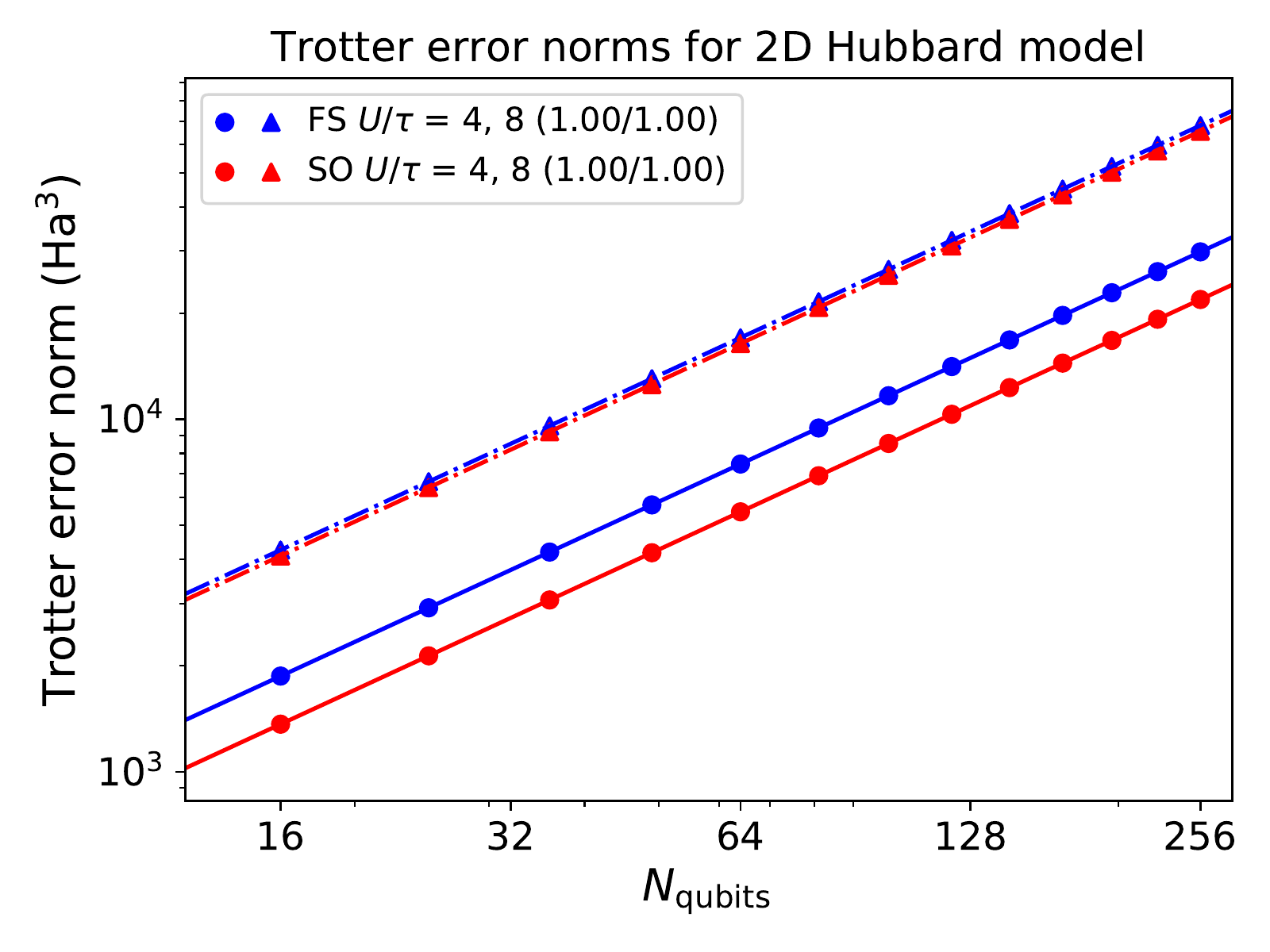}
\caption{
Trotter error norms for the Hubbard model in cubed Hartree (Ha$^3$). Lines show power law fits to the data, with the exponents of the fits shown in the caption for each system.
\label{fig:hubbard_trotter}}
\end{figure}

\begin{table*}[tb]
\centering
\begin{tabular}{| c | c | c | c | c | c |}
\hline
UEG System & System Logical Qubits & Wigner-Seitz Radius & $W_{FS}$ & $W_{SO}$  \\ \hline \hline
$2\times2\times2$ spinful  &  16  &  0.125  & 1.0e+07 & 9.2e+05 \\ \hline
$2\times2\times2$ spinful  &  16  &  0.25  & 2.1e+05 & 3.1e+04 \\ \hline
$2\times2\times2$ spinful  &  16  &  0.5  & 5.1e+03 & 1.1e+03 \\ \hline
$2\times2\times2$ spinful  &  16  &  1.0  & 1.4e+02 & 4.5e+01 \\ \hline
$2\times2\times2$ spinful  &  16  &  2.0  & 4.4e+00 & 2.0e+00 \\ \hline
$2\times2\times2$ spinful  &  16  &  4.0  & 1.6e-01 & 8.7e-02 \\ \hline
$2\times2\times2$ spinful  &  16  &  8.0  & 6.5e-03 & 3.9e-03 \\ \hline
$2\times2\times2$ spinful  &  16  &  16.0  & 3.1e-04 & 1.9e-04 \\ \hline
$2\times2\times2$ spinful  &  16  &  32.0  & 1.6e-05 & 1.1e-05 \\ \hline
$3\times3\times3$ spinful  &  54  &  0.125  & 1.9e+07 & 3.4e+06 \\ \hline
$3\times3\times3$ spinful  &  54  &  0.25  & 5.1e+05 & 1.4e+05 \\ \hline
$3\times3\times3$ spinful  &  54  &  0.5  & 1.6e+04 & 6.0e+03 \\ \hline
$3\times3\times3$ spinful  &  54  &  1.0  & 6.0e+02 & 2.6e+02 \\ \hline
$3\times3\times3$ spinful  &  54  &  2.0  & 2.5e+01 & 1.2e+01 \\ \hline
$3\times3\times3$ spinful  &  54  &  4.0  & 1.2e+00 & 5.8e-01 \\ \hline
$3\times3\times3$ spinful  &  54  &  8.0  & 6.6e-02 & 3.2e-02 \\ \hline
$3\times3\times3$ spinful  &  54  &  16.0  & 3.8e-03 & 1.8e-03 \\ \hline
$3\times3\times3$ spinful  &  54  &  32.0  & 2.3e-04 & 1.1e-04 \\ \hline
$4\times4\times4$ spinful  &  128  &  0.125  & 4.1e+08 & 6.8e+07 \\ \hline
$4\times4\times4$ spinful  &  128  &  0.25  & 1.1e+07 & 2.7e+06 \\ \hline
$4\times4\times4$ spinful  &  128  &  0.5  & 3.6e+05 & 1.2e+05 \\ \hline
$4\times4\times4$ spinful  &  128  &  1.0  & 1.3e+04 & 5.2e+03 \\ \hline
$4\times4\times4$ spinful  &  128  &  2.0  & 5.7e+02 & 2.3e+02 \\ \hline
$4\times4\times4$ spinful  &  128  &  4.0  & 2.8e+01 & 1.2e+01 \\ \hline
$4\times4\times4$ spinful  &  128  &  8.0  & 1.5e+00 & 6.4e-01 \\ \hline
$4\times4\times4$ spinful  &  128  &  16.0  & 8.7e-02 & 3.7e-02 \\ \hline
$4\times4\times4$ spinful  &  128  &  32.0  & 5.2e-03 & 2.2e-03 \\ \hline
$5\times5\times5$ spinful  &  250  &  0.125  & 4.7e+08 & 1.0e+08 \\ \hline
$5\times5\times5$ spinful  &  250  &  0.25  & 1.5e+07 & 4.6e+06 \\ \hline
$5\times5\times5$ spinful  &  250  &  0.5  & 5.7e+05 & 2.0e+05 \\ \hline
$5\times5\times5$ spinful  &  250  &  1.0  & 2.4e+04 & 8.9e+03 \\ \hline
$5\times5\times5$ spinful  &  250  &  2.0  & 1.2e+03 & 4.5e+02 \\ \hline
$5\times5\times5$ spinful  &  250  &  4.0  & 6.5e+01 & 2.5e+01 \\ \hline
$5\times5\times5$ spinful  &  250  &  8.0  & 3.8e+00 & 1.5e+00 \\ \hline
$5\times5\times5$ spinful  &  250  &  16.0  & 2.3e-01 & 8.8e-02 \\ \hline
$5\times5\times5$ spinful  &  250  &  32.0  & 1.4e-02 & 5.4e-03 \\ \hline

\end{tabular}
\caption{Uniform electron gas (UEG) Trotter error data, for both the fermionic swap (FS) and split-operator (SO) Trotter steps, for a range of systems in 3D with varying Wigner-Seitz radius and side length. $ W_X$ denotes the Trotter error norm for algorithm $X$. Hartree-Fock energies $E_\text{HF}$ serve as an upper bound on the true ground state energies for the systems considered. Energies are in units of Hartree (Ha), Trotter error norms are in cubed Hartree (Ha$^3$). For those systems for which $E_\text{HF}$ is positive, it is either exact or sufficiently large such that it is still a good proxy for $E_0$ for the purposes of relative precision.
\label{tab:rawdata_ueg3dsweep}
}
\end{table*}

\begin{table*}[tb]
\centering
\begin{tabular}{| c | c | c | c | c | c |}
\hline
Material System & System Logical Qubits & $W_{FS}$ & $W_{SO}$ & $E_\text{HF}$ \\ \hline \hline
Li $2\times2\times2$  &  16  & 1.3e+02 & 4.1e+01 & -3.0e+00 \\ \hline
Li $3\times3\times3$  &  54  & 4.7e+03 & 2.4e+03 & -4.7e+00 \\ \hline
Li $4\times4\times4$  &  128  & 3.3e+05 & 1.4e+05 & -5.1e+00 \\ \hline
Diamond $2\times2\times2$  &  16  & 9.7e-01 & 5.4e-01 & -1.5e+01 \\ \hline
Diamond $3\times3\times3$  &  54  & 3.0e+02 & 1.5e+02 & -2.5e+01 \\ \hline
Diamond $4\times4\times4$  &  128  & 6.9e+03 & 3.1e+03 & -2.8e+01 \\ \hline
Si $2\times2\times2$  &  16  & 3.1e-01 & 1.6e-01 & -5.2e+01 \\ \hline
Si $3\times3\times3$  &  54  & 6.6e+01 & 3.3e+01 & -7.7e+01 \\ \hline
Si $4\times4\times4$  &  128  & 1.2e+03 & 5.4e+02 & -8.8e+01 \\ \hline
LiH $2\times2\times2$  &  16  & 4.0e-01 & 2.4e-01 & -2.6e+00 \\ \hline
LiH $3\times3\times3$  &  54  & 1.4e+02 & 7.1e+01 & -3.8e+00 \\ \hline
LiH $4\times4\times4$  &  128  & 3.5e+03 & 1.6e+03 & -4.4e+00 \\ \hline
Graphite $2\times2\times2$  &  16  & 1.0e+02 & 1.4e+01 & -2.8e+01 \\ \hline
Graphite $3\times3\times3$  &  54  & 3.5e+03 & 4.1e+02 & -3.4e+01 \\ \hline
Graphite $4\times4\times4$  &  128  & 1.9e+05 & 2.9e+04 & -5.2e+01 \\ \hline

\end{tabular}
\caption{Trotter error data for both the fermionic swap (FS) and split-operator (SO) Trotter steps for a range of materials in 3D, including Hartree-Fock energies. $ W_X$ denotes the Trotter error norm for algorithm $X$. Energies are in units of Hartree (Ha), Trotter error norms are in cubed Hartree (Ha$^3$).
\label{tab:rawdata_materials}
}
\end{table*}

Trotter error norms in units of cubed Hartree (Ha$^3$)  
are included for the uniform electron gas in 2D in \tab{rawdata_ueg2d} and in 3D in \tab{rawdata_ueg3d}. We additionally plot the Trotter error norm for these systems in \fig{ueg_trotter}. 
\tab{hubbardtrotterdata} shows Trotter error norm data for the Hubbard model in 2D (with spin). The same data is plotted in \fig{hubbard_trotter} with a power law fit. 
We include the same data for a range of Wigner-Seitz radii in 3D in \tab{rawdata_ueg3dsweep} and for a series of materials in \tab{rawdata_materials}. For the materials, we include in the table the Hartree-Fock energies computed in PySCF \cite{sun2018pyscf} which we use for our relative precision energy target.

All numerics (for all systems and for the different Trotter steps) were performed using code which we have contributed to the open-source package OpenFermion \cite{openfermion2017}.

\section{T-count minimization}
\label{app:t_count}

Continuing from the discussion in \sec{resourceanalysis}, we present in this appendix our final numerics to compute the total number of T and Toffoli gates required for our simulations. We wish to minimize the total number of T and Toffoli gates needed to determine the ground state energy to precision $\Delta E \ge \Delta E_{TS} + \Delta E_{PE} + \Delta E_{HT}$ using phase estimation. $\Delta E_{TS}$ accounts for Trotter-Suzuki errors, $\Delta E_{PE}$ for phase estimation errors, and $\Delta E_{HT}$ for circuit synthesis errors. Throughout, we convert Toffoli to T gates at a cost of 2 T gates per Toffoli gate \cite{gidney2018efficient}.

The tradeoff between these error sources is non-trivial because the T-costs differ significantly as a function of the desired accuracy for each type of error. We discuss how to compute the T-cost as a function of these errors, and give the results of its numerical minimization subject to \eq{eq_constraint} for a variety of system sizes, in \sec{t_gates}. Our full objective function there was given in \eq{full_t_cost}, as
$$
(N_r N_{HT} + N_d) N_{PE} \approx  \frac{0.76 \pi \sqrt{W} }{\Delta E_{PE}\sqrt{\Delta E_{TS}}} \left( N_r \left[1.15\log_2\left(\frac{N_r \sqrt{W}}{\Delta E_{HT} \sqrt{\Delta E_{TS}}}\right)+9.2\right] + N_d \right),
$$
where $W$ is the Trotter error norm, $N_r$ is the number of arbitrary rotations per Trotter step, and $N_d$ is the number of ``direct'' T and Toffoli gates (those T/Toffoli gates which appear in the circuit before synthesis) multiplied by $N_{PE}$.

Data on the Trotter error norms $W$ for all systems and algorithms are given in \app{trotter_error_numerics}, and our approach to computing the numbers of arbitrary rotations $N_r$ and T/Toffoli gates which appear in the circuit before compiling arbitrary rotations (``direct'' T/Toffoli gates) $N_d$ is discussed in \sec{trottersteps} as well as in \app{costs_per_step}.   
We work with two total precision targets $\Delta E$ for all the systems we analyze. In the first, we consider a relative precision, with the total precision $\Delta E$ set as a fraction (0.5\%) of the ground state energy. This value is comparable to errors associated with the sign problem in quantum Monte Carlo \cite{tanatar1989ground,shepherd2012full}. The precision $\Delta E$ in this case scales with the system size. In the second case, we consider an absolute precision target set depending on the type of system, but independently of the particular system size. 

\begin{figure}[ht]
\centering
\subfloat[][]{
\includegraphics[width=0.4\textwidth]{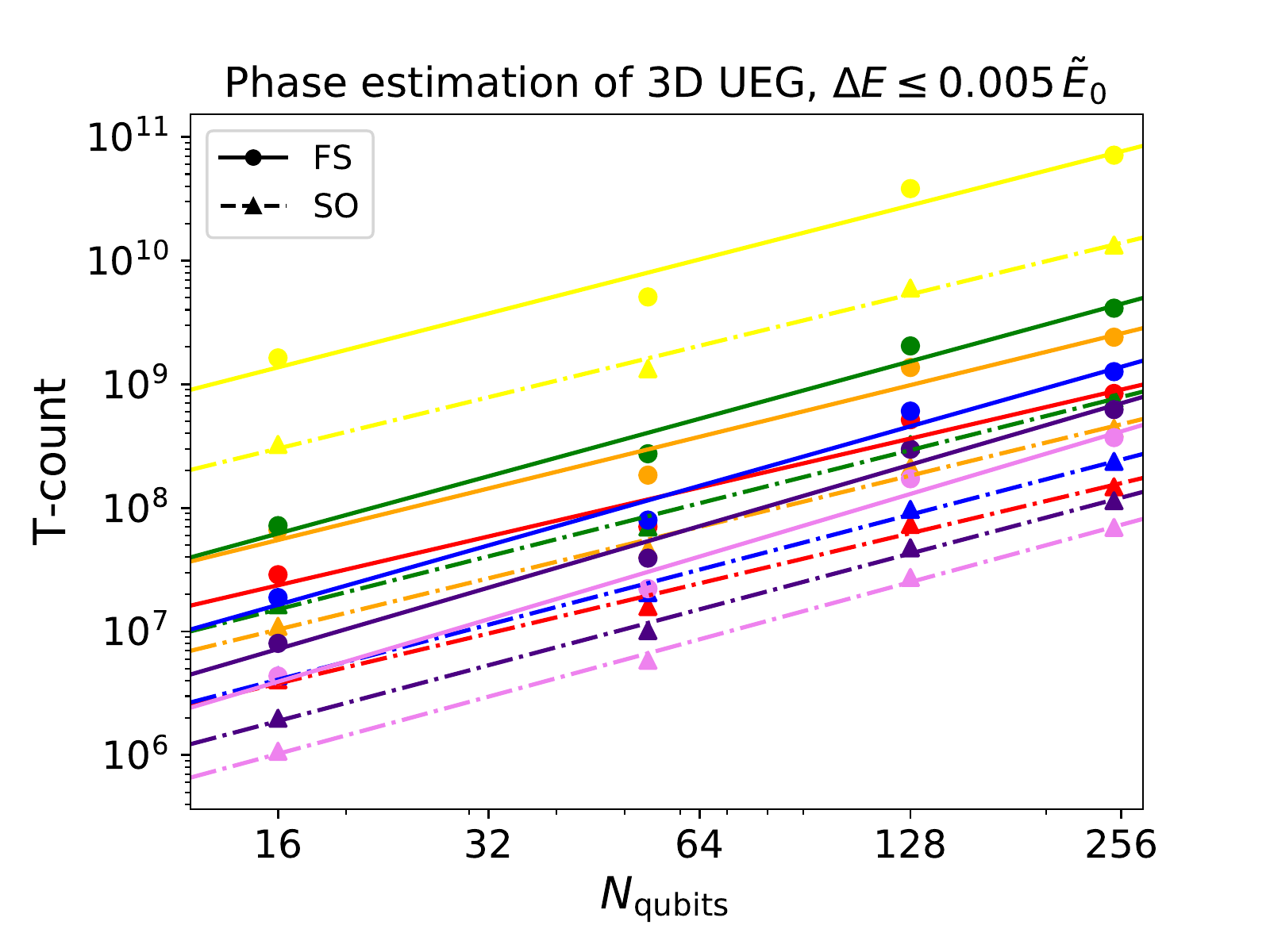}
\label{fig:TcountUEGsweep_rel}
}
\subfloat[][]{
\includegraphics[width=0.4\textwidth]{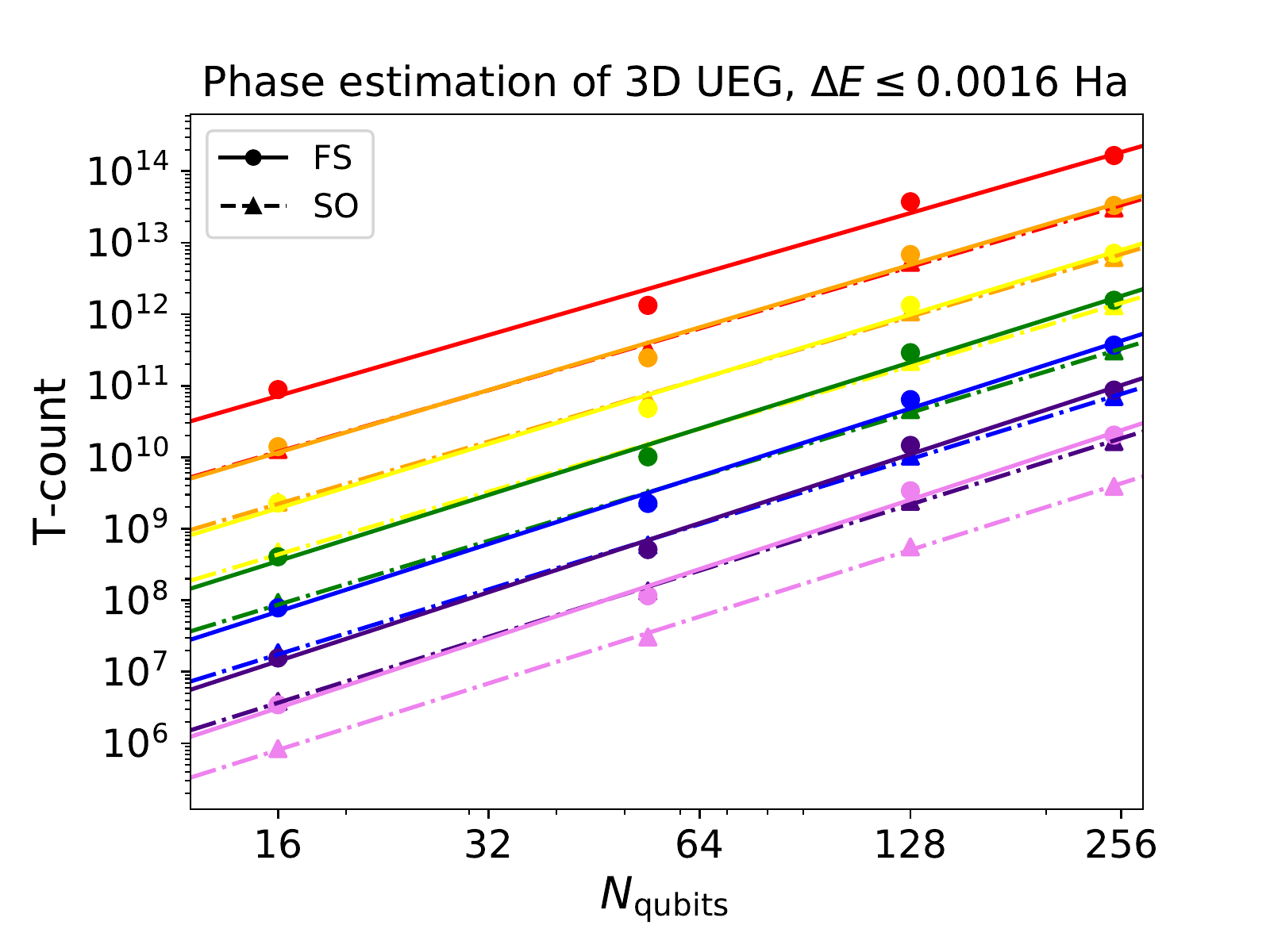}
\label{fig:TcountUEGsweep_abs}
}
\hspace{-0.5cm}
\subfloat{
\includegraphics[width=0.18\textwidth,clip,trim={6cm 1.5cm 5cm 2cm}]{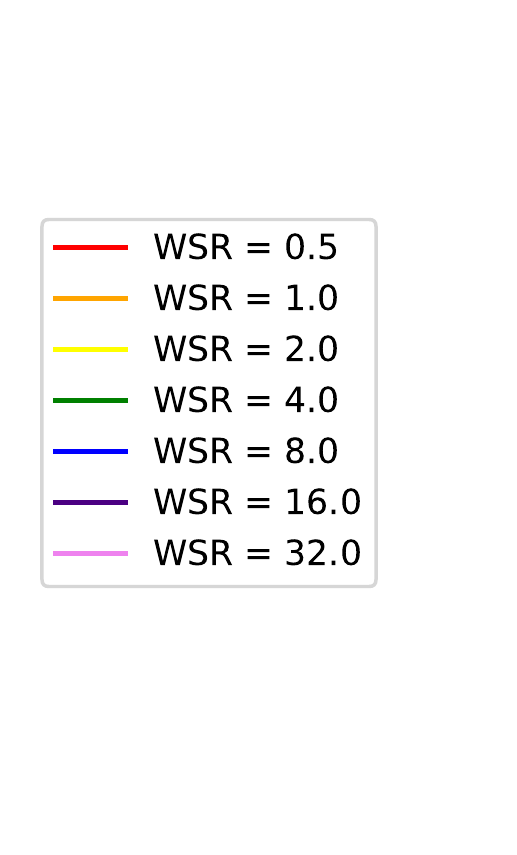}
\label{fig:TcountUEGsweep_legend}
}

\caption{Loose but rigorous upper bound on the number of T gates in the circuit which performs our Trotterized phase estimation simulation of the 3D uniform electron gas, for a range of values of the Wigner-Seitz radius, to \protect\subref{fig:TcountUEGsweep_rel} within half a percent of the total system energy and to \protect\subref{fig:TcountUEGsweep_abs} absolute precision $\Delta E = 0.0016$ Hartree. The colors of the rainbow (red through violet) correspond to logarithmically spaced Wigner-Seitz radii from 0.5 to 32 Bohr radii (0.5, 1, 2, $\ldots$, 16, 32). Lines show power law fits to the data. 14 ancilla qubits are used for Hamming weight phasing for all systems. \label{fig:TcountUEGsweep}}
\end{figure}

For the relative precision case for the jellium systems, since we cannot directly compute the ground state energy, we take $\Delta E$ to be 0.5\% of the system energy, computed as the number of electrons multiplied by the energy per electron using the correlation energy estimates of \cite{chachiyo2016simple}. In atomic units, the energy per electron takes the form
\begin{equation}
\frac{\tilde E_0(r_s)}{\eta} \approx \frac35 \left( \frac{9 \pi}{4}\right)^{2/3} \frac{1}{r_s^2} - \frac32 \left( \frac{9 \pi}{4}\right)^{1/3} \frac{1}{r_s} + a \ln\left(1 + \frac{b}{r_s} + \frac{b}{r_s^2}\right).
\end{equation}
We use the values $a = (\ln 2 - 1) / (2\pi^2)$ and $b = 20.4562557$ over the entire range of densities. This closely agrees with quantum Monte Carlo simulations in the low-density regime, and lower-bounds the magnitude of the ferromagnetic energy in the high-density regime. We then take $\Delta E$ to be 0.5\% of this value, after multiplying by the number of electrons ($\eta$ in the expression above). We use these energies in computing the T-cost for the uniform electron gas at a Wigner-Seitz radius of 10 Bohr radii in \fig{TcountUEG}, as well as for a range of Wigner-Seitz radii below in \fig{TcountUEGsweep}. For the material systems, we take as our relative precision target the Hartree-Fock energy $E_\text{HF}$ (given in \app{trotter_error_numerics}). We compute these using PySCF \cite{sun2018pyscf}. The true ground state energy is $E_0 \le \tilde E_0 = E_\text{HF}$ in general, and $E_\text{HF} < 0$ for all material systems considered, so choosing $\Delta E = 0.005 \,E_\text{HF}$ ensures that the precision better than 0.5\% of the magnitude of the ground state energy.

\begin{figure}[htb]
\centering
\includegraphics[width=0.5\textwidth]{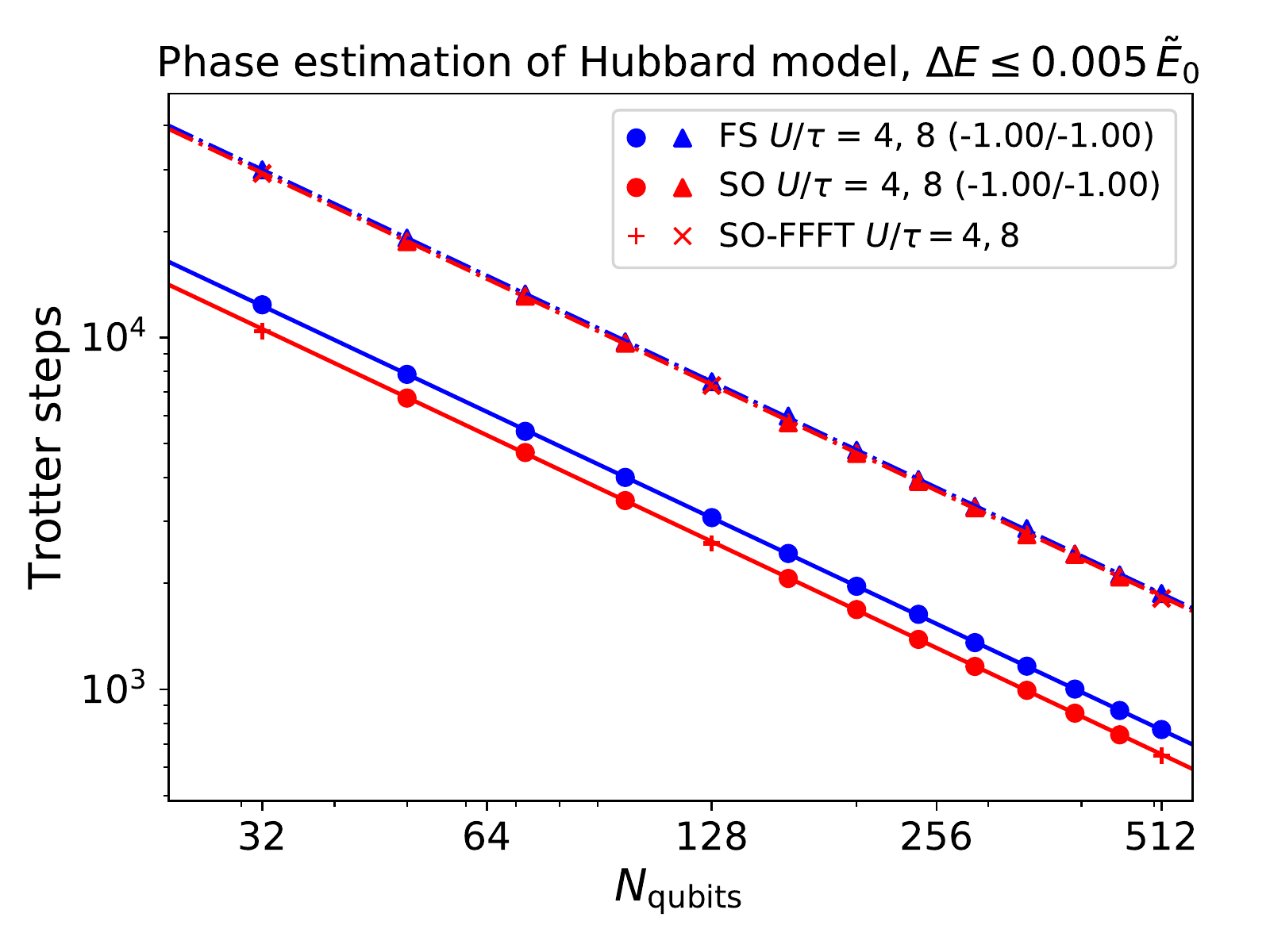}
\label{fig:TrotterstepsFHrel_2D}
\caption{\label{fig:Trotterstepsrel_2D}
Power-law fit to the number of Trotter steps used for phase estimation, with 14 ancilla qubits used for Hamming weight phasing, for 
the Hubbard model in 2D. The exponents of the fit are included in the caption. The fits intercept 1 well above 100,000 system qubits, far above the range of system sizes we consider.}
\end{figure}

Finally, for our relative precision target for the Hubbard model, we choose $1.02$ as an upper bound on the magnitude of the energy per site at $U/\tau=4$ and $0.74$ as an upper bound on the magnitude of the energy per site at $U/\tau=8$, following Table V of \cite{leblanc2015solutions}; again, this ensures that the precision tolerance is smaller than 0.5\% of the total ground state energy. For our absolute precision targets, we take $\Delta E = $ 0.0016 Hartree for jellium and the material systems (this precision is known as ``chemical accuracy''), and $\Delta E = \tau/100$ for the Hubbard model.

A possible criticism of the results of the optimization for the Hubbard model is that the number of Trotter steps decreases with system size in the relative precision case. Because of this, the scalings we present in \fig{TcountFHrel} do not hold in the asymptotic limit; rather, they only apply for a finite range of system sizes. 
We perform a power-law fit on the number of Trotter steps for the Hubbard model. This indicates that the system size at which the number of Trotter steps required goes to one is hundreds of thousands of spin-orbitals, even in the smallest case. We plot the number of Trotter steps with the power-law fit for the Hubbard model in \fig{Trotterstepsrel_2D}.

\end{document}